\documentstyle[11pt]{article}

\newtheorem{pro}{Proposition}

\begin{document}
\title{Invalidity of the relativity principle and a proposal of 
the twofold metric principle}
\author{Kosaburo Hashiguchi \\ Department of Information Technology \\
Faculty of Engineering, Okayama University \\
Tsushima, Okayama 700, Japan}
\maketitle

\noindent PACS index : 03.30 (relativity theory)

\noindent Telephone : +81-86-255-8821

\noindent Telefax : +81-86-255-9136

\noindent e-mail  :  hasiguti@kiso.it.okayama-u.ac.jp

\noindent Mailing address :

Kosaburo Hashiguchi

Department of Information Technology 

Faculty of Engineering, Okayama University

Tsushima, Okayama 700, Japan

\vspace{\bigskipamount}

\begin{abstract}
In this paper, we first show that all inertial
systems are not equivalent,
and the Lorentz transformation is not the
space-time transformation over two inertial systems moving with relative
constant velocity.
To do this, we consider imaginary signals travelling over any inertial
system $K$ with arbitrarily large velocities.  The travelling of an imaginary
signal over $ K $ is just a time lapse over $ K $.  Then we present
an example to show that all coordinate systems are not equivalent when
the related theory is restricted over tensor-based 
coordinate transformations, i.e.,
the genereal relativity principle is not valid.  Instead of the relativity 
principle, we propose the twofold metric principle which may be roughly
stated to assert that the set of equations $ H(v) $ describing the motion of
a material body with velocity $ v> 0 $ can be obtained from the 
corresponding set of equations $ H(0) $ for velocity $ v=0 $ by
replacing, in each differential equation in $H(0)$,
each infinitesimal time variable $ dt $
with $ dt / \beta(v) $, each maximal velocity-critical infinitesimal
length variable $ dr $ with  $ \beta(v) dr $, and each zero
velocity-critical infinitesimal length variable $ dx $ with $ dx $,
where $ \beta(v) =
1/ \sqrt {1- v^2/c^2} $. By depending on the twofold metric principle and
the energy-velocity equation, we derive $ \beta(v)mc^2 $,
the travelling distance of a muon with velocity $ 0.999c$, the twofold
Schwarzshild metric, the centennial procession of
planatery orbits and deflection of light.  We also present a reason why
the Michelson-Morley experiment is observed.  Several other topics are
also studied.
\end{abstract}

Key words : 

\noindent relativity principle, inertial system, Lorentz transformation,
time-space metric

\newpage

\sloppy
\section{Introduction}

An inertial system is a coordinate system in which the Newton equation
of motion holds.  The sepecial relativity principle asserts that all inertial 
systems are equivalent.  This principle plays key roles in [3] in which 
Einstein
constructs foundations of special relativity theory.  By depending on
the relativity principle and the constant light velocity principle,
Einstein develops arguments for supporting the validity of the Lorenz
transformation over two inertial systems moving with relative constant
velocity, observes time dilation and length contraction, and derives
transformation formulae of the elctromagnetic fields over two inertial
systems as above by studying the Maxwell equations from  
viewpoints  of the special
relativity principle.
Einstein develops general relativity theory in another significant
paper[5], which is based on the genareal relativity principle and tensor-based
time-space metrics.
Since its origin, relativity theory has attracted scientists and other
people theoretically and philosophically.  Now it seems that there do
not exist many people who have any doubts about the validity of the relativity
principle.  Although there have appeared "paradoxes" in the literature
such as the twin paradox and the garage paradox, it seems to be generally
acknowledged that these paradoxes are solved theoretically by being
given suitable explanations.

The main purpose of this paper is to show that
all inertial systems are not equivalent,
present a new principle called
the twofold metric principle and show how the new principle works well
for solving many problems.  We also show that the general relativity
principle asserting all coordinate systems are equivalent (w.r.t. 
tensor-based theory) is not valid by presenting an example
showing that when a proper 
coondinate system $K$ is transformed to another coordinate system 
$K'$, then the metric is changed so that important
principles such as the variational principle cannot be applied over $K'$,
and so we conclude that $K$ and $K'$ are not equivalent.

In the Lorentz transformation, the time at any point at rest over an
inertial system depends both on the time and the spacial coordinate
of the corresponding point at rest another inertial system.  We shall
show that the Lorentz transformation is not valid because it does
not preserve simultaneity, i.e., if the space-time transformation
from a coordinate system $ K $ to another coordinate system $ K'
$ is the nonidentity Lorentz transformation, then $ K $ or
$ K' $ is not synchrinized, i.e., $K$ or $K'$ is not an inertial
system.  
Especially we shall prove the following (see Theorem 1 in Section 4).

\vspace{2ex}

{\bf Fact A.}  Let $ K $ and $ K' $ be two inertial systems moving
with a relative constant velocity. The time at any point $ P $ at rest
on $ K $ depends only on the time at the corresponding point $ R $ at 
rest over $ K' $, and does not depend on the spacial coordinate
of $ R $.

\vspace{2ex}

One can see easily that if the Lorentz transformation is valid, then 
the following hold (see Remark 3 in Section 4) : (1) the flying of 
a small firework using a  small amount of energy over an inertial system
can correspond to a
travelling in an arbitrarily large diatance and an arbitrarily large time
lapse over another inertial system ; (2) a small diatance and the zero
time lapse over an inertial system can correspond to an arbitrarily large
diatance and an arbitrarily large time lapse, respectively, over 
another inertial system, and hence the (present) age of the universe is
infinite, and the whole space is also infinite.  We may also observe that
if two coordinate systems $K$ and $K'$ are related by 
the nonidentity Lorentz transformation,
then the light velocity is $c$ over both $K$ and $K'$, but the relation
of past and future over $K$ is different from the corresponding relation
over $K'$ (see Remark 1 in Section 2).
Thus in a sense, one may say $K$ and $K'$ are not equivalent.
From these observations, we cannot
help concluding the Lorentz transformation is not valid.
Morewever we shall present a more formal
proof showing invalidity of the Lorentz transformation.  To do this,
we introduce the notion of an imaginary signal  $ S(K,u) $ travelling
over an inertial system $ K $ with velocity $ u>0 $, where $ u $
can be arbitrarily large.  The travelling of $ S(K,u) $ from $ P $
at time $ t $ to point $ Q $ both at rest on $ K $ is just
a time lapse over $ K $ from time $ t $ at $ P $ to time 
$ t + l/u $ at $ Q $, where $ l $ is the distance between
$ P $ and $ Q $.  The travelling of $ S(K,u) $ from point $ P $
at time $ t $ to $ Q $ is in a sense "observable" because we
can considr by a Gedankenexperiment the corresponding continuous
time lapse over the line segment cennecting $ P $ and $ Q $.
Thus the notion of imaginary signals are meaningful mathematically
and physically, and the travelling of an imaginary signal
can be regarded to be a physical phenomenon which
really occurs in the universe.

The Lorentz transformations
are bijections and of the form of linear combinations of coordinates
and compose a group.  It seems that the light signals are
not sufficiently rich tools for analyzing whether or not the Lorentz
transformations are valid since the light velocity is finitely bounded.
Consider Achilles-tortoise running race problem.  If one does not notice
the fact that in certain cases, an infinite sum of time lapses $
\sum_{n=0}^{\infty} t_n $ is finite, then one cannot solve 
this problem.
Here we consider an arbitrarily small time lapse $ t_n
$ for a sufficiently large $ n $.  In the same way, if we depends only
on the travelling of light signals for analyzing the synchronization
problem over inertial systems, then we could not show that the Lorentz
transformation does not preserve simultaneity.  We may need the notion of
arbitrarily small time lapses, i.e., the travellings of imaginary signals
with arbitrarily large velocities to solve the synchronization
problem.

For any inertial
system $ K $ and any arbitrarily large real $ u >0 $, the
imaginary signal $ S(K,u) $ travels over $ K $ with 
velocity $ u $.
We develop arguments similar to the well known ones concerning the 
possibility of the existence of time machines , and establish Time
Independence Lemma (in short, TI Lemma, i.e., Lemma 3 in Section 4)
and Time Independence Theorem (in short, TI Theorem, i.e., Theorem 1
in Section 4).  In the Lorentz
transformation, imaginary signals with large velocities over a
coordinate system can go into past histories over another coordinate system.
TI Lemma and TI Theorem
imply that such phenomena never occur over inertial systems and Fact A
holds.  For any two
inertial systems $ K $ and $ K' $ moving with nonzero relative constant
velocity, the travelling of an imaginary signal $ S(K,u) $ over $ K
$ corresponds to a time lapse within which $ K $ and $ K' $ 
continues moving each other in the opposite directions.  The
travelling of $ S(K,u) $ corresponds to a physical phenomenon which
occurs in our universe.
We remark that one usually consider the notion of infinitesimal time lapses,
$ \lim _{ \triangle t \rightarrow 0} $, 
when he analizes  differential equations
describing the motions of material bodies or the travellings of
electromagnetic waves.  Without the notion of $ \lim_{\triangle t
\rightarrow 0} $, one can establish neither any theory of calculus nor
any control theory.
By depending on
TI Theorem, we study properties of inertial systems.  Some
of the results are the following :  (1) the Lorentz transformation is
not the space-time transformation over two inertail systems moving
with nonzero ralative constant velocity ; 
(2) all inertial systems are not equivalent
; (3) the Maxwell equations over an inertial system moving with
nonzero constant velocity in the vacuum space
are not of the same form as those over stationary inertial systems ;
(4) the velocity of a light signal measured over an inertial system
moving with nonzero constant velocity in the vacuum space 
is not constant but depends on its direction.

It is generally acknowledged that the expression
"stationary in the vacuum space" 
is meaningless since the special relativity principle holds.  However, in this 
paper, we understand that an inertail system $K$ is stationary in the vacuum
space (or, equivalently, stationary w.r.t. the universe) if there 
does not exist any material body of huge mass near $ K $ (i.e., with
(almost) zero gravitational potential)
and the Maxwell
equations over $K$ are of the same form as those in ''the vacuum space''
(the standard ones presented in any textbooks). Thus over a stationary
inertial system, the light velocity is $ c $ independent of its
direction. Since a light signal is a wave, its travelling is
in a sense independent from the source from which the signal is emmitted
and depends only on the space.
Thus light signals can be regarded to play roles as
instruments measuring the volocity of any material body in the space 
at least locally.
In the sequel, we also use often the words "
in the vacuum space" instead of "over an inertial system stationary
in the vacuum space".

Instead of the ralativity principle, we shall propose the twofold metric
principle.  We also propose the energy-velocity equation.  The
twofold metric principle may be roughly stated to assert that the set
of equations $ H(v) $ describing the motion of a material body
with velocity $ v> 0 $ can be obtained from the corresponding set
of equations $ H(0) $ for velocity zero by replacing, in each differential
equation in $H(0)$, each
infinitesimal time variable $ dt $ with $ dt / \beta(v) 
$, each maximal velocity-critical infinitesimal length variable $
dr $ with $ \beta(v) dr $, and each zero velocity-critical infinitesimal
length variable $ dx $ with $ dx $,
where $ \beta(u) = c/ \sqrt {c^2 - u^2} = 1/\sqrt{1- u^2/c^2}
$ for $ 0 \leq u < c $ (for details, see Section 5).
This principle implies that the total energy $ E $ of a
material body $ M $ plays a key role for determining the motion of $
M $, and it holds the larger $ E $ is, the
more slowly $ M $ responds to any outer force,
that is, the larger $ E $ is, the more "stubborn" $ M $ becomes.
This principle may present a unification of the correct parts of special
relativity and general relativity by the term "energy".
By depending on the energy-velocity
equation
and the twofold metric principle, we derive the following :
(5) $ E = \beta(v)mc^2 $ ; (6) the travelling distance of a muon with
velocity $ 0.999c $ ; (7) a modified version of the Schwarzshild
metric (called the twofold Schwartzshild metric); 
(8) the Maxwell equations and the light velocity over the 
space with a gravitional field ; (9) deflection and 
red shift of light due to gravity
; (10) an explanation for the Michelson-Morley experiment ; (11) showing
the general relativity principle asserting all coordinate syatems are
"equivalent" (at least w.r.t. the tensor-based laws of physics) is not
valid ; (12) remarks about Einstein's (field) equations and cosmology.  Our
calculated value of procession of planatery orbits depending on the
twofold Schwartzshild metric is the same as
the corresponding well known value
derived from the Schwarzshild metric.  However the twofold Schwartzshild
metric does not contain the points $ r= 2GM/c^2 $ as its singular ones, and
may imply that arguments for the existence of black holes depending on 
the points $ r=2GM/c^2 $ in the Schwartzshild metric are wrong.
We conclude the Schwartzshild
metric is an approximation of the twofold Schwartzshild metric in Section 5.
One may observe many two-fold pairs in the universe : (i) the space and the
time ; (ii) particle and antiparticle ; (iii) mass and energy ; (iv) hadron and
lepton ; (v) electricity and magnetism ; (vi) plus and minus electricity 
; (vii) north and south poles in magnetism ; (viii) potential and kinematic
energy, etc..

In developing arguments in  relativity theory, the following two
principles are
generally acknowledged to be valid (see, e.g., [14]).

\vspace{2ex}

{\bf Principle 1} \hspace{5mm}
  Space is isotropic, i.e., all spacial directions are
  equivalent.
  
\vspace{2ex}

{\bf Principle 2} \hspace{5mm}
  Space and time are homogeneous, i.e., no point of space or time
  is distinguished from others.  The origin of the coordinate system may be
  chosen arbitrarily, without affecting the measuring devices for space 
  and time.
  
\vspace{2ex}

The author is not certain whether these two principles are valid about
the entire universe, but we acknowledge that these princiles are valid
at least within the scope of the universe which this paper concerns.
We also acknowledge the validity of the following principle which is called
in this paper the weak constant light velocity principle.

\vspace{2ex}

{\bf Principle 3}  \hspace{5mm}
  A light signal over an inertial system stationary in the vacuum space
  proceeds into any direction with constant velocity $ c $.
  
\vspace{2ex}

Moreover the following four "principles" are also acknowledged to hold
in relativity theory.

\vspace{2ex}

{\bf Strong Constant Light Velocity Priniciple} \hspace{5mm}  
The velocity of a light signal measured over any inertial system 
  is $ c $ independent of its direction.
  
\vspace{2ex}

{\bf Special Relativity Principle} \hspace{5mm} All inertial systems are equivalent
w.r.t. the laws of physics.

\vspace{2ex}

{\bf General Relativity Principle 1} \hspace{5mm} All coordinate
systems are equivalent w.r.t. the laws of physics.

\vspace{2ex}

{\bf General Relativity Principle 2}  \hspace{5mm} All coordinate
systems are equivalent w.r.t. the tensor-based laws of physics.

\vspace{2ex}

We shall show that none of these four principles is valid.  To do this,
we shall show that Strong Constant Light Velocity Principle is not
valid by showing Fact A.  This implies invalidity of Special Relativity
Principle and General Relativity Principle 1.  To disprove General
Relativity Principle 2, we shall present one coordinate transformation
which transforms the Schwartzshild metric into a metric $ 
g'_{\mu\nu} $ such that from the metric $ g'_{\mu\nu} $,
one deduce procession of planatery orbits wrongly by applying the variational
principle.  Thus to apply the variational principle, we need a standard
coordinate syatem and the corresponding standard metric.

We acknowledge the validity of the following proposition due to 
Principles 1,2.

\vspace{2ex}

{\bf Proposition 1} \hspace{5mm}
All points which are at rest on an inertial system $ K $ are
equivalent w.r.t. the theory about $ K $, i.e., no point at rest
on $ K $ is distinguished from others at rest on $ K $ theoretically.
The origin of $ K $ may be chosen arbitrarily without affecting
the measuring devices for space and time.

\vspace{2ex}

Einstein [3] introduces the notion of synchronization of two clocks $ A $
and $ B $ which are at rest on two stationary points $ P $ and $ Q $ ,
respectively, on an inertial system $ K $, and have the 
same mechanism  as follows.  
Due to Strong Constant Light Velocity Principle,
Einstein admits that the time lapses for light signals to travel
from $ P $ to $ Q $ and to travel from $ Q $ to $ P $, respectively, 
are the same.  Now assume that a light signal is emitted at $P$ at
time $t_1$ on $A$ , then arrives and is reflected by a mirror at $Q$ at
time $t_2$ on $B$ , and finally returns at $P$ at time $t_3$ on $A$ .
Then $A$ and $B$ are said to be synchronized if it holds
$ t_2 - t_1 = t_3 - t_2 $.  Einstein
asserts that his synchronization relation is an equivalence relation.
We note that Einstein's synchronization
relation would be an equivalence relation only when the times at all points at
rest on $K$ proceed with the same rate since all clocks at rest on 
$K$ are assumed to be of the same mechanism.  This can be noted by
the following simple example.

\vspace{2ex}

{\bf Example 1} \hspace{5mm}
  We  say that the mechanisms of two clocks are equivalent if
the times shown by them proceed with the same speed (rate) (a more detailed
definition will be presented later).  Assume that there exist two
clocks $A$ and $B$ at points $P$ and $Q$, respectively, at rest
on the earth, and the following hold.
\begin{enumerate}
\item When the Greenwich mean time is 12 o'clock, January 1, 1993, the time
shown on $A$ is 12 o'clock, January 1, 1993 and the time shown on $B$
is $ 10^{-3} $ seconds, 12 o'clock, January 1, 1993.

\item When the Greenwich mean time is 13 o'clock, January 1, 1993, the time
shown on $A$ is 13 o'clock, January 1, 1993 and the time shown on $B$
is $ 10^{-3} $ seconds, 30 minutes, 12 o'clock, January 1, 1993.
\end{enumerate}

\vspace{2ex}

Thus the ratio of the speed (rate) of the time lapse on $B$ w.r.t. that on
$A$ is $ 1/2 $ (so that the machanisms of $A$ and $B$ are not
equivalent).  Assume that the time lapse measured by the Greenvich mean 
time within which a light signal travels from $P$ and $Q$ ( and
from $Q$ to $P$ also) is $ 2 \times 10^{-3} $ seconds.  Assume also that
a light signal is emitted from $P$ at time $ t = 12 $ o'clock, January
1, 1993 on $A$, arrives and is reflected at $Q$ at time $t'$ on $B$,
and finally returns at $P$ at time $ t'' $ on $A$.  Then
it holds $ t' = 2 \times 10^{-3} $ seconds, 12 o'clock, January 1, 1993, 
$ t'' = 4 \times 10^{-3} $ seconds, 12 o'clock, January 1, 1993, and
$ t' - t = t'' - t' $.  But if a light signal is emitted from
$P$ toward $Q$ at time $ t_0 \neq t $ on $A$, then the above relation
clearly does not hold, and it is impossible to make two clocks
$A$ and $B$ synchronized.  For two clocks to be synchronized,
it is necessary that their mechanisms are equivalent.

\vspace{2ex}

In the sequel, we assume that on each inertial
system $K_0$, there exists an imaginary clock $C_0(P)$ to each point $P$
at rest on $K_0$ indicating the time at $P$ so that the expression
"at time $t$ at $P$" means that the time shown on $C_0(P)$ is $t$, and 
"the speed of the time lapse at $P$" coincides with the speed of the
time lapse shown on $C_0(P)$.  Since an inertial system is a coordinate
system in which the Newton equation of motion holds, we 
acknowledge that all clocks
$\in \{C_0(P) \mid P $ is a point at rest on $K_0 \}$ are synchronized
(a new definition of synchronization will be presented later).  In this
paper, a coordinate system is for use of descriptions of events occurring
in the universe by being given corresponding changes of coordinates of
events.  To each coordinate system $K_1$, we assume that to each point
$P$ at rest on $K_1$, there exists an imaginary clock $C_1(P)$ indicating
the time at $P$ as in case of inertial systems.  In some cases, all
imaginary clocks associated with a coordinate system may not be
synchronized.  We say that a coordinate system is synchronized if all
imaginary clocks associated with the system are synchronized. 
We shall introduce the
notion of clock-coordinate systems as follows.  Let $K$ be a coordinate system.
Then a clock-coordinate system (in short, a cc system) of $K$ 
is a pair $ H = <K,C> $ such that
$C$ is a bijection from the set $ A = \{ P \mid P $ is a point at 
rest on $ K \} $  to a set of imaginary clocks $B$  such that
for each $ P \in A$, $C(P) $ indicates the time at $P$.
Since all cc systems of $K$ are
equivalent, we call each of them the cc system of $K$.
$H$ is said to be synchronized if all clocks in $B$ are synchronized,
i.e., if $K$ is synchronized.

Instead of Strong Constant Light Velocity Principle, we acknowledge
the validity of the following proposition since due to Proposition 1, 
we can choose the origin of $K$ arbitrarily.

\vspace{2ex}

{\bf Proposition 2} \hspace{5mm}
  The time lapse within which a light signal travels from a point
$P$ to a point $Q$ both of which are at rest on an inertial system
$ K $
is uniquely determeined if the distance between $P$ and $Q$ and the
direction from $P$ and $Q$ are fixed.

\vspace{2ex}

We note that it is acknowledged that the
universe continues expanding, and one might assert that the distance
between two points which are at rest on the above $K$ continues  
changing as the time passes.  But if one acknowledges the validity of this
assertion, then one can deduce that the Newton equation of motion
does not hold on $K$ by noting the weak constant light
velocity principle.  We also note that the distance between
any two points which are at rest on the earth coordinate system has
been almost unchanged within a very long time period independent of
the expansion of the universe.  Moreover in the Lorentz
transformation over two inertial systems moving with relative constant
velocity, it is acknowledged that the coordinate of a point which is at rest
on one of the systems and the distance between two points at rest on
the system are constant independent of the time lapse on the system.
In this
paper, we acknowledge that an inertial system is a coordinate system
which will be used for approximate descriptions of events occurring in a rather
small local space and a local time period in that space.  In Section 6,
we shall study the expansion of the universe by depending on Hubble's law,
but it will be left open
to present coordinate syatems for describing our entire expanding
universe.

Synchronization is indispensable for analyzing properties of events
occurring on an inertial system as the following two examples indicate.

\vspace{2ex}

{\bf Example 2} \hspace{5mm}
Consider the following situation.
  
\begin{enumerate}
\item Two persons $A$ and $B$ live at New York and Tokyo, respectively.
\item $A$ and $B$ have two clocks $C$ and $D$ , respectively, whose mechanisms
are equivalent.
\item When the Greenwich mean time is 12 o'clock, March 1, 1993, the time
shown on $C$ is 11 o'clock, April 1, 1991 and the time shown on $D$ is
10 o'clock, May 1, 1992.
\item An air mail is sent from $A$ at time 9 o'clock, June 1, 1991 on
clock $C$, and arrives at $B$ at time 8 o'clock, July 8, 1992 on clock
$D$.
\end{enumerate}

If they assert that the time lapse in which the mail is carried from
$A$ to $B$ is 371 days and 23 hours ( = the time difference shown by their
clocks), then this assertion is nonsense.

\vspace{2ex}

We also need the notion of bias about two clocks. A detailed definition
will be given later.

\vspace{2ex}

{\bf Example 3} \hspace{5mm}
  Let $A$ be a clock indicating the french standard time, and
$B$ be a clock indicating the Japanese standard time.  Then their
mechanisms are equivalent, but $A$ and $B$ are not synchronized, and
the bias of $A$ w.r.t. $B$ is $- 8$ hours in the winter.

\vspace{2ex}

We  again remark in Example 2 that if the bias of $C$ w.r.t. $D$ is two weeks,
then $A$ and $B$ may assert that the time lapse needed for the delivery
of the mail is negative.
To decide the delivery time of the mail,
it may hold that two clocks are synchronized or their mechanisms
are  equivalent and the bias between them is known.

Since we do not acknowledge the validity of Strong Constant Light
Velocity Principle, we shall present our new definition of synchronization
which is theoretical rather than feasible as follows.  Let $K$ be a
coordinate system, $P$ and $Q$ be two points at rest on $K$, and
$A$ and $B$ be two clocks at rest on $P$ and $Q$, respectively.
Let $a$ and $l$  denote the direction from $P$ to $Q$ and the
distance between $P$ and $Q$, respectively.  If $K$ is an inertial system,
by Proposition 2, it holds
that the velocities of a light signal travelling from $P$ to $Q$
and from $Q$ to $P$, respectivrly, are uniquely determined if
$a$ and $l$ are determined.  Let
these velocities be $r$ and $s$, 
respectively.  Now consider time $t_1$ on $A$, and
let a light signal be emitted from $P$ at time $t_1$ on $A$,
arrive and be reflected at $Q$ at time $t_2$ on $B$, 
return at $P$ at time $t_3$ on $A$, and finally arrive at $ Q $ at
time $ t_4 $ on $ B $.  The time lapse $ tl(A,B,t_1)$ of
$ A $ w.r.t. $ B $ at $ t_1 $ on $ A $ is defined by :
$ tl(A,B,t_1) = t_3 - t_1 $.  Similarly $ tl(B,A,t_2)$ is
defined by : $ tl(B,A,t_2) = t_4 - t_2 $.  Clock $ A $ ( $
B $, respectively) is said to be punctual if for any time $
t_{10} $  on $ A $, it holds $ tl(A,B,t_{10}) = 
l/r + l/s $ (for any time $ t_{20} $ 
 on $ B $, it holds $ tl(B,A,t_{20}) =  l/r+l/s$,
respectively).  The mechanisms of $ A $ and $ B $ are said to be
equivalent if both $ A $ and $ B $ are punctual.  The bias
$ b(A,B)$ of $ A $ w.r.t. $ B $ is defined by :
$ b(A,B) = l/r -(t_2 - t_1)$.  Note that $ b(A,B) = l/r -
(t_4 - t_3) = t_3 - t_2 -l/s $, and $ b(A,B)$ is independent from
the choice of $ t_1 $ when $A$ and $B$ are punctual.  
Clocks $ A $ and $ B $ are said
to be synchronized if the mechanisms of $ A $ and $ B $
are equivalent and the bias of $ A $ w.r.t. $ B $ is zero.
In Appendix 1, we present a proof showing the new synchronization
relation is an equivalence relation.

 Now we shall introduce the contents of the following sections briefly.
 The paper consists of nine sections and five appendecies.
 Section 2 presents properties of the Lorentz transformation.  
 Section 3 presents
 an electromagnetic example which seems contradictory to
 Special Relativity Principle by
 observing the Maxwell equations about a stationary charged material body
 and the corresponding moving one. Sections 4-5 are the main parts of this
 paper.
 Section 4 presents properties
 of inertial systems.  It introduces the notion of imaginary signals
 over an inertial system, and presents TI Lemma (Lemma 3) and TI Theorem
 (Theorem 1).  By depending on
TI Theorem, we deduce that neither Special Relativity Principle nor
Strong Constant Light Velocity Principle is valid, and the Lorentz
 transformation is not the space-time transformation over two
 inertial systems moving with nonzero relative constant velocity.
In Section 5, we introduce a new principle called the twofold metric
principle, and the energy-velocity equation.  By depending on the
energy-velocity equation and the new principle, we derive the results
(5)-(12) explained briefly in the beginning part of this section.  We also
conclude by depending on the new principle that
the existence of black halls can not be predicted from the Schwarzshild
metric, and the Schwartzshild metric is an approximation of the twofold
Schwartzshild metric.  Section 6 presents arguments about the expansin
of the universe, which contains the assertion that the (present) age of
the universe is greater than the Hubble time $H_0^{-1}$.
Section 7 presents a new observation about the
dark matter problem.
Section 8 presents arguments for
concluding that one interpretation of the principle of equivalence in general
relativity theory is not sufficiently valid.
The final section
 presents concluding remarks.

 \section{The Lorentz transformation}

 By a coordinate system in the sequel of this section, 
 we mean a Cartesian coordinate
 system $K$ in which the time lapses at all points at rest on $K$ are also
described.  For each coordinate system $K$, a coordinate on $K$ is
a quadruple, $ (x,y,z,t) $, such that $x$, $y$ and $z$ are the x-axis,
y-axis and z-axis coordinates of some point $P$ on $K$ and $t$
is a time at $P$.  $K$ can be used for describing the behavior of any
event occurring on $K$ by denoting changes of corresponding coordinates.
For example, travellings of electromagnetic waves can be described
on $K$.  In the sequel, each coordinate system will be denoted often 
by $K,K',K''$ or $K_0$, and $t,t',t''$ and $t_0$ will be often used
for variables denoting times on $K,K',K''$ and $K_0$, respectively.
When $K,K',K''$ or $K_0$ is a Cartesian coordinate system, then
a-axis, a'-axis, a''-axis  or $a_0$-axis for $ a \in \{ x,y,z \} $ 
means the corresponding axis of $K,K',K''$ or $K_0$, respectively,
in the standard way.
To each coordinate system $K$, we define the clock-coordinate
system $ H = <K,C> $ (in short, the cc system ) of $K$ as in
Section 1.  Thus for each point $P$ at rest on $K$, $C(P)$ is an imaginary
clock indicating the time at $P$. The set of all coordinates on $K$
will be denoted by $CO(K)$.  An inertial system in the sequel means
a synchronized Cartesian coordinate system moving with gravitational
potential (almost) zero in the vacuum 
space with constant
velocity.  Let $K$ and $K'$ be two Cartesian coordinate
systems such that their corresponding axises are parallel, the origins
of $K$ and $K'$ coincide at the initial time $ t = t' = 0 $, and the
origin of $K'$ moves in the direction of the positive x-axis of $K$
with constant velocity $v > 0$.  In this situation, we say that $K$ and
$K'$ are two parallel Cartesian coordinate systems with 
relative constant x'-axis
velocity $v$.  When $K$ and $K'$ are inertial systems, we say
in this situation that $K$ and $K'$ are two parallel inertial systems with
relative constant x'-axis velocity $v$.
The space-time transformation from $K$
to $K'$ is a bijection $ {\cal T} $ from $CO(K)$ to $CO(K')$ such that
for each $ (x,y,z,t) \in CO(K)$, ${\cal T}(x,y,z,t) = (x',y',z',t') $ 
means that if $ (x,y,z) $ is a spacial coordinate at a point $P$ on 
$K$, then $(x',y',z') $ is the corresponding spacial
coordinate at $P$ on $K'$ at time $t$ at $P$ on $K$, and $t'$ is 
the corresponding time at 
$P$ on $K'$.  $ (x',y',z',t') $ is interpreted in a sense as the
coordinate corresponding to $(x,y,z,t)$ at $P$ on $K'$ measured by 
an observer who is at rest on
$K'$. ${\cal T}$ is called the Lorentz transformation (from $K$ to $K'$)
if for any $ (x,y,z,t) \in CO(K)$, $(x',y',z',t') =
{\cal T}(x,y,z,t) $ satisfies the following, where $ \beta(v) =
c/\sqrt{c^2 - v^2}=1/\sqrt{1-v^2/c^2}$ .

\[ x' = \beta(v)(x-vt), \; t' = \beta(v)(t-vx/c^2), \; y' = y, \; z' = z \]

The inverse $ {\cal T}^{-1} $ of ${\cal T}$ satisfies the following.
\[ {\cal T}^{-1}(x',y',z',t') = (x,y,z,t) \] 
\[  x = \beta(v)(x' + vt'), \; t = \beta(v)(t' + vx'/c^2), \; y = y', \; z = z' \]

Einstein[3] asserts that the following statement is true.

\vspace{2ex}

{\bf Statement 1} \hspace{5mm}
  Let $K$ and $K'$ be two parallel inertial systems with
relative constant x'-axis velocity $v > 0$. 
 Then the Lorentz
transformation from $K$ to $K'$ is the space-time transformation
from $K$ to $K'$.

\vspace{2ex}

In Section 4, we shall conclude that Statement 1 is false by showing that
if the space-time transformation over two parallel coordinate systems
$ K $ and $ K' $ with constant x'-axis velocity $ v> 0 $ is
the Lorentz transformation, then $ K $ or $ K' $ is not synchronized
, i.e., $ K $ or $ K' $ is not an inertial system.

The following properties of the Lorentz transformation are well known
and can be proved easily (see, e.g., [12,14,15]).

\vspace{2ex}

{\bf Fact 1} \hspace{5mm}
 Let $K$ and $K'$ be two parallel Cartesian coordinate systems with relative
constant x'-axis  velocity $ v > 0 $ such that the space-time
transfomation from $K$ to $K'$ is the Lorentz transformation.  Then the
following hold, where $ \beta(v) = c/\sqrt{c^2 - v^2} $.

\begin{enumerate}
\item Let $R$ be a point at rest on the x'-axis of
$K'$ whose x'-coordinate is $x'_1$
.  Consider time $t'_1$ and time $t'_2$ ($ t'_1 < t'_2) $ at $R$ on
$K'$.  Let $ (x_1,t_1) $ and $(x_2,t_2) $ be the pairs of 
x-coordinate and time on $K$ corresponding to $ (x'_1,t'_1) $
and $(x'_1,t'_2) $, respectively.  Then it holds $ t_2 -
t_1 = \beta(v) (t'_2 - t'_1) $.

\item Let $P$ be a point at rest on the x-axis of
$K$ whose x-coordinate is $x_1$.
Consider time $t_1$ and time $t_2$ $(t_1 < t_2)$ at $P$ on $K$.  Let $ 
(x'_1,t'_1) $ and
$ (x'_2,t'_2) $ be the pairs of x'-coordinate and time on $K'$
corresponding to $ (x_1,t_1) $ and $(x_1,t_2) $ , respectively.
  Then it holds $ t'_2 - t'_1 = \beta(v) (t_2 - t_1)$.

\item Let $R$ and $S$ be two points at rest on the x'-axis of $K'$
whose x'-coordinates are $x'_1$ and $x'_2$ $( x'_1 < x'_2) $,
respectively.  Consider time $t'_1$ at $R$ and time $t'_2$ at $S$ on
$K'$ such that the pairs of x-coordinate and time corresponding
to $ (x'_1,t'_1) $ and $ (x'_2,t'_2) $ are $ (x_1,t_1) $
and $ (x_2,t_1)$, respectively, on $K$ for some $ x_1,x_2,t_1
$.  Then it holds
$ x'_2 - x'_1 = \beta(v) (x_2 - x_1) $.

\item Let $P$ and $Q$ be two points at rest on the x-axis of $K$ whose
x-coordinates are $x_1$ and $x_2$ $(x_1 < x_2) $, respectively.
Consider time $t_1$ at $P$ and time $t_2$ at $Q$ on $K$ such that the pairs
of x'-coordinate and time corresponding to $ (x_1,t_1) $ and
$ (x_2,t_2)$ on $K'$ are $ (x'_1,t'_1) $ and $ (x'_2,t'_1)$,
respectively, for some $ x'_1,x'_2,t'_1 $.  Then it holds
$ x_2 - x_1 = \beta(v) (x'_2 - x'_1) $.

\item Let $R$ be a point at rest on the x'-axis of $K'$.  Then the
velocity of $R$ measured on $K$ w.r.t. the x-axis of $K$ is $v$.

\item Let $P$ be a point at rest on the x-axis of $K$.  Then the
velocity of $P$ measured on $K'$ w.r.t. the x'-axis of $K'$ is $ -v $.

\end{enumerate}

\vspace{2ex}

In Fact 1, (1) and (2) are called time dilations and (3) and (4)
are called length contractions.  Note that time dilation occurs both in
(1) and (2).  At this moment, we may acknowledge that this phenomenon
is odd since any time lapse over any inertial system should be a net time
lapse, and in (1), the speed (rate) of time lapse over $ K' $ is slower than
that over $ K $, and in (2), the converse holds.  We also acknowledge that
Hafele-Keating experiment concerning the differences of time lapses
$\tau _B - \tau_A $ and $\tau_C - \tau_A $ on clocks carried on jet
flights and stationary on the earth, respectively, introduced in Section
8 (see [1]) seems a strong evidence to the invalidity of time dilations
(1) and (2) above since (1) and (2) may imply $\tau_B - \tau_A $ almost equals
$\tau_C - \tau_A$.  We shall conclude
Hafele-Keating experiment suggests
that the speed of the time lapse over a material body moving
with high velocity in the universe is smaller than that over a material
body moving with low velocity in the universe (see also Example 5 in
Section 3 about the life time of muons).

The following remark may be simple, but present a note about the
equivalence of coordinate systems.

\vspace{2ex}

{\bf Remark 1} \hspace{5mm} Let $ K $ and $ K' $ be two parallel
coordinate systems with relative constant x'-axis velocity $ v>0 $.  Assume
that the space-time transformation from $ K $ to $ K' $ is the
Lorentz transformation.  Let $ P $ and $ Q $ be two points
at rest on the x-axis of $ K $ with x-coordinates $ x_1 $ and $
x_2 $, respectively $ (x_1 < x_2 ) $.  Let $ H=<K,C>$ and $ J=<K',D>$
be the cc systems of $ K $ and $ K' $, respectively.  Consider a time
$ t $ over $ K $. Let $ R $ and $ S $ be two points at rest
on the x'-axis of $ K' $ which coincide with $ P $ and $ Q $,
respectively, at time $ t $ over $ K $.  Let $ t' $ be the time
shown on $ D(S) $ when $ S $ and $ Q $ coincide.  Let $ t'_1
$ be the time shown on $ D(R) $ when $ R $ and $ P $ coincide.
Then it holds

\[ t' = \beta(v)(t-v x_2 /c^2), \; t'_1 = \beta(v)(t-v x_1 /c^2),\;
t'_1 > t'   \]

Thus over $ K $, at time $ t $, the time at each point at rest on
the line segment connecting $ P $ and $ Q $ is $ t $ while
the corresponding times at any two distinct points at rest on the line
segment connecting $ R $ and $ S $ are distinct (one is a past
time and the other is a future time).  Thus in a sense, we may say that
the past-future relations over $ K $ and $ K' $ are different,
and $ K $ and $ K' $ are not equivalent.

\vspace{2ex}

In Appendix 2, we present remarks about twin paradox.

\section{The Maxwell equations}

The Maxwell equations in the vacuum space are expressed as follows.

\[ \nabla \cdot D(\vec{r},t) = \rho(\vec{r},t), \;
\nabla \cdot B(\vec{r},t) = 0 \]
\[ \nabla \times H(\vec{r},t) - \displaystyle{\frac {\partial D(\vec{r},t)}
{\partial t}}
 = i(\vec{r},t) , \;
\nabla \times E(\vec{r},t) + \displaystyle{\frac {\partial B(\vec{r},t)}
{\partial t}}
 = 0 \]

First we shall present an electromagnetic example which seems to be a strong
evidence to the invalidity of 
Special Relativity Principle.

\vspace{2ex}

{\bf Example 4} \hspace{5mm}
Let $K$ and $K'$ be two parallel inertial systems with relative constant
x'-axis velocity $ v > 0 $, and $ H = <K,C>  $ and $ J = <K',D> $
be the cc systems of $K$ and $K'$, respectively.
Assume that $K$ is
stationary in the vacuum space (or in the space very close to the
surface of the earth for certain small cases).  Let an electrically
charged material body $M$ with $q$ coulombs be at rest on the origin
of $K'$.  We consider 
the origin $O$ of $K$.  Since the distance between $M$ and $O$ changes
as $t$ changes, for each time $ t $ $(>0) $ on
$C(O)$,  it holds at $O$

\[ \displaystyle{\frac {\partial E_x}{\partial t}  \neq 0 } \]

Then due to the Maxwell equations, it holds at $O$

\[\frac{1}{c^2} \frac{\partial E_x}{\partial t} = \frac{\partial B_z}
{\partial y} - \frac{\partial B_y}{\partial z} \neq 0 \]

We call this case
the moving charge example.  Conversely we consider the stationary
charge example as follows.  Let $M$ be at rest on the origin of $K$
.  Then for any point $P$ at rest on the x-axis of $K$ whose x-coordinate
is $ x $ $( \neq 0) $, it holds

\[ \displaystyle{\frac{\partial E_x}{\partial t} = 0     } \]

Due to the Maxwell equations, $\displaystyle{\frac{\partial B_z}{\partial y}
- \frac{\partial B_y}{\partial z} = 0}$ at any point, particularly
at the origin $O'$ of $K'$.

We may also
consider the following situations.  Consider the moving charge example.
In this case, for a sufficiently large integer $ m \geq 1 $ and a real $ l>0
$, we consider $ 2m+1 $ electrically charged material bodies
$ M_{-m},M_{-m+1}, \cdots, M_{-1},M_0,M_1,M_2,\cdots,M_m $,
each being  with $ q $
coulombs, at rest points on the x'-axis of $ K' $ whose x'-coordinates
are $ -ml,(-m+1)l,\cdots,-l,0,l,2l,\cdots,ml $, respectively.  We also
consider a mariner's compass $X$ at rest on a point $P'$ whose x'-coordinate
and y'-coordinate are zero and whose z'-coordinate is $ b>0 $.
Then due to Biot-Savart's law, these $ 2m+1 $ electrically charged
material bodies may produce the nonzero magnetic field over $K$ and $K'$, and 
the needle of 
$X$ will be parallel to the y'-axis of $ K' $.  On the other hand,
in the stationary charged example, we consider correspondingly
$ 2m+1 $ electrically charged material bodies at rest on the x-axis of
$ K $ and the marinar's compass $X$ at rest on the z-axis of $ K $.
Then these $ 2m+1 $ electrically
charged material bodies do not produce any nonzero magnetic field
over $K$ and $K'$, and
the direction of the needle of $X$ will be arbitrary.

Thus we conclude that $K$ and $K'$ may  not be
equivalent, and these observations may provide a strong evidence
suggesting the invalidity of Special Relativity Principle.  In Section 5, from
Subprinciple 3, we shall develop arguments from which the above
phenomena (and the Michelson-Morley experiment) may be explained reasonably.

\vspace{2ex}

Einstein[3] asserts that the following statement is true.

\vspace{2ex}

{\bf Statement 2} \hspace{5mm}
Let $K$ and $K'$ be two parallel inertial systems with relative 
constant x'-axis  velocity $v > 0 $. 
Let $ (E_x,E_y,E_z),
(B_x,B_y,B_z), (E'_{x'},E'_{y'},E'_{z'}) $ and $(B'_{x'},B'_{y'},B'_{z'}) $ be
the electric field and the magnetic field measured on $K$, and
those measured on $K'$, respectively.  Then the following hold.

\[ \frac {1}{c^2} \frac { \partial E'_{x'} }{\partial t'} = 
\frac{\partial B'_{z'}}{\partial y'} - \frac{\partial B'_{y'}}{\partial z'},\;
  \frac{\partial B'_{x'}}{\partial t'} = - \left( \frac{\partial E'_{z'}}{\partial y'} -
  \frac{\partial E'_{y'}}{\partial z'} \right)  \] 
\[ \frac {1}{c^2} \frac { \partial E'_{y'} }{\partial t'} = 
\frac{\partial B'_{x'}}{\partial z'} - \frac{\partial B'_{z'}}{\partial x'},\;
  \frac{\partial B'_{y'}}{\partial t'} = - \left( \frac{\partial E'_{x'}}{\partial z'} -
  \frac{\partial E'_{z'}}{\partial x'} \right)  \] 
\[  \frac{1}{c^2} \frac{\partial E'_{z'}}{\partial t'} = \frac{\partial B'_{y'}}{\partial x'}
- \frac{\partial B'_{x'}}{\partial y'}, \; \frac{\partial B'_{z'}}{\partial t'}
  = - \left( \frac{\partial E'_{y'}}{\partial x'} - \frac{\partial E'_{x'}}{\partial y'}
  \right) \] 
\[  E'_{x'} = E_x, \; B'_{x'} = B_x \]  
\[ E'_{y'} = \beta(v)(E_y - vB_z) , \;  B'_{y'} = \beta(v)(B_y + vE_z/c^2) \] 
  \[ E'_{z'} = \beta(v)(E_z + vB_y),\; B'_{z'} = \beta(v)(B_z - vE_y/c^2) \]

\vspace{2ex}

In Section 4, we shall conclude that Statement 2 is false.

\vspace{2ex}

{\bf Example 5} \hspace{5mm}
Let $K$ and $K'$ be two parallel inertial systems with relative constant
x'-axis velocity $v = 0.999c $, and $K$ be stationary w.r.t. the universe
(or w.r.t. the earth since we consider here a small space).  It
is well known that the mean life time of a muon $A$ at rest on
the earth is $ \tau \approx 2.20 \times 10^{-6} s $, and a muon $B$ moving
with velocity $ v = 0.999c $ can travel in a distance of about $
14.8 km $ in the space very close to the surface of the earth.
Thus the life time $\tau '$ of muon $B$ is

\[ \tau ' \approx 49.2 \times 10^{-6} s \approx \tau \beta(v) \]

Now let $K$, $K'$ and $v$ be as above.  Assume
that muon $A$ is at rest on $K$ and muon $B$ is at rest on $K'$.  Then
their life times are quite distinct w.r.t. the earth coordinate
system, and this fact may seggest the speed of the time lapse over
$B$ is slower than that over $A$ by factor $ 1 / \beta(v) $.  
Here we acknowledge
the time lapses of the life of muon over $ K $ and $ K' $,
respectively, should be the net time lapses over $ K $ and $ K' $.
Thus we conclude that these observations present a strong evidence suggesting
the invalidity of Special Relativity Principle.
Note that the time lapses on $A$ and $B$
are measured by their own proper clocks, respectively.  In Subsection 5.1,
we shall present arguments for deriving the above $14.8km$ by depending
on the twofold metric principle.

\section{Inertial systems}

In this section, we establish Time Independence Lemma (TI Lemma, i.e., Lemma
3) and Time Independence Theorem (TI Theorem, i.e., Theorem 1)
and study properties of inertial systems.  
 Throughout this sectin, an inertial system means a 
 synchronized Cartesian coordinate system which
moves in the vacuum space with constant velocity.  
Over any inertial system $K$ and any arbitrarily large
real $u>0$, we consider an imaginary signal $S(K,u)$ whose velocity 
is $u$ over $K$.  This means that for any two distinct points $P$ and
$Q$ at rest on $K$ and any time $ t $ at $ P $, if the signal
$S(k,u)$ is emitted from $ P $ to $ Q $ at time $ t $, then
$ S(K,u) $ arrives at $ Q $ at time $ t + l/u $, where $ l $
is the distance between $ P $ and $ Q $.
This notion  is unambiguous and mathematically 
well defined since the distance $ l $
is unambiguously defined and $ K $ is assumed to be synchronized.
For other words, the travelling of imaginary signal $ S(K,u) $
from $ P $ to $ Q $ over $ K $ is just a time lapse
from a past time $ t $ at $ P $ to a future time $ t +
l/u $ at $ Q $.
Thus for any points $P_0$ and $Q_0$ at rest on the line segment
connecting $P$ and $Q$, we consider the time lapse from time $t + l_1 /u$
at $P_0$ to time $t + l_1 /u + l_2 /u $ at $Q_0$, where $l_1$ and $l_2$
are the distances between $P$ and $P_0$ and $P_0$ and $Q_0$, respectively.
One may say that the travelling of imaginary signal
$ S(K,u) $ from $ P $ at time $t$ at $P$ to $ Q $ at time $t+ l/u$ at $Q$
over $K$
is in a sense "observable"
since by a Gedankenexperiment, one can imagine
this travelling occurs from a past time at $ P $ to a
future time at $ Q $ over $ K $ continuously.
By depending on TI Theorem,
we shall show that Strong Constant Light Velocity Principle is
not valid, the Lorenz transformation is not the space-time
transformation over two inertial systems moving with relative constant
velocity, and all inertial systems are not equivalent, i.e.,
Special Relativity Principle is not valid.  

Now we shall proceed to establishing TI Lemma and TI Theorem.
We shall recall the definition of synchronization due to
Einstein[3].  Let $ K $ be a Cartesian coordinate system, and $ 
P $ and $ Q $ be two distinct points at rest on $ K $. Let $ 
H = <K,C>$ be the cc system of $ K $.  Assume that a light signal
$ L $ is emitted at $ P $ at time $ t_1 $ on $ C(P) $, then
 arrives and is reflected at $ Q $ at time $ t_2 $ on $ C(Q)
$, and finally returns at $ P $ at time $ t_3 $ on $C(P)$.  Then $ C(P) $
and $ C(Q) $ are said to be synchronized due to Einstein[3] if it 
holds $ t_2 - t_1 = t_3 - t_2 $.  Here Einstein acknowledges the
validity of Strong Constant Light Velocity Principle, and in
this definition, when $ L $ arrives at $ Q $ at time $ t_2
$ on $ C(Q) $, it is acknowledged that the corresponding time on
$ C(P) $ is also $ t_2 $.  Since we do not acknowledge the validity
of Strong Constant Light Velocity Principle, in the above event,
we acknowledge only that the time lapse $ t_2 - t_1 $ (the time lapse
$ t_3 - t_2 $, respectively) is uniquely detemined if the direction
from $ P $ to $ Q $ and the distance between $ P $ and
$ Q $ (the direction from $ Q $ to $ P $ and the distance between
$ P $ and $ Q $, respectively) are fixed.  Thus we acknowledge,
as in the definition of our new syncronization relation in Section 1,
that in the above event, if $ t_2 - t_1 $ and $ t_3 - t_2 $
are the net time lapses 
over $ K $ in which $ L $ travels from $ P $ to $ Q 
$ and $ L $ travels from $ Q $ to $ P $, respectively,
then, when $ L $ arrives at $ Q $ (when $ L $ finally
returns at $ P $, respectively), the corresponding time on 
$ C(P) $ is $ t_2 $ (the corresponding time on $ C(Q) $ is
$ t_3 $, respectively). From all these arguments, we acknowledge
that the following fact holds.

\vspace{2ex}

{\bf Fact 2} \hspace{5mm}
Let $ K $ be an inertial system, and $ H = <K,C>$ be the cc
system of $ K $.  Let $ P,Q,R,S $ be four points at 
rest on $ K $.  Assume that (i) the direction from $ P $ to
$ Q $ is the same as that from $ R $ to $ S $, (ii) a light
signal $ L_1 $ is emitted at $ P $ toward $ Q $ at time
$ t_1 $ on $ C(P) $ and arrives at $ Q $ at time $t_2 $ on $
C(Q) $ ($ t_2 > t_1 $), and (iii) another light signal $ L_2 
$ is emitted at $ R $ toward $ S $ at time $ t_3 $
on $ C(R) $ and arrives at $ S $ at time $ t_4 $ on $ C(S)
$ ($ t_3 < t_4 $).  Then the following hold.

\begin{enumerate}
\item When $ L_1 $ arrives at $ Q $ at time $ t_2 $ on $
C(Q) $, the corresponding time on $ C(P) $ is also $ t_2 $, and $
t_2 - t_1 $ is the net time lapse on $K$ within which $ L_1 $ travels
from $P$ to $Q$.  
\item It holds $ t_2 - t_1 = t_4 - t_3 $ iff the distance between
$ P $ and $ Q $ is the same as that between $ R $
and $ S $.
\end{enumerate}

\vspace{2ex}

{\bf Remark 2} \hspace{5mm}
(1) In Fact 2, we do not acknowledge the validity of
Strong Constant Light Velocity Principle, but (1) and (2) in the fact
would be, a fortiori, true if Strong Constant Light Velocity Principle
would be valid.  

(2) To understand Fact 2 more precisely, we shall 
present the following explanation using imaginary signals.
In Fact 2-(1), 
the sentence ''the corresponding time on $C(P)$ is also $t_2$'' means
the following : (i) for any large $u >0$, if the signal $S(K,u)$
is emitted from $Q$ at time $t_2$ on $C(Q)$ to $P$, then  $S(K,u)$
arrives at $P$ at a time $t_5$ on $C(P)$ with $t_5 > t_2$ ; (ii) for any 
small $\epsilon >0$, there exists a sufficiently large
$u(\epsilon)>0$ such that when the signal $S(K,u(\epsilon))$ is emitted
from $Q$ to $P$ at time $t_2$ on $C(Q)$, then  $S(K,u(\epsilon))$ arrives
at $P$ at a time $t_6$ on $C(P)$ with $t_2 < t_6 < t_2 + \epsilon $.
If we consider two clocks $A$ and $B$ at rest Paris and New York, respectively,
whose mechanisms are equivalent and which are synchronized, then the above
(i) and (ii) clealy must hold about $A$ and $B$.

\vspace{2ex}

We  note here that  the Lorentz
transformation from a coordinate system $K$ onto another coordinate 
system $K'$
asserts that if an imaginary signal $ S(K,u)$ with
$ u $ being sufficiently large is emitted over $ K $ to the positive
x-axis direction of $ K $, then $ S(K,u)$ can reach a past history
over $ K' $.  
TI Lemma presented in the sequel
implies that this phenomenon never occurs over inertial systems.

Before presenting TI Lemma and TI Theorem,
we shall study properties of the space-time transformation over
two inertial systems moving with relative constant velocity (Lemmas 1,2,
Proposition 3).  We recall that two coordinate (or inertial) systems
$K$ and $K'$ are
said to be two parallel coordinate (or inertial) systems with relative 
constant x'-axis
velocity $v \geq 0$ if their corresponding axises are parallel, the origin
of $K$ and $K'$ coincide at the initial time $t=t'=0$, and the origin of
$K'$ moves in the direction of the positive x-axis of $K$ with constant
velocity $v$ (see Section 2).
In the sequel from Lemma 1 to Proposition 3, let $K$ and $K'$ 
be two parallel inertial systems with relative constant x'-axis 
velocity $v > 0$.  Let $ H = <K,C> $ and $ J = <K',D> $ be the
cc systems of $K$ and $K'$, respectively, and we acknowledge that $H$
and $J$ are synchronized, respectively. 
Moreover we assume that the velocity of 
a light signal over $K$ is $c$ independent of its direction.
Let $ {\cal T} $ be the space-time transformation from 
$ K $ to $ K' $.  For any point $ P $ at rest on $ K $ 
and time $ t $ on $ C(P) $, we write $ {\cal T}(P,t) =R $ and $ 
{\cal T}(x_1,t)=(x'_1, t')$ if $ x_1 $ is the x-coordinate of $ P $,
$ R $ is the point at rest on $ K' $ which coincides with
$ P $ at time $ t $ on $ C(P) $ and at time $ t' $ on $
D(R) $, and $ x'_1 $ is the x'-coordinate of $ R $.  Here we note that 
$(x_1,t)$ depends only on $(x'_1,t')$ since one can choose the origins
of $K$ and $K'$ arbitrarily due to Principle 2 (see also the proofs of
Lemma 1 and Proposition 3 below).  In this
situation, we also write $ {\cal T}^{-1}(R,t')=P $ and 
${\cal T}^{-1}(x'_1,t')=(x_1,t)$.

\vspace{2ex}

{\bf Lemma 1} \hspace{5mm}
Let $ R $ and $ S $ be two points at rest on $ K' $ whose
x'-coordinates are $ x'_1 $ and $ x'_2 $, respectively.  Consider
times $ t'_1 $ and $ t'_2 $ ($ t'_1 < t'_2 $) on $ D(R) $ and
$ D(S) $.  Let $ P_1, P_2, Q_1 $ and $ Q_2 $ be four points at
rest on $ K $ whose x-coordinates are $ x_1,x_2,x_3 $ and $ x_4
$ and let $ t_1,t_2,t_3 $ and $ t_4 $ be times on $ C(P_1),C(P_2),
C(Q_1) $ and $C(Q_2) $, respectivley, such that the following hold.

\[ {\cal T}^{-1}(R,t'_1)= P_1, \; {\cal T}^{-1}(R,t'_2)= P_2, \; {\cal T}^{-1}(S,t'_1)= Q_1,
\; {\cal T}^{-1}(S,t'_2)= Q_2 \]
\[ {\cal T}^{-1}(x'_1,t'_1)=(x_1,t_1), {\cal T}^{-1}(x'_1,t'_2)=(x_2,t_2), 
{\cal T}^{-1}(x'_2,t'_1)=(x_3,t_3),  {\cal T}^{-1}(x'_2,t'_2)=(x_4,t_4)  \]

Then it holds $ x_2 - x_1 = x_4 - x_3 $ and $ t_2 - t_1 = t_4 - t_3 
$.

{\bf Proof.}  We consider two inertial systems $ K'_0 $ and $ K'_1
$ both of which are equivalent to $ K' $ and whose origins 
coincide with $ R $ and  $ S $, respectively.  Here we
acknowledge the validity of Principle 2.  Let $ O'_0 $ and
$ O'_1 $ denote the origins of $ K'_0 $ and $ K'_1 $,
respectively, and the time on $ D_0(O'_0) $ 
corresponding to time
$ t'_1 $ on $ D(R) $ be zero, and in the same way, let the time
on $ D_1(O'_1) $ corresponding to time $ t'_2 $ on $ D(S) $
be zero, where $ < K'_0, D_0 > $ and $ < K'_1,D_1 > $ are
the cc systems of $ K'_0 $ and $ K'_1 $, respectively.
In the same way, we consider two inertial systems $ K_0 $
and $ K_1 $ both of which are equivalent to $ K $, whose
cc systems will be denoted by $ < K_0, C_0 > $ and 
$ < K_1, C_1 > $, respectively, and whose origins coincide with
$ P_1 $ and $ Q_1 $, respectively.  Let $ O_0 $ and $ O_1 $
denote the origins of $ K_0 $ and $ K_1 $, respectively.
We also let the time on $ C_0(O_0) $ corresponding to 
time $ t_1 $ on $ C(P_1) $ be zero, and the time on $ 
C_1(O_1) $ corresponding to time $ t_3 $ on $ C(Q_1)  $
is zero.  Since $ K', K'_0 $ and $ K'_1 $ are equivalent
and are distinct only about their origins, and the same situations
hold for $ K, K_0 $ and $ K_1 $, we conclude that the
assertion in the lemma holds due to Principle 2.  $\Box$

\vspace{2ex}

As in the proof of Lemma 1, we can prove the following lemma by
choosing the proper origins.

\vspace{2ex}

{\bf Lemma 2} \hspace{5mm}
Let $ R, S, T $ and $ U $ be four points at rest on $ K'
$ whose x'-coordinates are $ x'_1,x'_2,x'_3 $ and $ x'_4 $,
respectively, and the following hold : $ x'_2 - x'_1 = x'_4 - x'_3 > 0 $, and
the (y',z')-coordinate of $R$ (the (y',z')-coordinate of $T$, respectively)
is the same as that of $S$ (is the same as that of $U$, respectively).
Let $ t' $ be a time on $ K' $.  Let $ P_1,P_2,Q_1,Q_2,
x_1,t_1, x_2, t_2, x_3, t_3, x_4 $ and $ t_4 $ be such that $
P_1 = {\cal T}^{-1}(R,t'), P_2 = {\cal T}^{-1}(S,t'), 
Q_1 = {\cal T}^{-1}(T, t'),
Q_2 = {\cal T}^{-1}(U,t'), (x_1,t_1)= {\cal T}^{-1}( x'_1,t'), 
(x_2,t_2)= {\cal T}^{-1}(x'_2,t'), 
(x_3,t_3)= {\cal T}^{-1}(x'_3,t') $ and $ (x_4,t_4)= {\cal T}^{-1}(x'_4,t') $.
Then it holds $ x_2 - x_1 = x_4 - x_3 $ and $ t_2 - t_1 =
t_4 - t_3 $.

\vspace{2ex}

We shall prove the following proposition by depending on Lemmas 1,2.

\vspace{2ex}

{\bf Proposition 3} \hspace{5mm}
There exist positive constants
$ \alpha_1,\gamma_2,\alpha_3, \gamma_3, \gamma_4 $ and constants
$\gamma_1, \alpha_2, \alpha_4$ such that 
the following hold : for any $ (x,y,z,t)
\in CO(K) $ and $(x',y',z',t') = {\cal T}(x,y,z,t) $,
it holds $ x' = \alpha_1 x + \gamma_1 t, \; t' = \alpha_2 x + \gamma_2 t,
\; x= \alpha_3 x' + \gamma_3 t' $ and $ t= \alpha_4 x' + 
\gamma_4 t' $.

{\bf Proof.}  Let $ R $ be a point at rest on $ K' $ whose 
x'-coordinate is $ x' $.  Consider time $ t' $ on $ D(R) $,
and let $ P,x $ and $ t $ be such that $ P = {\cal T}^{-1}(R,
t') $ and $ (x,t)= {\cal T}^{-1}(x',t') $.  We consider three
cases.

Case (1) : $ x' = 0 $.  We shall present the proof for the case $ 
t' > 0 $.  The case $ t' =0 $ is trivial and the case $ t' < 0 $
is symmetric.  Consider the set $ A $ of points $ Q $ at rest
on $ K $ such that $ Q $ coincides with $ R $ at some time 
$ t_1 $ on $ C(Q) $ with $ 0 \leq t_1 \leq t $.  Then from Lemma 1,
one can see that there exist positive constants $ \gamma_3 $ and
$ \gamma_4 $ such that $ x= \gamma_3 t' $ and $ t= \gamma_4 t' $.

Case (2) : $ x' > 0 $.  We shall present the proof for the case $t
\geq 0 $.  The case $t<0$ can be handled in the same way.
Consider the set $ B $ of points $ S $ at rest on
$ K' $ such that the (y',z')-coordinates of $ S $ is the same
as that of $ R $ and the x'-coordinate $ x'_1 $ of $ S $
satisfies $ 0 \leq x'_1 \leq x' $.  Then due to Lemma 2, one can
see that there exists a positive constant $ \alpha_3 $ to which
the following holds : $ x = vt + \alpha_3 x' $.  From Case (1),
it holds $ vt = \gamma_3 t' $.  Thus it holds  

\[   x = \alpha_3 x' + \gamma_3 t'  \;\;\;\;\;\;\;\; (1)   \]

Now we consider three subcases in order to
show $ t= \alpha_4 x' + \gamma _4 t'$.

Case (2.1) : $ t' = 0 $.  Consider the set $ X $ of points $
Q_1 $ at rest on $ K $ such that the (y,z)-coordinate of
$ Q_1 $ is the same as that of $ P $, and the x-coordinate 
$ x_1 $ of $ Q_1 $ satisfies $ 0 \leq x_1 \leq x $.  Then
due to Lemma 2, one can see that there exist a nonzero constant
$ \delta_1 $ and a constant $ \alpha_4 $ to which the following holds
: $ x= \delta_1 x' $ and (2) $ t = \alpha_4 x' $.  From the consistency
of $ {\cal T} $, it must hold $ \delta_1 = \alpha_3 $ from (1)
in Case (2).

Case (2.2) : $ t' > 0 $.  Let $ P_1 $ be the point at rest on $ K $
which coincides with $ R $ at time zero on $ 
D(R) $, and let $ (x_1,t_1)$ be such that $ (x_1,t_1)= {\cal T}^{-1}(x',0)$.
Then from Case (2.1), it holds $ t_1 = \alpha_4 x'$.  Now consider the set
$ Y $ of points $ Q_2 $ at rest on $ K $ such that $ Q_2 $
coincides with $ R $ at time $ t'_1 $ on $D(R) $ with $ 
0 \leq t'_1 \leq t' $.  Then due to Lemma 1 and Case (1), 
it holds that $ t= t_1 + \gamma_4 t'
= \alpha_4 x' + \gamma_4 t' $.  

Case (2.3) : $ t' < 0 $.  In this case, let $ P_1 $ and $ (x_1,t_1)$
be as in Case (2.2).  Note that $ x_1 >0 $ since $ x' > 0 $.
Let $ Z $ be the set of points $ Q_3 $ at rest on $ K $ which
coincides with $ R $ at time $ t'_1 $ on $ D(R)$ with
$ t' \leq t'_1 \leq 0 $.  By Lemma 1 and Case (2.2), one can see that 
$ t- t_1 = \gamma _4 t' $.  By Case (2.1), it holds $ t_1 =
\alpha _4 x' $.  Thus $ t= \alpha_4 x' + \gamma _4 t' $.

Case (3) : $ x' < 0 $.  This case can be handled as Cases (1),(2)
by considering the subcases of $ t' =0 $, $ t' > 0 $
and $ t' < 0 $, and using the result in Case (1).

By changing the roles of $ K $ and $ K' $, one can see that
there exist positive constants $ \alpha_1, \gamma_2 $ 
and constants $ \gamma_1, \alpha _2 $ to which the 
following hold : $ x' = \alpha_1 x
+ \gamma_1 t, \; t' = \alpha_2 x + \gamma_2 t $.  This completes the proof
of Proposition 3.  $\Box$

\vspace{2ex}

Before presenting TI Lemma, we note the following two remarks.

\vspace{2ex}

{\bf Remark 3} \hspace{5mm}
The Lorentz transformation has the following contradiction within itself.
Let $ K $ and $ K' $ be two parallel Cartesian coordinate systems
with relartive constant x'-axis velocity $ v > 0 $.  We assume that $ c-v $
is very small and $ \beta(v) $ is very large.  Assume that a person 
$ B $ travelling in a spaceship with velocity $ v $ is at rest
on the x'-axis of $ K' $, and $ B $ is spreading his hands on the
x'-axis so that the x'-coordinates at the left and the right hands
of $ B $, $ x'_1 $ and $ x'_2 $, satisfy $ x'_2 - x'_1 = 170 
cm $.  Assume also that a person $ A $ is at rest on $ K $ whose
z-coordinate $ z $ satisfies $ z= - 100 $ m.  Person $ A $ is also
assumed to spread his hands so that the x-coordinates $ x_1 $ and
$ x_2 $ of the left and the right hands of $ A $ satisfy $ x_2
- x_1 = 170 $ cm.  Now consider time $ t' $ over $ K' $, and at this
time, $ B $ is just over $ A $, and let $ x_3 $ and $ x_4 $ be
the corresponding coordinates to $ x'_1 $ and $ x'_2 $ and $ t_1
$ and $ t_2 $ be the times corresponding to $ t' $, respectively.
We acknowledge that this phenomenon
would be physically feasible as a Gedankenexperiment
if Special Relativity Principle is valid
since $ A $ and $ B $ exist in the same
universe.  (To visualize the situation, we may imagine that $A$ is at
rest on the surface of the moon and $B$ is flying in the space very
close to the surface of the moon).
If the Lorentz transformation is valid, then the following hold.

\[ x_4 - x_3 = \beta(v)(x'_2 - x'_1), \; t_2 - t_1 = 
\beta(v)v(x'_2 - x'_1)/c^2 \]

These equations are clearly wrong
since $ \beta(v) $ can be arbitrarily large, and $ x_3,\; x_4,\; t_1 $ 
and $ t_2 $ are coordinates measured over $ K $.  If the Lorentz
transformation is valid, one could not estimate the (present) age of the
universe since the above $t_2-t_1$ may become arbitrarily large.
Thus we have
discovered an inconsistency of the Lorentz transformation.

\vspace{2ex}

{\bf Remark 4} \hspace{5mm}
Let $ K $ and $ K' $ be two parallel inertial systems with x'-axis
velocity $ v>0 $.   Let $ B $ be a spear at rest on the x'-axis
of $ K' $.  Thus $ B $ continues flying over $ K $ with velocity
$ v>0 $.  Let $ R $ and $ S $ denote the end points of $ B $
with $ x'_1 < x'_2 $, where $ x'_1 $ and $ x'_2 $ are the x'-coordinates
of $ R $ and $ S $.  Let $ u_1, u_2 >0 $ be reals.  Assume that
an imaginary signal $ S(K,u_1) $ over $ K $ is emitted from $ R $
to $ S $, and another imaginary signal $ S(K',u_2) $ over
$ K' $ is emitted from $ S $ to $ R $ when $ S(K,u_1)
$ arrives at $ S $.  Then we acknowledge that while the travellings
of $ S(K,u_1) $ and $ S(K',u_2) $ over $ K $ and $ K' $,
respectively, $ B $ continues flying over $ K $ with
velocity $ v>0 $.  Note that we observe this phenomenon by a
Gedankenexperiment without using any measuring devices so that we
do not need any energy for changing the observation from the travelling of
$ S(K,u_1) $ to the travelling of $ S(K',u_2) $.

\vspace{2ex}

The following lemma will be called Time Independence Lemma (in short,
TI Lemma).

\vspace{2ex}

{\bf Lemma 3} \hspace{5mm}
Let $ K $ and $ K' $ be two parallel inertial
systems with relative constant x'-axis velocity $ v>0 $.  Then
for any event $ E $ occurring over $ K $ and $ K' $, while
$ E $ is occurring, the whole $ K' $ continues moving into the
positive x-axis direction of $ K $.  Particularly the
following hold.

\begin{enumerate}
\item For any real $ u>0 $ and two distinct points $ P $ and $ Q
$ at rest on $ K $, the whole $ K' $ continues moving into the
positive x-axis direction of $ K $ while imaginary signal
$ S(K,u) $ over $ K $ travels from $ P $ to $Q $.

\item For any real $ u>0 $ and two distinct points $ R $ and $ S $
at rest on $ K' $, the whole $ K' $ continues moving into the
positive x-axis direction of $ K $ while imaginary signal
$ S(K',u) $ over $ K' $ travels from $ R $ to $ S $.

\item Let $ H=<K,C> $ and $ J=<K',D> $ be the cc systems of 
$ K $ and $ K' $, respectively.  Let $ R $ and $ S $
be two points at
rest on the x'-axis of $ K' $ whose x'-coordinates are $
x'_1$ and $x'_2$ ($ x'_1 < x'_2$), respectively.
Let $ u_1,u_2 >0 $ be sufficiently large reals.  Let $ t_1 $ be a time
over $ K $, and consider the following situation.
 \begin{enumerate}
 \item $R$ coincides with  point $P_0$ at rest on the x-axis of $K$  
 at time $ t_1 $ on $ C(P_0) $. 

 \item Imaginary
signal $ S(K,u_1) $ is emitted over $ K $ from $ P_0 $ at time $ t_1
$ on $ C(P_0) $ and arrives at $ Q $ at time $ t_2 $ on $ C(Q)
$, where $ Q $ is the point at rest on the x-axis of $ K $ which
coincides with $ S $ when $ S(K,u_1) $ arrives at $ S $.

 \item Imaginary signal $ S(K',u_2) $ is emitted 
over $ K' $ from $ S $ at time $ t_2 $ on $ C(Q) $
and arrives at $ R $ at time $ t_3 $ on $ C(P_1) $, where $ P_1 $ is
the point at rest on the x-axis of $ K $ which coincides with 
$ R $ when $ S(K',u_2) $ arrives at $ R $.
 \end{enumerate}

Let $ x_0 $ and $ x_1 $ be the x-coordinates of $ P_0 $ and $ P_1
$, respectively.  Then it holds $ x_0 \leq x_1 $.

\item  Let $ H,J,R,S,x'_1,x'_2 $, and $ u_1,u_2 >0$ 
be as in (3)
except $ x'_2 < x'_1 $.
Let $ t_1 $ be a time
over $ K $, and consider the following situation.
 \begin{enumerate}
 \item $R$ coincides with  point $P_0$ at rest on the x-axis of $K$  
 at time $ t_1 $ on $ C(P_0) $. 

 \item Imaginary
signal $ S(K,u_1) $ is emitted over $ K $ from $ P_0 $ at time $ t_1
$ on $ C(P_0) $ and arrives at $ Q $ at time $ t_2 $ on $ C(Q)
$, where $ Q $ is the point at rest on the x-axis of $ K $ which
coincides with $ S $ when $ S(K,u_1) $ arrives at $ S $.

 \item Imaginary signal $ S(K',u_2) $ is emitted 
over $ K' $ from $ S $ at time $ t_2 $ on $ C(Q) $
and arrives at $ R $ at time $ t_3 $ on $ C(P_1) $, where $ P_1 $ is
the point at rest on the x-axis of $ K $ which coincides with 
$ R $ when $ S(K',u_2) $ arrives at $ R $.
 \end{enumerate}
Let $ x_0 $ and $ x_1 $ be the x-coordinates of $ P_0 $ and $ P_1
$, respectively.  Then it holds $ x_0 \leq x_1 $.
\end{enumerate}

{\bf Proof.}  (1) and (2) are clear by definition of imaginary signals and
the fact that $ K $ and $ K' $ are synchronized, respectively.
(3).  Assume that the situations in (3) hold.  We shall present three 
proofs of (3) whose main ideas are almost the same and rely on our
observation that 
the time passes from the past to the future over any inertial system
(so that (1) and (2) hold).

The first proof of (3).
Assume that $ x_1 < x_0 $.  We note that when imaginary signal
$ S(K,u_1) $ is emitted from $ R $, $ R $ coincides with $ P_0 $,
and $ x_1 < x_0 $ implies that when imaginary signal $
S(K',u_2) $ arrives at $ R $ and $ P_1 $, it holds
$ R $ is to the
negative direction from $ P_0 $ on the x-axis of $ K $.  Then we note that
while $ S(K,u_1) $ travels form $ P_0 $ to $ Q $, the whole $
K' $ has moved into the positive x-axis direction of $ K $
even when $ u_1 $ is arbitrarily large.  Thus while $ S(K',u_2) $
travels from $ S $ to $ R $, at least a part of $ K' $ has
moved into the negative x-axis direction of $ K $.  Particularly the
time at $ R $ over $ K' $ passes from a future time to a past time
while $ S(K',u_2) $ travels from $ S $ to $ R $.  We 
note that the travelling of $ S(K',u_2) $ corresponds to a (positive) 
time lapse over $ K' $.  Thus we conclude that these observations
produce a contradiction to (2) since $ K' $ is synchronized.

The second proof of (3).  Assume that $ x_1 < x_0 $.  As in the first
proof, it holds that while $ S(K',u_2) $ travels from $ S $ to
$ R $, at least a part of $ K' $ containing $ R $ has moved
into the negative x-axis direction of $ K $.  Now as a Gedankenexperiment,
we consider the travelling of a person instead of imaginary
signals.  Assume that a person $ A $ is at rest at $ R $ on $ K'
$, and another person $ B $ travels from $ R $ to $ S $
over $ K $ with velocity $ u_1 $, arives at $ S $ at time $
t_2 $ on $ C(Q) $, travels from $ $ S to $ R $ over $ K' $
with velocity $ u_2 $ and arrives at $ R $ at time $ t_3 $
on $ C(P_1) $.  We note that the travelling of $ B $ corresponds
to merely a time lapse over $ K $ and $ K' $, and we observe only
successive time lapses over $ K $ and $ K' $.
Assume that from time $ t_3 $ on $ C(P_1) $, $ B $ continues his
travel with velocity $ u_2 $ over $K'$ and arrives at a point $ T $ at rest
on the x'-axis of $ K' $ whose x'-coordinate is $ x'_0 $ with
$ x'_0 < x'_1 $.  Assume that $ B $ arrives at $ T $ at time
$ t_4 $ on $ C(P_2) $, where $ P_2 $ is the point at rest on the
x-axis of $ K $ with x-coordinate $ x_2 $ which coincides with $T$ when
$S(K',u_2)$ arrives at $T$.
Without loss of
generality, we may assume that $ u_2 > v $ and $ x_2 < x_1 $.
Now assume that $ B $ begins his travel from $ T $ to $ R $
over $ K $ with velocity $ u_3 >0 $ such that $ B $ arrives
at $ R $ at a time $ t_5 $ on $ C(P_3) $ with $ t_5 > t_1 $,
where $ P_3 $ is the point at rest on the x-axis of $ K $
which coincides with $ R $ when $ B $ arrives at $ R $.
Since $ R \neq T $ and $ x_2 < x_1 $, clealy such $ u_3 $
exists.  Thus during the time period while $ A $ moves from
$ P_0 $ to $ P_3 $ over $ K $, $ B $ travels from $ R $
to $ S $, $ S $ to $ R $, $ R $ to $ T $ and $ T $
to $ R $.
Since $ x_1 < x_0 $, we conclude that at least a part of the 
travelling of $ B $ occurs in a universe which is different from
the universe where $ A $ lives since in the A's universe, during
the travelling of $ B $, $ K' $ continues moving into the
positive x-axis direction of $ K $.  But this is impossible since
$ S(K,u_1), S(K',u_2) $ and $ S(K,u_3) $ travel over $ K $ and $ K' $
(travellings from  past times to  future times over $ K, K' $
and $ K $, respectively), and $ K $ and $ K' $ exist in the
A'universe.

The third proof of (3). To visualize the situation more concretely,
we consider the travellings of imaginary signals between Andromeda galaxy
and the sun with certain idearization.  As a Gedankenexperiment, we
assume that Andromeda galaxy is at rest on the x-axis of an inertial
system $ K $ and the sun is at rest on the x'-axis of another inertial
system $ K' $ so that $ K $ and $ K' $ are two parallel inertial
systems with relative constant
x'-axis velocity $ v > 0 $.  By simplicity, we assume
that Andromeda galaxy is at rest at a point $ P_0 $ on the x-axis of $ K
$ with x-coordinate $ x_0 $ and the sun is at rest at a point $ S $
on the x'-axis of $ K' $ with x'-coordinate $ x'_2 $.  Let $ H=<K,C)$
and $J=<K',D>$ be the cc systems of $ K $ and $ K' $, respectively.
Assume that a light signal $ X $ is emitted from $ P_0 $ at time
$ t_0 $ on $ C(P_0) $ and arrives at $ S $ at time $ t'_0 $
on $ D(S) $.  We assume that $ t'_0 $ is a time in the year of 1996
over $ K' $.  Let $ u_1 $ and $ u_2 $ be as above.  Assume that
imaginary signal $ S(K,u_1) $ is emitted from $ P_0 $ at time $ t_1 $
on $ C(P_0) $ and arrives at $ Q $ at time $ t_2 $ on $ C(Q) $
and at time $ t'_2 $ on $ D(S) $,
where $ Q $ is as above.  We consider the case where $ t'_0 < t'_2 $.
As above, let $ R $ be the point at rest on the x'-axis of $ K' $
which coincides with $ P_0 $ at time $ t_1 $ on $ C(P_0) $.
Since $ X $ is a light signal and we assume $ u_1 > c $, it holds
$ t_0 < t_1 $.  Let $ L $ denote the line segment connecting
$ R $ and $ S $ over $ K' $.  For each point $ V $ at rest
on $ L $, Let $ C(V,u_1) $ and $ D(V,u_1) $ be times shown on $
C(V) $ and $D(V) $ when $ S(K,u_1) $ arrives at $ V $.  We acknowledge that
the time in the universe proceeds from the past to the future, and for
each point $ V $ at rest on $ L $, after $ S(K,u_1) $ arrives at
$ V $, all times $ t $ at $ V $ over $ K $ with $ t<
C(V,u_1) $ and all times $ t' $ at $ V $ over $ K' $ with $ t'
< D(V,u_1) $ disappeared in the universe and only such times $ t>
C(V,u_1) $ at $ V $ over $ K $ and such times $ t' >
D(V,u_1) $ at $ V $ over $ K' $ can occur at $ V $.  We
admit that the universe continues aging, and $ K $ and $ K' $
have the common space.  Now assume that imaginary signal $
S(K',u_2) $ is emitted from $ S $ at time $ t'_2 $ on $
D(S) $ and at time $ t_2 $ on $ C(Q) $, and arrives at $ R $ at
time $ t'_3 $ on $ D(R) $ and at time $ t_3 $ on $ C(P_1) $,
where $ P_1 $ is as above.  For each point $ V $ at rest over
$ L $, let $ C(V,u_2) $ and $ D(V,u_2) $ be times shown on
$ C(V) $ and $ D(V) $ , respectively, when $ S(K',u_2) $ arrives at $ V $.
From the above observation, we acknowledge that for each point $
V $ at rest on $ L $, it holds $ C(V,u_2) > C(V,u_1)
$ and $ D(V,u_2) > D(V,u_1) $.  Particularly it holds $
C(R,u_2) > C(R,u_1) $.  This implies that it holds $ x_1 > x_0 $.
Moreover we may observe the following.
Let $ E_1 $ and $ E_2 $ denote the travellings of
imaginary signals $ S(K,u_1) $ and $ S(K',u_2) $ from
$ P_0 $ to $ S $ over $ K $ and from $ S $ to $ P_0 $
over $ K' $, respectively.  Let $ E_1 + E_2 $ denote the 
composite event in whch $ E_1 $ firstly occurs and $ E_2 $
successively occurs.  By induction, for $ n \geq 2 $, let
$ n(E_1 + E_2) $ denote the composite event in which $
(n-1)(E_1 + E_2) $ firstly occurs and $ E_1 + E_2 $ successively
occurs. As above, assume it holds $ x_1 < x_0 $.  Then one can
see easily that for any sufficiently large $ n>0 $, in the event
$ n(E_1 + E_2) $, the n-th travelling of $ S(K',u_2) $ could
arrive at $ P_0 $ at a time $ t_6 $ on $ C(P_0) $
with $ t_6 < t_0 $.  This is clearly a contradiction since we admit
that Andromeda galaxy at time $ t_0 $ disappeared in the universe,
and does not exist at any place in the universe of the present age.
This completes the third proof of (3).

(4) can be proved in the same way.  This completes the proof of Time
Independence Lemma.  $\Box$

\vspace{2ex}

{\bf Remark 5} \hspace{5mm}
In special relativity, the possibility of the existence of
time machines has been discussed very often.  One of the typical
arguments presents a conclusion that over an inertial system $ K $,
one would go back to a past history over $ K $ by depending on the
existence of time machines whose velocity can be arbitrarily large
(see, e.g., [15]).  In the above proof of TI Lemma, if (3) does not hold,
then imaginary
signal $ S(K',u_2) $ can go into the past histories over both
$ K $ and $ K' $, which, we conclude, is impossible.  We admit the time passes
from the past to the future and the "positive" direction of time's arrow
is one of the most fundamental property of the universe [8,10] (see also
Appendices 3,4).

\vspace{2ex}

The following theorem will be called Time Independence Theorem (in short,
TI Theorem).

\vspace{2ex}

{\bf Theorem 1} \hspace{5mm}
Let $ K $ and $ K' $ be two parallel inertial systems with relative
constant x'-axis  velocity $ v > 0 $.  Let $ {\cal T} $
be the space-time transformation from $ K $ to $ K' $ as in
Proposition 3.  Then it holds $ \alpha_2 =
\alpha_4 =  0 $, and Fact A in Section 1 holds.

{\bf Proof.}  Assume that it holds $ \alpha_2 \neq 0 $ 
or $ \alpha_4 \neq 0 $.
Since $ {\cal T} $ is a bijection, it holds $ \alpha_2 
\alpha_3 + \alpha_4 \gamma_2 =0 $ and $ \alpha_2 =0 $ implies $ \alpha_4 = 0
$.  In the same way, $ \alpha_4 =0 $ implies $ \alpha_2 = 0
$.  Thus we may assume that $ \alpha_2 \alpha_4 
\neq 0 $.  We consider the case where
$ \alpha_2 < 0 $, and depend on (3) of TI Lemma.
For the case $ \alpha_2 > 0 $, it suffices to apply (4) of TI Lemma.
Let $ H=<K,C> $ and $ J=<K',D> $ be the cc systems of $ K
$ and $ K' $, respectively.  Let $ R $ and $ S $ be two
points at rest on the x'-axis of $ K' $ whose x'-coordinates
are $ x'_1 $ and $ x'_2 $ with $ x'_1 < x'_2 $.
Put $ l= x'_2 - x'_1 >0 $.
Let $ u_1, u_2 >0 $ be sufficiently large
reals.  
Consider time $ t>0 $ over $ K $, and assume that the origin $
O $ of $ K $ coincides with $ R $ at time $ t $ on
$ C(O) $, and imaginary signal $ S(K,u_1) $ over $K$ is
emitted from $ O $ to $ S $ at time $ t $ on $ C(O) $,
and arrives at $ S $ at time $ t+a $ on $ C(Q) $
with $ a $ being very small, where $ Q $ is the point
at rest on the x-axis of $ K $ which coincides with $ S $
when $ S(K,u_1) $ arrives at $ S $.  Let $ t'_1 $
and $ t'_2 $ be times on $ D(R) $ and $ D(S) $, respectively, such that
$ O $ and $ R $ coincide at time $ t'_1 $ on $ D(R) $,
and $ Q $ and $ S $ coincide at time $ t'_2 $ on $ D(S) $.  Assume also
that imaginary signal $ S(K',u_2) $ is emitted from $ S $ at time
$ t'_2 $ on $ D(S) $ and arrives at $ R $ at time $ t'_3 $
on $ D(R) $.
One can see easily that the x-axis coordinate $ x_2 $
of $ Q $ satisfies $ x_2 = \alpha_3 l + \gamma_3(t'_2 - t'_1) $ since
$0= \alpha_3 x'_1 + \gamma_3 t'_1$,
and the following hold. 

$ t'_1 = \gamma_2 t, \; t'_2 = \alpha_2(\alpha_3 l + 
\gamma_3(t'_2 - t'_1))+ \gamma_2(t+a)$

$\simeq \alpha_2(\alpha_3 l + \gamma_3(t'_2 - t'_1)) + \gamma_2 t, 
\; t'_2 - t'_1 \simeq 
\alpha_2(\alpha_3 l + \gamma_3(t'_2 - t'_1))$,

$(t'_2 - t'_1)(1- \alpha_2 \gamma_3) \simeq \alpha_2 \alpha_3 l, \; 
t'_3 - t'_1 = t'_2 +l / u_2 - t'_1 $

$\simeq t'_2 - t'_1 \simeq \alpha_2 \alpha_3 l/(1- \alpha_2 \gamma_3) <0 $

Here it holds $ \gamma_3 > 0 $ due to Proposition 3.
Let $ P_1 $ be the point at rest on the x-axis of
$ K $ which coincides with $ R $ when $ S(K',u_2) $ arrives
at $ R $ at time $ t'_3 $ on $ D(R) $.  Let $ x_1 $ be the
x-coordinate of $ P_1 $.  Then the following hold.

$ 0= \alpha_3 x'_1 + \gamma_3 t'_1, \; x_1 = \alpha_3 x'_1 + 
\gamma_3 t'_3 \simeq 
\alpha_3 x'_1 + \gamma_3(t'_1 + \alpha_2 \alpha_3 l/(1- \alpha_2 \gamma_3)) $

Thus  

$ x_1 - 0 \simeq \alpha_2 \alpha_3 \gamma_3 l/(1- \alpha_2 \gamma_3) 
<0, \;  x_1 < 0 $

This is a contradiction to (3) of TI Lemma.  Thus $ \alpha_2 =
\alpha_4 = 0 $.  For the case of $\alpha_2 >0$,
we apply (4) of TI Lemma.  Here we remark
that $x_2= - \alpha_3 l + \gamma_3 (t'_2 - t'_1)$ and it holds $ 1-
\alpha_2 \gamma_3 > 0 $ since when $u_1 $ is sufficiently small and
$\gamma_2 a > \alpha_2 \alpha_3 l $, it must
hold $t'_2 - t'_1 = (- \alpha_2 \alpha_3 l + \gamma_2 a)/(1- \alpha_2
\gamma_3) > 0 $.
Now Fact A is clear.
This completes the proof of Theorem 1 (TI Theorem).  $\Box$

\vspace{2ex}

From TI Theorem and Proposition 3, the following corollary is clear.

\vspace{2ex}

{\bf Corollary 1} \hspace{5mm} (1) Let $H,J,R,S,x'_!,x'_2,u_1,u_2,x_0$
and $x_1$ be as in (3) of TI Lemma.  Then it holds $x_0 < x_1$.

(2) Let $H,J,R,S,x'_1,x'_2,u_1,u_2,x_0$ and $x_1$ be as in (4) of TI
Lemma.  Then it holds $x_0 < x_1$.

\vspace{2ex}

{\bf Definition 1} \hspace{5mm}
Let $ K $ and $ K' $ be two parallel inertial systems with relative
constant x'-axis velocity $ v>0 $.  Let $ R $ and $ S $ be two distinct
points at rest on $ K' $.  Let $ u_1,u_2 $ be sufficiently large
reals.  Let $ E_1,E_2,E_3 $ and $ E_4 $ denote the travellings
of imaginary signals $ S(K,u_1),S(K,u_1),S(K',u_2) $ and $ S(K',u_2) $,
respectively, from $ R $ to $ S $ over $ K $, from $ S $
to $ R $ over $ K $, from $ R $ to $ S $ over $ K'
$, and from $ S $ to $ R $ over $ K' $, respectively.
For $ i_1,i_2 \in \{1,2,3,4 \} $, let $ E_{i_1} + E_{i_2} $ denote the 
composite event in which $ E_{i_1} $ firstly occurs, and $ E_{i_2} $
successively occurs, where we refer to Fact 2-(1).  
By induction, define $ E_{i_1} + \cdots + 
E_{i_m}  = (E_{i_1} + \cdots + E_{i_{m-1}}) + E_{i_m} $ for $ i_j \in
\{1,2,3,4 \} $.

\vspace{2ex}

{\bf Corollary 2} \hspace{5mm}
Let $ K $ and $ K' $ be two parallel inertial systems with relative
constant
x'-axis velocity $ v>0 $.  Let $ u_1, u_2 >0 $ be sufficiently large, and
let $ E_1, E_2, E_3 $ and $ E_4 $ be as in Definition 1.  For any
$ i_1, i_2, \cdots, i_m, j_1, j_2, \cdots, j_m \in \{1,2,3,4\} $, if
$ (j_1, \cdots,j_m) $ is a permutation of $ (i_1,\cdots,i_m) $, then
it holds $ E_{i_1} + \cdots + E_{i_m} \equiv E_{j_1} + \cdots + E_{j_m} $,
i.e.,  the time lapse experienced at $ R $ while $ E_{i_1} + \cdots
+ E_{i_m} $ occurs is the same as the tima lapse experienced at $ R $
while $ E_{j_1} + \cdots + E_{j_m} $ occurs.  Thus the law of superposition
holds about the time lapses over $ K $ and $ K' $.

\vspace{2ex}

Let $ K $ and $ K' $ be two parallel coordinate system with
relative constant x'-axis velocity
$ v> 0 $ and the space-time transformation between $ K $ and $ K' $
be the Lorentz transformation.  Then the law of superposition as in
Corollary 2 does not hold (see (3) of Apendix 3). In Appendix 4, we
present other odd properties of the Lorentz transformation.

The following theorem also holds.

\vspace{2ex}

{\bf Theorem 2} \hspace{5mm}
\begin{enumerate}
\item The Lorentz transformation is not the space-time transformation
over two inertial systems moving with nonzero constant relative
velocity.
\item Strong Constant Light Velocity Principle is not valid.
\item All inertial systems are not equivalent, i.e., Special Relativity
Principle is not valid.
\item Statement 2 is false.
\end{enumerate}

{\bf Proof.}  (1) is clear from TI Theorem.  Let $ K $ and $ K' $
be two parallel inertial systems with relative constant x'-axis velocity
$ v>0 $.  We assume the velocity of a light signal over
$ K $ is $ c $ independent from its direction.
Now it suffices to prove (2)
since (2) implies the Maxwell equations over $ K' $ are not of the
same form as those over $ K $, and
(3)-(4) follow from (2).
Let $
H = <K,C>$ and $ J = <K',D>$ be the cc systems of $ K
$ and $ K' $, respectively.
Let $ P $ and
$ R $ denote the origins of $ K $ and $ K' $, respectively.
Let $ t >0 $ and $ Q $ be the point at rest on the x-axis of
$ K $ whose x-coordinate is $ ct $.  Let a light signal $ L_1
$ be emitted from $ P $ at time $ 0 $ on $ C(P) $ and 
arrive at $ Q $ at time $ t $ on $ C(Q) $.  Let $ S $
be the point at rest on the x'-axis of $ K' $ which coincides
with $ Q $ when $ L_1 $ arrives at $ Q $.  Let $ t'_1
$ be the time shown on $ D(S) $ when $ S $ coincides with
$ Q $.  By TI Theorem, it holds $ t'_1 = \gamma_2 t $. 
Let a light signal $ L_2 $ be emitted at $ Q $ at time
$ t $ on $ C(Q) $, and arrive at $ R $ at time $ t'_2 
$ on $ D(R) $.  Let $ P_0 $ be the point at rest on the
x-axis of $ K $ which coincides with $ R $ when $ L_2
$ arrives at $ R $.  Let $ (x_2,t_2) $ be the pair of
x-coordinate and time on $ K $ at $ P_0 $ corresponding to
$ (0,t'_2) $ at $ R $.  Then it  holds $ t_2 - t< t $ 
since $ v<c$ and $ ct >
x_2 > 0$.  By $ {\cal T} $, it holds $t'_2 - t'_1 = \gamma_2(t_2 -t)
< \gamma_2 t =t'_1$.  This implies that the velocities of $ L_1 $
and $ L_2 $ over $ K' $ are distinct, completing the proof.
$\Box$

\vspace{2ex}

We have shown that TI Lemma and TI Theorem hold.  This implies that
Strong Constant Light Velocity Principle is not valid as stated in
Theorem 2-(2).  Thus we must acknowledge that the result discovered
in the Michelson-Morley experiment is not due to Strong Constant
Light Velocity Principle, but due to other causes.  We shall present
a subprinciple of the twofold metric principle in Subsection 5.3
which may explain why the Michelson-Morley
experiment is observed.

We present the following properties of the Lorentz transformations.

\vspace{2ex}

{\bf Theorem 3} \hspace{5mm}
Let $ K $ and $ K' $ be two parallel coordinate systems with relative
constant x'-axis  velocity $ v>0 $ such that the space-time
transformation from $ CO(K)$ onto $CO(K')$ is the Lorentz transformation.
Let $ H=<K,C> $ and $J=<K',D>$ be the cc systems of $ K $ and 
$ K' $, respectively.
Let $ P $ and $ Q $ be two points at rest on the x-axis of $ K $
whose x-coordinates are $ x $ and $ x+ut $, respectively, where $
t>0 $ and $ u \neq 0 $.  Let $ t_1 $ be a time on $ C(P) $ and
assume that an imaginary signal $ S(K,u) $ is emitted from $P$ at time 
$ t_1 $ on $ C(P) $ and arrives at $ Q $ at time $ t_2 =
t_1 + t $ on $ C(Q) $.  Here if $ u<0 $, then the signal $ S(K,u)
$ travels into the negative direction of the x-axis of $ K $
so that it holds $ x+ut < x $.  Let $ R $ and $ S $ be two points
at rest on the x'-axis of $ K' $ such that $ R $ and $ S
$ coincide with $ P $ and $ Q $, respectively, at time
$ t_1 $ on $ C(P) $ and at time $ t_2 $ on $ C(Q) $,
respectively.  Let $ (x'_1,t'_1) $ and $ (x'_2,t'_2) $ be the pairs
of x'-coordinates and times at $ R $ and $ S $, respectively,
corresponding to $ (x,t) $ and $ (x+ut,t_1 +t) $ at $ P
$ and $ Q $, respectively.  Let $ u'(u) $ denote $ (x'_2 -
x'_1)/(t'_2 - t'_1) $ when $ t'_2 - t'_1 \neq 0 $, and
$ t'(u) $ denote $ t'_2 - t'_1 $.  Then the following hold, where
$ \beta(v) = c/\sqrt{c^2 - v^2} $.

\begin{enumerate}
\item $ u'(c) = c$, $ u'(-c)= -c $, $ u'(v)= 0 $ and $u'(u) \neq u$
for all $ u \not \in \{c,-c \} $.
\item $ t'(u) = \beta(v) t(1- uv/c^2) $ and $t'(u) < 0 $ if $ u> c^2/v $.
\item $ u'(u) = c^2(u-v)/(c^2 - uv) $ if $ u \neq c^2/v $.
\item $ \displaystyle{ \frac{ d u'(u)}{du} = \frac{c^2(c^2 -v^2)}
{(c^2 - vu)^2} > 0 } $ if $ u \neq c^2/v $.
\item $ \displaystyle{ \frac{d t'(u) }{du} = - \beta(v) vt/ c^2 <0 } $.
\end{enumerate}

{\bf Proof.}  By the Lorentz transformation, the following hold.

(6) $ x'_1 = \beta(v)(x- v t_1) $, $ x'_2 = \beta(v) (x+ ut -v(t_1 + t)) $

(7) $ t'_1 = \beta(v)(t_1 - vx/c^2) $, $ t'_2 = \beta(v)(t_1 + t - v(x+ut)
/c^2) $

By (6) and (7), the assertions can be proved easily. $\Box$

\vspace{2ex}

From (1) in the above theorem, one observes any velocity $u'$ except
$c$ and $-c$ over the x'-axis of $K'$ is observed velocity $u$ over
the x-axis of $K$ with $u \neq u'$.

In the rest of this section, we shall study properties of inertial systems
by depending on TI Theorem.
Throughout the rest of this section, let $K$ and $K'$ 
be two parallel inertial systems with relative constant x'-axis 
velocity $v > 0$.  Let $ H = <K,C> $ and $ J = <K',D> $ be the
cc systems of $K$ and $K'$, respectively.  Then $H$
and $J$ are synchronized, respectively, by definition. 
Moreover we assume that the velocity of 
a light signal over $K$ is $c$ independent of its direction.
Let $ {\cal T} $ be the space-time transformation from 
$ K $ to $ K' $.   From Proposition 3, TI Lemma and TI Theorem, the 
following theorem and corollaries hold.

\vspace{2ex}

{\bf Theorem 4} \hspace{5mm}
Let $P$ and $Q$ be two points at rest on $K$, and consider time
$t_1$ on $C(P)$ and $C(Q)$.  Let $R$ and $S$ be two points 
at rest on $K'$ which
coincide with $P$ and $Q$, respectively, at time $t_1$ on $C(P)$ and
$C(Q)$.  Let $t'_1$ and $t'_2$ be the corresponding times at $R$
and $S$ on $D(R)$ and $D(S)$, respectively. Then it holds $ t'_1 =
t'_2 $.

\vspace{2ex}

{\bf Corollary 3} \hspace{5mm}
For any point $P$ at rest on $K$, if a point $R$ at rest on $K'$
coincides with $P$ at time zero on $C(P)$, then the corresponding 
time shown on $D(R)$ is also zero.

\vspace{2ex}

{\bf Corollary 4} \hspace{5mm}
For any point $R$ at rest on $K'$ and for any point $P$ at rest on $K$,
the ratio of the speed of the time lapse at $R$ on $K'$ w.r.t.
that at $P$ on $K$ is constant.

\vspace{2ex}

In Proposition 3, one can see that $ \gamma_1 = 
- \alpha_1 /v $ since the origin $ O'$ of $ K' $ coincides
with point $ P $ at rest on the x-axis of $ K' $ at time $ t $
on $ C(P)$ such that the x-coordinate of $ P $ is $ vt $.
Thus from Propositions 1,3  and TI Theorem,
the following theorem holds.

\vspace{2ex}

{\bf Theorem 5} \hspace{5mm}
There exist positive constants $ \alpha _1, \alpha _2
,\alpha _3,\alpha _4 $ such that the space-time transformation ${\cal T}$ from
$K$ to $K'$ satisfies the following.

For any coordinate $ (x,y,z,t) $ on $K$, the corresponding coordinate
$ (x',y',z',t') = {\cal T}(x,y,z,t)$ on 
$K'$ is such that $ x' = \alpha _1(x - vt)
,y' = \alpha _2 y, z' = \alpha _3 z $ and $ t' = \alpha _4 t$.

\vspace{2ex}

Note that the validity of Theorem 5 presents solutions to the twin
paradox and the garage paradox.
Instead of Strong Constant Light Velocity Principle, the corresponding
theorem is presented in Appendix 5.

\section{The twofold metric principle}

In Section 4, we show that all inertial systems are not equivalent, and
the Lorentz transformation is not the space-time transformation over
two inertial systems moving with relative nonzero constant velocity.  We
acknowledge that an inertial system $ K $ is stationary in the vacuum
space if there does not exist any material body with huge mass near $
K $, and the Maxwell equations of the standard form hold over $ K $.
Thus over a stationary inertial system, the light velocity is $ c $
independent of its direction.  The velocity of any inertial system $
K' $ may be determined by its relative velocity w.r.t. any stationary
inertial system which may coexist over $ K' $.
In general relativity, one often asserts that any physical phenomenon can
be described by a set of equations over an arbitrarily chosen coordinate
system.  However this assertion obviously contains certain misleading
feature.  For example, let $ K $ be a stationary inertial system over
which the Maxwell equations of the standard form hold.  Let $ K' $ be
a Cartesian coordinate system such that $ t' =t, y' = y, z' = z  $
but $ x' = 2x $.  Clearly the Maxwell equations of the standard form
do not hold over $ K' $.  We may describe any physical phenomenon $
E $ in one of the simplest forms of a set of equations only when we 
choose one of the most appropriate coordinate systems for $ E $.

In this section, we shall present the twofold metric principle, which
may replace the relativity principle and is described as a combination of
four subprinciples, and solve several well known
problems appearing in general relativity by depending on the twofold
metric principle and a newly introduced equation named the
energy-velocity equation.
These problems are related with gravity or nonzero velocities
w.r.t. the universe.  Throughout this section, we often use the equality
sign $=$ instead of the approximation sign $ \approx $ when
the conrext is clear.  Thus an equation $ A=B $ often means $A$ and
$B$ are approximately equal.
We acknowledge that the time (the most "fundamental" time) at
point at reat over a stationary inertial asystem 
in the universe (or at least in a part of the unverse
which this paper concerns) proceeds with the same speed (rate).  We call this
time the global time.  We also admit that for any material body $ B $
with rest mass $ m $, the total energy $ E_T $ of $ B $ is the same
over any coordinate system, and the time at $ B $
proceeds more slowly than the global time by factor $  
1/ \beta(v) $, where $ E_T = \beta(v) mc^2 $, and $ \beta(u) =
c/ \sqrt{c^2 - u^2}=1/ \sqrt{1-u^2/c^2}
$ for $ 0 \leq u <c $.  These observations may be
explained more in detail in the sequel.

\vspace{2ex}

{\bf 5.1  Subprinciple 1.} \hspace{5mm} In this subsection, we study
the motion of a material body $ B $ in a space where the gravitational
potential can be negligible.
The velocity of $ B $ can be related with its total
energy and its rest mass energy as can be seen in the following.
Due to the experimental results, we acknowledge that the total energy of
$ B $ moving with velocity $ v>0 $ and with rest mass $ m>0 $
is $ \beta(v)mc^2 $.  Due to this result, we acknowledge  the following
equation is fundamental.

\vspace{2ex}

{\bf The energy-velocity equation (in short, the ev-equation)} :

\[  \frac{d ((E_0/E(v))^2)}{d(v^2)} = - \frac{1}{c^2}  
\; \; \;\;\;\;\; (1)  \]

\vspace{2ex}

Here $ E_0 $ is the rest energy of
$ B $ with mass $ m>0 $ and $ E(v) $ is the total energy of $
B $ moving with velocity $ v \geq 0 $.  By integration, it holds

\[ \left (\frac{E_0}{E(v)}\right)^2 = \int_0^{v^2}\left(-\frac{1}{c^2}
\right)d(u^2) = a - \frac{v^2}{c^2}  \;\;\;\;\;\;\; (2) \]

Since $ E(0)= E_0 $, it holds $ a= 1 $.  Thus

\[  E(v) = \beta(v) E_0  \; \; \;\;\;\;\; (3) \]

When $ v $ is small, it holds $ \beta(v) \approx 1+ v^2/2c^2 $
and $ mv^2/2 \approx (\beta(v) - 1)E_0 \approx E_0 v^2/2c^2$.
Here $mv^2/2$ is the kinematic energy of $B$ at velocity $0\leq v
<< c$.
Hence

\[  E_0 = mc^2 \; \; \;\;\;\;\; (4) \]

From the ev-equation, we have derived (3) and (4).  
We shall propose our
first subprinciple which presents a new space-time transformation for the
case where an outer force $ \vec{F} = ( F_x,F_y,F_z) $
works over  $B$ moving  with velocity $ \vec{v} = (v_x,
v_y,v_z) $ with mass $ m>0 $ and with gravitational potential being almost
zero.  We say that the direction of $ \vec{F} $ is a maximal
velocity-critical direction , and the direction of any vector which is
perpendicular to $\vec{F}$
is a zero velocity-critical direction.
We remark that if we remove $ \vec{F} $, then $ B $ continues
moving in the direction of $ \vec{v} $ with the same velocity.  About the
travelling of light, we shall admit that any direction is a maximal
velocity-critical one.  The reason of this acknowledgement will be presented
in Subsection 5.3.

\vspace{2ex}

{\bf Subprinciple 1}  \hspace{5mm}  
Let $ K $ be a stationary inertial system,
and $ B,m,\vec{F} $ and $ \vec{v} $ be as above.  Let 
$ K' $ be a Cartesian coordinate system such that $ x' = \beta(v) x,
y' = \beta(v) y, z' = \beta(v) z $ and $ t' = t/ \beta(v) $, where
$ v= \sqrt {v_x^2 + v_y^2 + v_z^2 } $.  Then it holds

\[  m\left(\frac{ d^2x' }{ dt'^2 }, \frac{ d^2y' }{ dt'^2 },
\frac{ d^2z' }{ dt'^2 }\right) = (F_x,F_y,F_z)  \; \;\;\;\;\;\; (5) \]

Thus it holds

\[  m \left( \frac{ d^2x }{ dt^2 }, \frac{ d^2y }
{ dt^2 }, \frac{d^2z}{dt^2} \right) =  \vec{F}/\beta(v)^3  
\; \;\;\;\;\;\; (6) \]

Moreover it holds the time at $B$ with velocity $\vec{v}$ proceeds more
slowly than the global time by factor $1/\beta(v)$.

\vspace{2ex}

In the above subprinciple, it holds

\[ \frac{ d^2x' }{ dt'^2 }= \frac{ d}{ dt' } \left( \frac{ dx' }{ dx}
\frac{dx}{dt}\frac{dt}{ dt' } \right)
 = \frac{d}{dt} 
\left( \frac{dx}{dt} \right) \frac{dt}{ dt' } \beta(v) ^2
= \beta(v) ^3 \frac{d^2x}{dt^2} \;\;\;\;\;\;\; (7) \]

\noindent , etc.  Equations (5) and (6) can be interpreted in such a way
that in order to increase the velocity of $ B $ from $ \vec{v} $
to $ \vec{v} + \vec{\alpha} dt $, where $ \vec{\alpha} =
(d^2x/dt^2,d^2y/dt^2,d^2z/dt^2) $, it needs
an outer force $ \beta(v) ^3 \vec{F} $ working over $ B $
during a time period $ dt $, which is greater than the ordinary
force (in Newtonian mechanics)
$ \vec{F} $ by factor $ \beta(v) ^3 $.  Note that in (5) and (6),
the infinitesimal length variable $(
dx',dy',dz') $ into a maximal velocity-critical direction ( $ = \;
the \; direction \; of \; \vec{F} $) is replaced by $
\beta(v) (dx,dy,dz) $.

By depending on Subprinciple 1, we shall derive (3) in a different way.
We consider a stationary coordinate system $ K $ and all the values
are measured w.r.t. $ K $.
Assume that
while the velocity of $ B $ increases from zero to $ \vec{v} $,
an outer force $ \vec{F}(\vec{u}) $ works over $ B $ when $
B $ is at velocity $ \vec{u} $.  We put $ u = | \vec{u} | $
and $ v= | \vec{v} | $, so on.  Then $E_{(v)}= (\beta(v) - 1) E_0 $
can be calculated as follows, where $ \vec{u} $ is the velocity w.r.t.
$ K $ and is not affected by Subprinciple 1.

\[ E_{(v)} = \int_0^v m \vec{u} \cdot \vec{du'} = \int_0^v m \vec{u} \cdot 
\left(\frac{d^2x'}
{dt'^2},\frac{d^2y'}{dt'^2},\frac{d^2z'}{dt'^2} \right) dt \]

\[ = \int_0^v m \vec{u} \cdot \beta(u) ^3 \vec{du} = m \int_0^v 
\beta(u) ^3 udu \;\;\;\;\;\; (8) \]

Here we note that the length variable $ \vec{u} $
remains the same as that over a stationary
inertial system while the length of $ (dx',dy',dz') $ into a maximal
velocity-critical direction is multiplied by factor $ \beta(u) $.

We put $ \sqrt {1- u^2/c^2} = \cos\theta $.  Then $u/c = \sin \theta$
and $ du/d \theta = c \cos \theta $.  Thus

\[ E_{(v)} = m \int_0^v \beta(u) ^3 udu = m \int_0^
{\theta(v)}\frac{c\sin\theta c
\cos \theta d \theta}{\cos ^3 \theta}  \]
\[  = mc^2\int_0^{\theta(v)}
\frac{\sin\theta d \theta}{\cos ^2 \theta}=
mc^2(\beta(v) - 1) \; \;\;\;\;\;\; (9) \]

This result is consistent with (3) and (4) since it must hold 
$E(v) = \beta(v)mc^2
= E(0) + E_{(v)} $.
We note that in the above
calculation, $ du $ may be sometimes negative, and even in this case,
the calculation is valid.  We note that in order to obtain $(
\beta(v) -1)mc^2 $ in (9), we need $ \beta(u) ^3 $ exactly.
We also remark that we can obtain the
corresponding space-time transformations over $ K,K', K_0 $ and
$ K'_0 $, where $ K_0 $ and $ K'_0 $ are the inertial systems such
that $ x_0 = x- v_z t, y_0 = y - v_y t, z_0 = z - v_z t, t_0 = t
/\beta(v),
x'_0 = \beta(v) x_0, y'_0 = \beta(v) y_0, z'_0 = \beta(v) z_0 $ and
$ t'_0 = t/ \beta(v) $.  Thus the space-time transformation 
$ {\cal T} $ from $ K $
onto the inertial system $ K(u) $ where $ B $ is at rest with velocity
$ \vec{u} $, the coordinate $(t_u,x_u,y_u,z_u) $ of $ {\cal T} $ 
corresponding to coordinate $ (t,x,y,z) $ may be described as
$ x_u = x - u_x t, \; y_u = y - u_y t, \; z_u = z - u_z t $ (Galileo
transformation), $t_u=t/\beta(u)$ 
and the equation of motion is changed from the
Newton equation of motion to the form as (6).

Subprinciple 1 can be applied to the well known fact of muon's mean
life time.  The mean life time of a muon $ A $ at rest on the earth
is $ t_0 \approx 2.20 \times 10^{-6} s $, and a muon $ B $ moving with
velocity $ v= 0.999c $ can travel in a distance $ l $ of about
$ 14.8 km $ in the space very close to the surface of the earth.
By depending on Subprinciple 1, the distance $ l $ can be calculated
as follows.

\[  l= \int_0^{t_0} vdt' =
v \int_0^{t_0} \beta(v) dt  = \beta(v) v t_0  
\;\;\;\;\;\; \; (10) \]

Now we shall show that well known arguments in relativistic mechanics
for deriving (3) and (4) have  defects.  Let us trace the
arguments briefly (for more details, see, e.g., [14]).  We consider
a particle $ B $ with mass $ m>0 $ moving over a stationary
inertial system $ K $ with velocity $ \vec{v} $.
The well known arguments define the volocity four-vector,
the four-acceleration, and
the four-momentum, and conclude the following equation 
holds (see, e.g., [14]).

\[ \frac{d}{dt}(\beta(v)mc^2) = \vec{F} \cdot \vec{v} \; \;\;\;\;\;\;
(11)   \]

Here $ \vec{F} \cdot \vec{v} $ means the work done over $ M $ by the
force during a unit interval over $ K $.  By integration,

\[  E(v) = \beta(v) mc^2 = \int_0^{t(v)} F(\vec{u}) \cdot \vec{u} dt 
+ constant    \; \;\;\;\;\;\; (12) \]

In the well known arguments, it is acknowledged that "constant" in (12)
iz zero.  But this is clealy wrong since when $ v=0 $, it should hold 
$ \beta(0) mc^2 = mc^2 = \int_0^0 \vec{F} \cdot \vec{v} dt + constant 
= constant $.  
Thus it must
hold $constant = mc^2 $.  But under this condition, it must hold
$ \int_0^{t(v)} \vec{F} \cdot \vec{u} dt =(\beta(v) -1) mc^2 $ and this 
holds only in the case where $ \vec{F} = \vec{K} / \beta(u) ^3 $,
where $ u $ is the velocity of $ B $ and $ \vec{K} $ is the
outer force working over $ B $.  Since in the well known arguments,
the term $ \vec{K} / \beta(u) ^3 $ does not appear, we conclude that
the well known arguments are wrong.

\vspace{2ex}

{\bf 5.2  Subprinciple 2.}  \hspace{5mm} In this subsection, we propose Subprinciple
2 concerning gravity.  By depending on this new subprinciple, we shall
derive a modified version of the Schwarzshild metric 
(called the twofold Schwartzshild metric) which is almost
the same, but the points $ r= 2GM /c^2 $ is not its singular points.

We consider a gravitational field $ {\cal F} $ such that at any point $ P $,
the (gravitational) potential at $ P $ is denoted by $ \Phi(P) $ , where
$ \Phi(P) $ is the work done over a material body $ M_0 $ with unit mass
while $ M_0 $ is carried from infinity to point $ P $ due to the
force of $ {\cal F} $.  Let $ B $ be a material body which is at point
$ P $.  We define the gravity-based velocity (in short, the 
G-velocity) of $ B $, $ v(\Phi(P)) $, by : $(\beta
(v(\Phi(P))) -1)c^2 = \Phi(P) $.  Thus $ v(\Phi(P)) $ depends only on
$ \Phi(P) $, independent from the mass of $ B $.  $ v(\Phi(P))
$ will be denoted by $ v_G $ if the context is clear.  The
(gravitational) potential energy $ E_G $ of $ B $
is $ m \Phi(P) $.  We also consider the kinematic energy $
E_K $ of $ B $ due to $ {\cal F} $.  
Thus we conclude that the total energy $ E_T $
of $ B $ is $ E_T = mc^2 + m \Phi(P) + E_K $.  We define 
the K-velocity $ v_K $ and the T-velocity
$ v_T $ of $ B $ by $ (\beta(v_K) -1)mc^2 = E_K $ and
$ \beta(v_T)mc^2 = E_T $.
We say that $ B $ moves in the semi-eternal mode if it holds
$ E_G = E_K $.  We note planetes rounding the sun move in the
semi-eternal mode since its potential and kinematic energies are (almost)
the same so that they keep rounding the sun without falling into the sun
or runnig away from the sun.  If we consider a stone stationary on the 
surface of the earth, then its potential enrgy is greater than its 
kinematic energy.  From our theory, if two stones $ A $ and $ B $
with the same mass $ m $
are stationary in the sky at height $ h_1 $
and $ h_2 $, respectively, $ 0 < h_1 < h_2 $, then the G-energy of
$ A $ is greater than that of $ B $ since we need positive enrgy
to raise $ A $ to height $ h_2 $, and to cancel the work done for 
carrying $ B $ to height $ h_1 $ by gravity.  We also note that (i)
at $ h $ being infinity, the effect caused by gravity is zero, (ii) the
rounding velocity of the earth is smaller than that of Mercury, and (iii)
the gravitational force works always over $ A $ and $ B $ so that we need
certain forces to make $A$ and $B$ be stationary at height $h_1$ and
$h_2$, respectively.
This is a distinction from classical potential theory in which the
potential at $ h_1 $ is smaller than that at height $ h_2 $.  Here
we admit it holds $ \Phi(P(h_1)) = \Phi(P(h_2)) + E $,
where $ \Phi(P(h_i)) $ is the potential energy of height $ 
h_i (i=1,2) $, and $ E $ is the energy needed to raise a
particle of unit mass from point $ P(h_1) $ to point $ P(h_2) $.
We also note that the total energy of a stone
$ C $ at height $ h_1 $, whose mass is $ m $, which is fallig freely
and whose velocity at height $ h_1 $ is $ v>0 $, is greater than the
total energy of $ A $ since their potential enrgies are the same but
the kinematic energy of $ C $ is greater than that of $ A $.

We say that a physical phenomenon or a property of a physical phenominon
is macro if it is sufficiently large w.r.t. the uncertainty principe.

\vspace{2ex}

{\bf Subprinciple 2} \hspace{5mm}  Let $ {\cal F} $ be a 
gravitational field and
$ \Phi $ be as above.  Let $ E $ be any macro physical phenomenon
or a macro property of any physical phenemenon which is descrived
by a set of equations $ H(0) $ when $ E $ occurs in the space where
the gravitational potential is (almost) zero.  Let $ K $ be a
coordinate system
such  that it is stationary over $ {\cal F} $, and 
the direction of each spacial axis is a macximal velocity-critical one or a
zero velocity-critical one.  Then when $ E $ occurs
in $ {\cal F} $ without any other outer forces, the behavior of $ E $
can be described by the set of equations $ H(v_T(P)) $
which is obtained from $ H(0) $ by replacing, in each differential equation
in $H(0)$,
any infinitesimal time variable $ dt $ with $ dt/ \beta(v_T(P)) $,
any infinitesimal maximal velocity-critical direction
length variable $ dr $ with
$ \beta(v_T(P)) dr $, and any infinitesimal zero velocity-critical direction
length variable $ dr' $ with $ dr' $, where $ v_T(P) $
is the T-velocity at point $ P $ concerning $ E $.

\vspace{2ex}

Now we shall study the Schwarzshild metric.  Let $ B $ be a material
body moving with constant velocity $ v \geq 0 $ in the space with
(gravitational) potential (almost) zero.  
As explained in the muon's
example, we acknowledge that the time $ \tau $ at $ B $ proceeds
more slowly by factor $ 1/ \beta(v) $
than the time $ t $ at a stationary material body.  Thus
about $ B $, the following equation holds.

\[  c^2 d \tau ^2 = c^2 d t^2
-(d x^2 + d y^2 + d z^2)  \;\;\;\;\;\; \; (13) \]

Now consider a material body $ A $ of huge mass $ M $ (e.g., the
sun), and the space-time near $ A $.  Let $ B $ be a material body with
mass $ m<< M $.  Assume that an outer force working over $ B $
is only the gravitational force due to $ A $, and $ B $ rounds
$ A $ due to the gravity of $ A $ and $ B $ (like the planets).
Instead of a Cartesian coordinate system, we consider a polar coordinate
system.  We choose $ x^0 $ to be a time variable $ t $, and $ x^1,
x^2,x^3 $ be the polar coordinates $ r, \theta $ and $ \phi $ in
the standard way.  The equation corresponding to (13) is

\[ c^2 d \tau ^2 = c^2 d t^2 -
(d r^2 + r^2 d \theta ^2 + r^2 \sin^2 \theta d \phi ^2)  
\;\;\;\;\;\; \; (14) \]
\[ 1= \frac{d t^2}{d \tau ^2} - \frac{1}{c^2}\left(\frac{d
r^2}{d \tau ^2} + r^2 \frac{d \theta ^2}{d \tau ^2}+ r^2
\sin ^2 \theta \frac{d \phi ^2}{d \tau ^2}  \right)
\; \; \; \; \; \; \; (15) \]

Due to Subprinciple 2, the equation over $ {\cal F} $ corresponding
to (15) is

\[ 1= \frac{ d t^{'2}}{d \tau ^2} - \frac{1}{c^2}\left(\frac
{d r^{'2}}{d \tau ^2} + r^2 \frac{d \theta ^{' 2}}{d \tau ^2}
+r^2 \sin ^2 \theta \frac{d \phi ^{' 2}}{ d \tau ^2}
\right)    \;\;\;\;\;\;\; (16)  \]

In order to apply Subprinciple 2 to (16), we need the following differential
equations (17), where we consider the case $ r $ is sufficiently
large (as in the case of planatary orbits).
By applying the variational principle to (16), we have

\[ \frac{d t'}{d \tau} = b = constant, \;
 \frac{d r'}{d \tau} \approx d = constant \]
\[ r^2\frac{d \theta '}{d \tau} = f = constant, \;
   r^2 \sin ^2 \theta \frac{ d \phi '}{d \tau} = h = constant  
   \;\;\;\;\;\;\; (17)  \]

\noindent where we neglect the term $ dI/dr'$ 
 for obtaining $ d r' / d \tau \approx d = constant $.
Here $ I $ is the righthand side of (16).
By Subprinciple 2, we admit that $ \tau $ is proportional
to the proper time at $ B $, and $ d\tau ' = d\tau $ since 
$ d \tau $ should be determined by the metric.  Since the direction of $ r $ is
a maximal velocity-critical one and the directions of $
\theta $ and $ \phi $ are zero velocity-critical ones, in (17), we
replace $ d t' $ by $ dt / \beta(v_T) $, $ d r' $ by $ 
\beta(v_T) dr$, $r d \theta ' $ by $ rd \theta $
and $ r\sin ^2 \theta d \phi ' $ by $ r\sin ^2\theta d \phi $.  (Here $ v_T $ 
is the T-velocity of $ B $).  In order to hold these relations,
(16) becomes the following.

\[ c^2 d \tau ^2= c^2dt^2 / \beta(v_T)  - (\beta(v_T) d r^2  + r^2
d \theta ^2  + r^2 \sin ^2
d \phi ^2 ) \;\;\;\;\;\;\;  (18)   \]

The metric (18) will be called the twofold Schwartzshild metric.

We shall calculate $ \beta(v_T) $ in the case $ 0 \leq v_T << c $.  First
we calaulate $ \Phi(P) $ as follows.  Assume that 
$ B $ is carried from infinity to a point $ P $ whose
distance from $ A $ is $ r $.  Then due to Subprinciple 1,
during the travelling of $ B $, the equation describing the motion
of $ B $ at any point $ Q $ is

\[ m \frac{d^2 w'}{d t^{'2}}= \beta(u) ^3 
m\frac{dw^2}{dt^2} = \frac{GmM}{w^2} \]
\[ m\frac{d^2w}{dt^2}= \frac{1}{\beta(u) ^3}\frac{ G
m M}{w^2}  = F(w) \; \;\;\;\;\;\;   (19) \]

Here $ w $ is the distance between $ A $ and $ Q $, and $
\beta(u) mc^2 $ is the total energy of $ B $ at $ Q $, i.e. it holds
$ (\beta(u) -1)mc^2 = E_G + E_K =
2 \int _w ^{\infty} F(x)dx $.  We assume that
$ \beta(u)  \approx 1 $ and acknowledge that the following holds.

\[  (\beta(v_G) -1)mc^2  \approx
\int _{\infty}^r\frac{G m M}{w^2}(-dw) = \frac{G m M}{r}, \;
\beta(v_G) = 1+ GM/c^2r 
 \; \;\;\;\;\;\; (20)  \]

Here $ v_G= v(\Phi(P)) $ is the G-velocity of $ B $ and when $
v_G <<c $, it holds $ (\beta(v_G) -1)mc^2 \approx m v_G ^2 /2 $
and $ v_G ^2 \approx 2GM /r $.  We also note that the direction of the
gravitational force due to $A$ is opposite to that of $r$.
By the note above, we
admit the total energy of $ B $ is $ mc^2 + E_G + E_K
=mc^2 + 2 E_G \approx mc^2 + 2 G m M /r $.  Then $ (\beta(v_T) -
1)mc^2 \approx 2 G m M /r  $.  Thus $ \beta(v_T)
\approx 1 + 2 G M /c^2r $.
Thus we obtain the following
metric which will be called the twofold approximate
Schwarzshild metric.  When the context is clear, it will be called simply
the twofold Schwartzshild metric.

\[  c^2 d \tau ^2 = c^2 dt^2/(1+ 2 G M /c^2r) - ((1+ 2 G M /c^2r)r^2
+r^2 d \theta ^2 + r^2 \sin ^2 \theta d \phi ^2)  
\;\;\;\;\;\; \; (21)  \]

We recall that the well known Schwarzshild metric is of the following
form.

\[  c^2 d \tau ^2 = \left( 1- \frac{2 G M}{c^2r}
\right) c^2 d t^2 - \left(\frac{d r^2}{1- 2 G M /
c^2r} + r^2 d \theta ^2 + r^2 \sin ^2 \theta d
\phi ^2 \right)  \; \;\;\;\;\;\; (22)   \]

In the case $ v_T<<c $, the twofold approximate Schwartzshild metric
and the Schwartzshild metric are almost the same, and the former can
be used to solve such problems as centennial procession of planatary
orbits and light deflect as the latter can be used.  Note also that the former
does not have the points $ r= 2 G M /c^2 $ as singular points,
but the latter does.  We note the following Taylor expansion for $ x
\geq 0 $.

\[ 1 /(1+x) = 1 -x + x^2 - x^3 + \cdots   \;\;\;\;\;\;\; (23) \]

If we put $ x= 2GM/c^2r $, then $ g_{00} = 1- 2GM/c^2r $ in the
Schwartzshild metric is an approximation of $ g'_{00} = 1/(1+ 2GM/c^2r)
$ in the twofold Schwartzshild metric to the first order.  Tensors
are related with multilinear transformations, and it may be difficult
to obtain any approximations correct to a higher order by tensor calculus.
Particulaly it may be difficult to deduce such a solution $ 1/(1+
2GM/c^2r) $ more complicated than $ 1- 2GM/c^2r $ by tensor calculus.

One may note the following equation for $x \geq 0 $..

\[ 1/(1-x) = 1+x-x^2 + x^3 - \cdots    \;\;\;\;\;\;\; (24) \]

This equation is not correct for $ x=1 $.  Thus the assertion $
g'_{11} = 1+ 2GM/c^2r $ in the twofold Schwartzshild metric is an approximation
of $ g_{11} = 1/(1- 2GM/c^2r) $ in the Schwartzshild metric may be wrong.
We note that the Schwartzshild metric is deduced almost purely mathematically
except the boundary conditions (1),(2) presented in Subsection 5.7.
The boundary conditions seem very normal physically, and it may be incorrect
that the solution (= the Schwartzshild metric) has singular points
$ r= 2GM/c^2 $.  Moreover in deriving the Schwartzshild metric, $
g_{00} = 1- 2GM/c^2r $ is firstly deduced and $ g_{11}
$ is deduced by the relation $ g_{00} g_{11} =c^2 $.  It seems very 
difficult in this tensor-based calculation to deduce $ g_{11} $
firstly.  Hence we conclude the Schwartzshild metric is an approximation
to the first order of the twofold Schwartzshild metric for $ 2GM/c^2r
<< 1 $, and the latter metric is more right than the former metric.
These arguments seem to show a limitation of tensor calculus. 
More comparisons between these metrics will be
presented in Subsection 5.7.

In (18), $ c d \tau $ is the distance for light to travel within
a time period $ d \tau $.  The time variable $ \tau $ may be
regarded to denote the time proportional to
the proper time lapse over $ B $ (recall (14)).  It is admitted
the motion of $ B $ is under a world line called a geodesic,
and it holds

\[  \tau _{ab} = \int _{\tau _a}^{\tau _b}d \tau \sqrt
{t'^2 / \beta(v_T)  -(\beta(v_T) r'^2 + r^2 \theta'^2 + r^2
\sin ^2 \phi'^2) /c^2}  \;\;\;\;\;\; \; (25)  \]

\noindent where $ \beta(v_T) \approx 1+ 2 G M /c^2r $, and
$ t',r',\theta' $ and $\phi' $ stand for $ dt/d \tau, dr/
d \tau, d \theta/d \tau $ and $ d \phi/
d \tau $, respectively.  A geodesic takes an extremum path.  This
law may be associated with the following Fermat's law, where $ n $
and $ ds $ correspond to $ 
\sqrt{t'^2 / \beta(v_T)  -(\beta(v_T) r'^2 + r^2 \theta'^2 + r^2
\sin ^2 \phi'^2) /c^2}  $ and $ d \tau $, respectively, in (25).

\vspace{2ex}

{\bf Fermat's law.}  \hspace{5mm} When a light signal $ L $ 
travels from point $ P $
to point $ Q $, then the following value is an extremum value for any
other path which $ L $ may take from $ P $ to $ Q $.

\[  W= \int _P^Q nds   \; \;\;\;\;\;\; (26)  \]

Here $ n $ is the refractive index.

\vspace{2ex}

Now we shall turn to the twofold Schwzrtzshild metric in more general
cases.  Assume that a material body $ B $ with mass $ m $ is moving over
a gravitational field $ {\cal F} $ with total energy $ \beta(v_T) 
mc^2 $.  Let $ \vec{F} =(F_x,F_y,F_z) $ be the composition of gravitational
forces working over $ B $, and $ \Phi(P) m $ be the potential
energy of $ B $ due to $ {\cal F} $.  We consider two
cases.

Case (1) : $ \vec{F} \neq (0,0,0) $.  In this case, we choose a polar
coordinate $ r, \theta $ and $ \phi $ in such a way that the
direction of $ r $ is that of $ - \vec{F} $, and we calculate $ r $
and a mass $ M $ by the following equations.

\[ \frac{GmM}{\beta(v_T) ^3 r^2} = | \vec{F} |, \; \int_r^{\infty}
\frac{GmMdu}{\beta(v_u) ^3 u^2} = \Phi(P) m   \;\;\;\;\;\;\; (27) \]

Here $ (\beta(v_u) -1)mc^2 = \int_u^{\infty}GmMdw/ (\beta(v_w)^3
w^2) $.  Then we can regard $ {\cal F} $ as the gravitational field 
produced by a material body $ D $ with mass $ M $ and with 
distance $ r $ from $ B $.  Here we assume $ m<< M $.
Then we may have the following metric.

\[ c^2d \tau ^2 = c^2 d t^2 / \beta(v_T) - (\beta(v_T) dr^2 +
r^2 \sin ^2 d \theta ^2 + r^2 \phi ^2)   \;\;\;\;\;\;\; (28) \]

Case (2) : $ \vec{F} = (0,0,0) $.  In this case, it seems
that any direction is a maximal velocity-critical one.  Thus due to
Subprinciple 2, any infinitessimal length variable $ dx $ 
should be replaced
by $ \beta(v_T) dx $.  Thus in this case, we consider a Cartesian
coordinate system $ (x,y,z) $, and may have the following metric.

\[ c^2 d \tau ^2 = c^2dt^2/ \beta(v_T) - \beta(v_T)(dx^2 + dy^2 + dz^2)   
\;\;\;\;\;\;\; (29) \]

Now we shall study the metric for freely falling material bodies
in a gravitational field.  We consider the case where a material body
$ A $ of huge mass $ M $ exists and a small material body $ B
$ with mass $ m $ freely falls into $ A $.  From our theory,
the metric depends on the total energy of $ B $ (or equivalently on
the T-velocity of $ B $) rather than the acceleration of $ B $.
Thus assume that at time $ t<0 $, $ B $ is at rest over $ A $
at height $ h $ and at time $ t=0 $, $ B $ begins falling freely
into $ A $.  At time $ t>0 $, we admit the total energy of
$ B $ is $ mc^2 + E_G + E_K = mc^2 + GmM/r + \int_h^r GmM du/
u^2 = mc^2 + GmM/r + GmM(1/r - 1/h) $.  Here $ r $ is the distance 
between $ B $ and $ A $ ($0<r<h$).  Now the T-velocity $ v_T $ satisfies

\[ (\beta(v_T) -1)c^2 = GM(2/r - 1/h)    \;\;\;\;\;\;\; (30) \]

Thus $ v_T $ depends not only on $ r $ but also on $ h $.  Now the
corresponding twohold Schwartzshild metric may be as follows.

\[ c^2 d \tau ^2 = \frac{c^2 dt^2}{\beta(v_T)} - \beta(v_T) dr^2 \; \; \; 
\;\;\;\; (31)  \]

In genareal relativity, the corresponding metric is given as follows by
putting $ \theta = \pi/2 $ and $ d \phi / dt =0 $ in the Schwartzshild
metric.

\[ c^2(1-2GM/r)\left(\frac{dt}{d \tau}\right) ^2 - 
(1- 2GM/r)^{-1}\left(\frac{dr}{d \tau}\right) ^2 = c^2    \;\;\;\;\;\;\; (32)\]

From this metric, one concludes in general relativity that the proper time
experienced by the particle at rest at $ r=r_0 $ and falling freely is
given by (where $ \tau =0 $ when $ r=r_0 $)

\[ \tau = \frac{1}{\sqrt{2GM}}\int_r^{r_0}\left(\frac{r_0r}{r_0 - r}
\right) ^{1/2} dr     \;\;\;\;\;\;\; (33) \]

\noindent which is finite (see [6]) as $ r \rightarrow 2GM/c^2 $.  
On the other hand, one also concludes in general relativity
that the coordinate time $ t $
for falling from $ r_0 $ to $ r=2n + \epsilon $ ($n= GM/c^2, \;
 \epsilon >0 $) is 
calculated by

\[ t_{\epsilon} = \left(\frac{r_0 - 2n}{2nc^2}\right) ^{1/2}
\int_{2n + \epsilon }^{r_0}\frac{r^{3/2}dr}{(r-2n)(r_0 - r) ^{1/2}} \]
\[ > \left(\frac{r_0 - 2n}{2nc^2}\right) ^{1/2} \frac{(2n)^{3/2}}
{r_0^{1/2}} - \int_{2n + \epsilon}^{r_0} \frac{dr}{r - 2n}
\rightarrow \infty \; as \; \epsilon \rightarrow 0   \;\;\;\;\;\;\; (34) \]

\noindent since $ \displaystyle{\int_{2n + \epsilon}^{r_0}
\frac{dr}{r-2n} = ln \frac{r_0 - 2n}{\epsilon} } $ (see [6]).  These
calculations in general relativity
seem wrong to the author since in (31), $ \beta(v_T) $ is always positive
and $ \beta(v_T) \approx 1/(1- 2GM/r) $ is an approximation in the case
$ 2GM/r << 1 $.  It is also very odd that $ \tau $ is finite
and the corresponding $ t $ is infinite.  The above relation 
between $ \tau $ and $ t $ may be regarded to be a relation like
the one between $ d = \sum_{n=0}^{\infty} a_n $ and $e= \sum_{n=0}
^{\infty} b_n $, where $ a_i =1 $ and $ b_i = 1/2^i $ for all
$ i $.  The infinite time and a finite time are fundamentally different,
and this kind of relation between $ \tau $ and $ t $ may be
impossible.

Now we shall show that General Relativuty Principle II is wrong. We shall
study the twofold Schwartzshild metric in case $ \theta = 
\pi /2 $.

\[ c^2 d \tau ^2 = \frac{c^2 dt^2}{\beta(v_T)} - (\beta(v_T) dr^2 + r^2
d \phi ^2)    \;\;\;\;\;\;\; (35) \]

Here we recall the centennial procession of planatary orbits.  
This metric precisely predicts the centennial procession of 
Mercury's orbit.  We consider the following
coordinate transformation.

\[ t' = t/\sqrt{\beta(v_T)}, \; r' = r, \; \phi ' = \phi   \;\;\;\;\;\;\; (36)  \]

This transformation is made in the range of $ r $ where the twofold
Schwartzshild metric is valid.  Then the corresponding metric over the
coordinate system $ (t',r', \phi ') $ is as follows.

\[ c^2d \tau ^2 = c^2 d t^{' 2} - (\beta(v_T) dr^2 + r^2 
d \phi ^2)     \;\;\;\;\;\;\; (37) \]

One can see this metric predicts the centennial procession of
Mercury's orbit wrongly by applying the variational principle even with
the aid of the transformation.
Thus under certain coordinate transformations,
some principles such as the variational principle cannot be applied
over the new transformed coordinate system.  Hence we conclude General
Relativity Principle II is invalid.
Any coordinate trnasformation $ {\cal T} $ from a coordinate system $ K
$ to another coordinate system $ K' $
should preserve the properties observed over $ K $ so that $ K $ and
$ K' $ should be equivalent at least in the following meaning.

(1) Any property observed over $ K $ should be observed over $ K' $
with aid of $ {\cal T} $.

(2) Any property observed over $ K' $ shoud be observed over $ K $
with aid of $ {\cal T} ^{-1} $.

In the above coordinate transformation,
the transformed metric is not equivalent to the twofold Schwartzshild
metric since in the Schwartzshild metric, the procession of Mercury'
orbit is $ 3GM \pi (1/r_1 + 1/r_2)/c^2 $ (see [6]) which cannot be obtained
from the calculated value by (37) with transformation (36).  Here
$ r_1 $ and $ r_2 $ are the values of $ r $ at aphelion and
perihelion, and under transformation (36), $ 3GM \pi
(1/ r_1 + 1/ r_2)/c^2 $ is not changed since $ r' = r $ and $ \phi '
= \phi $.

\vspace{2ex}

{\bf 5.3  The Maxwell equations over a gravitational field.} \hspace{5mm}
Let $ {\cal F} $
be a gravitational field and $ \Phi $ denote the corresponding
(gravitational) potential function as in Subsection 5.2.  We shall study
the Maxwell equations over ${\cal F}$.
We admit any direction
is a maximal velocity-critical one since we observe the following.
Consider the earth and the gravity field over the earth.  Let $ O $
denote the center of the earth and the origin of a polar coordinate
system $ (r,\theta,\phi) $ describing coordinates of points around the 
earth.  Then over the surface of the earth, the r-direction may be
clearly a maximal velocity-critical one.  On the other hand,
the $ \theta $-direction and $ \phi $-direction might be
zero velocity-critical ones since into these directions, the gravitational
potentila does not change.  However if we admit this is right, then
the light velocity into these directions calculated in the sequel
would become $ c $ 
(as in the space with gravitational potential (almost) zero) w.r.t.
the corresponding coordinate system stationary in the universe.  
Since the earth is not
stationary and the Michelson-Morley experiment is observed, we acknowledge
that the $ \theta $-direction and $ \phi $-direction are also
maximal velocity-critical ones, and so are all directions.  We admit the light
velocity is completely governed by the gravitational field (perphaps
except very special cases).
We also
note if we remove gravity, then the light velocity would become $
c $ changing from $ c/ \beta(v_G) ^2 $ which will be obtained
in the sequel.  We shall show the light velocity at point
with G-velocity $ v_G $ is $ c/ \beta(v_G) ^2 $ and if we remove
$ {\cal F} $, then the velocity may become $ c $.  On the other
hand, in the twofold Schwartzshild metric (18), if we remove the gravity
of $ B $, then $ A $ will continue moving with velocity $
\vec{v} $.  Thus we admit any direction is a maximal velocity critical
one in the Maxwell equations.  These observations presents a distinction 
between the motion of material
bodies and the travelling of light ( comparing Subprinciles 1,2).
Moreover we admit a photon does not have kinematic energy since its
mass is zero.  Thus we propose the following subprinciple.

\vspace{2ex}

{\bf  Subprinciple 3.} \hspace{5mm} 
Let $ {\cal F} $ and $ \Phi $ be as above,
and consider a point $ P $ whose gravitational potential is $
\Phi(P) $.  Then the Maxwell equations at $ P $ are of the
following form.

\[  
\nabla' \cdot D(x,y,z,t) = \rho(x,y,z,t), \;
\nabla' \cdot B(x,y,z,t) = 0 \]
\[ \nabla' \times H(x,y,z,t) - \displaystyle
 {\frac {\partial D(x,y,z,t)}{\partial t'}}
 = i(x,y,z,t)   \]
\[  \nabla' \times E(x,y,z,t) + \displaystyle{\frac {\partial 
B(x,y,z,t)}{\partial t'}}
 = 0    \; \;\;\;\;\;\; (38)  \]
 
Here $ \nabla' = (\partial /\partial x',\partial /\partial y',
\partial /\partial z'), \; \partial x' = \beta(v_G) \partial x, \;
\partial y' = \beta(v_G) \partial y, \; \partial z' = \beta(v_G)
\partial z $, and $ \partial t' = \partial t/ \beta(v_G) $ with G-velocity
$ v_G= v(\Phi(P)) $ at $P$.  The coordinate
system $ (t,x,y,z) $ is a standard coordinate system $ K $ which is
stationary over $ {\cal F} $.  This means for each point $ P $
stationary over $ {\cal F} $, the corresponding spacial 
coordinate $ (x(P),y(P),z(P)) $ is constant independent of time,
and the time $t$ at $ K $ proceeds with the same speed (rate) as the global
time.  Thus if $ {\cal F} $ continues moving, then $ K $
continues moving together with $ {\cal F} $ (as in the case of
the gravitational field over the earth).

\vspace{2ex}

We consider the case where $ \rho(x,y,z,t)=i(x,y,z,t)=0 $, and
calculate the light velocity at $ P $.  To do this, we consider the
travelling of a plane electromagnetic wave in the case where $ E $
and $ B $ depend only on $ z $ and $ t $.  From (38),

\[  \frac{\partial E_z(z,t)}{\partial z'}= \frac
{\partial B_z(z,t)}{\partial z'}= \frac{\partial E_z(z,t)}{\partial t'}
=\frac{\partial B_z(z,t)}{\partial t'} =0  \;\;\;\;\;\; \; (39)   \]

For $ x,y$-components of $ E(z,t) $ and $ B(z,t) $, it holds

\[  - \frac{\partial B_y(z,t)}{\partial z'} - \epsilon_0 \mu_0
\frac{\partial E_x(z,t)}{\partial t'}= 0  \]
\[ \frac{\partial B_x(z,t)}{\partial z'} - \epsilon_0 \mu_0 
\frac{\partial E_y(z,t)}{\partial t'}= 0   \]
\[ -\frac{\partial E_y(z,t)}{\partial z'} + \frac{\partial B_x(z,t)}
{\partial t'} = 0    \]
\[ \frac{\partial E_x(z,t)}{\partial z'} + \frac{\partial B_y(z,t)}
{\partial t'} = 0   \; \;\;\;\;\;\; (40)  \]

From the relations $ \partial z' = \beta(v_G) \partial z $ and $ \partial
t' = \partial t/ \beta(v_G) $, (40) are of the following form.

\[  - \frac{\partial B_y(z,t)}{\partial z} - \beta(v_G) ^2 \epsilon_0
\mu_0 \frac{\partial E_x(z,t)}{\partial t} = 0  \]
\[ \frac{\partial B_x(z,t)}{\partial z} - \beta(v_G) ^2 \epsilon_0 \mu_0
\frac{\partial E_y(z,t)}{\partial t}= 0  \]
\[ - \frac{\partial E_y(z,t)}{\partial z} + \beta(v_G) ^2 
\frac{\partial B_x(z,t)}
{\partial t} =0  \]
\[ \frac{\partial E_x(z,t)}{\partial z} + \beta(v_G) ^2 \frac
{\partial B_y(z,t)}{\partial t} = 0    \;\;\;\;\;\; \; (41)  \]

Hence

\[  \frac{\partial ^2 E_x(z,t)}{\partial z^2} - \beta(v_G) ^4 \epsilon_0
\mu_0 \frac{\partial ^2 E_x(z,t)}{\partial t^2}= 0   
\;\;\;\;\;\; \; (42)  \]

The solution of (42) is

\[  E_x(z,t) = E_0 \cos k(z \mp \frac{c}{\beta(v_G) ^2} t)   
\;\;\;\;\;\; \; (43) \]

Here $ c= 1/\sqrt {\epsilon_0 \mu_0} $ is the light velocity at
any place where the gravitational potential is (almost) zero.  Thus
the light velocity at point $ P $ with gravitational potential
$ \Phi(P) $ is $ c/ \beta(v_G) ^2 $, where $ ( \beta(v_G) -1)c^2 =
\Phi(P) $.

Now we shall solve the problem of light deflection
due to gravity by depending on $ c/ \beta(v_G) ^2 $ and Huygens's
principle in a similar way as Einstein[5] does.  Assume that light
$ L $ passes close to the sun with minimum distance $ r_0 $ from
the sun and with orthogonal angle at time $ t=0 $, and eventually
is deflected as $ L $ recedes from the sun.  For time $ t \geq
0 $, let $ \phi(t) $ be the angle which is constructed by the 
horizontal line $ L_1 $ on which $ L $ passes at $ t=0 $ and the center
of the sun lies,
and
the line $ L(t) $ on which $ L $ passes at $ t >0 $ and which crosses
through the center of the sun.  The angle $ \phi(t) $ will be
denoted by $ \phi $ often.

Let $ K $ be a plane which has the x-axis and the y-axis and whose
origin is the center of the sun.  We choose the x-axis of $ K $ as
$ L_1 $ and at time $ t=0 $, $ L $ is at point $(r_0,0) $
and at time $ t >0 $, $ L $ is at point $(x,y) $ with $ y>0 $.
Consider time $ t>0 $, and we shall calculate the deflection angle
$ \delta \theta $ made within a small change of time from $ t $ to $ t+
\delta t $.  As an approximation, we consider this problem over the
line $ L_2 $ such that $ L_2 $ is orthogonal to $ L_1 $ and
$ L_2 $ intersects with $ L_1 $ at point $ (r_0,0) $.  Let
$ L $ be at point $ P $ whose angle as above w.r.t. $ L $
is $ \phi $ and whose distance from the origin is $ r $.  Then the
$(x,y)$-coordinate of $ P $ is $ x= r_0, \; y= r_0 \tan \phi $
and $ r= \sqrt{x^2+y^2}= x/ \cos \phi $.  The velocity
$ \gamma $ of $ L $ at $ P $ is $ c/ \beta(v_G) ^2 $, where 
from Subsection 5.2, equation (20),
$ 1/ \beta(v_G) ^2 
\approx 1- 2GM /c^2r $.
Thus $ \gamma =(1- 2 G M /c^2r)c
$.  Due to Huygens's principle, $ \delta \theta $ is calculated
by

\[ \delta \theta = \sin \delta \theta = \frac{\partial \gamma}
{\partial x}= \frac{2 G M c}{c^2r^2}\frac{1}{2}\frac
{2x}{r} =\frac{2 G M}{c}\frac{x}{r^3}  \; \;\;\;\;\;\; (44) \]

This deflection
$ \delta \theta $ is the deflection when $ L $ travels within the
distance $ c/ \beta(v_G) ^2 $.  Thus when $ L $ travels within
the diatance $ dy $, the corresponding deflection $ d \theta
$ is

\[ d \theta = \frac{2 G Mx}{cr^3}\frac{dy}{c/ \beta(v_G) ^2}
\approx \frac{2 G M x dy}{c^2r^3} = \frac{2 G M r_0 dy}
{c^2 r_0 ^3/ \cos ^3 \phi}  \]
\[ = \frac
{2 G M \cos ^3 \phi}{c^2 r_0 ^2}\frac{r_0 d \phi}
{\cos ^2 \phi}=\frac{2 G M \cos \phi d \phi}{c^2 r_0}   \; \;\;\;\;\;\; (45) \]

Then the total deflection is approximately

\[  \theta = \int_0^{\frac{\pi}{2}} \frac{2 G M \cos d \phi}
{c^2 r_0} = \frac{2 G M}{c^2 r_0},\; \;
2 \theta = \frac{4 G M}{c^2 r_0}   \; \;\;\;\;\;\; (46)  \]

This value is the same as the well known one which is derived from the
Schwartzshild metric.  We also note that in solving the problem of radar
sounding in general relativity, it is concluded that the light velocity
is $ c/ \beta(v_G)^2 \approx c(1- 2GM/c^2r) $ (see [6]).

From (43), we can also present a reason why the Michelson-Morley
experiment is observed.  Consider a small horizontal circle $ C $
which is stationary in a space very close and parallel to the surface
of the earth.  Then the potential energy $ \Phi(P) $ at
any point $ P $ at rest over $ C $ is $ G M_0 /r $,
where $ r $ is the distance between the center of the earth and
$ C $, and $ M_0 $ is the mass of the earth.  Due to (43),
the velocity of light travelling horizontally over $ C $ is
$ c/ \beta(v_G) ^2 $ independent of its direction, where
$ (\beta(v_G) -1)c^2 = G M_0 /r $.  Thus the Michelson-Morley experiment
can be observed.  The moving charge examples in Example 4 in Section 3
can be explained in the same way.  From these observations, one may say
that the gravitational force is more fundamental than the
electromagnetic force.

Now we shall turn to the red shift problem of light.  It is observed
that a photon is affected by a gravitational field, which causes
a drop in frequency and an increase in wavelength (red shift) ;
traditionaly stated in terms of a red shift parameter, $ z= \Delta
\lambda / \lambda $.  Consider a material body $ A $ of huge mass
$ M $ and the gravitational field $ {\cal F} $ near $ A $ whose
gravitational potential function is denoted by $ \Phi $.
By (43),
the light velocity $ c(v_G) $ at point $ P $ with
G-velocity $ v_G $ is $ c/ \beta(v_G) ^2 $, where 
$ v_G^2/c^2 = 2 G M /c^2r $ in case $ v_G <<c $
and $ r $ is the distance between
$ A $ and $ P $.  Now consider electromagnetic waves $ L $
are continued to be emitted from hydrogen atoms at the surface $ S $
of $ A $.  Due to Subprinciple 2, the time at $ S $ proceeds more
slowly by factor $ 1/ \beta(v_0) $ than the time at point with G-velocity
(almost) zero, where $ v_0 ^2/c^2 = 2 G M /c^2r_0 $ and $ r_0 $ is the
radius of $ A $.  Here we admit the kinematic energy of hydrogen can be
negligible since due to the nuclear explosion, hydrogen may be regarded
to be floating over $ S $.
One can see that this implies the frequency $
\nu_0 $ of $ L $ at $ S $ is smaller than the frequency
$ \nu(0) $ correcponding to G-velocity zero by factor $ 1/ \beta(v_0)
$.  We can see the following hold.

\[  \nu(0) \lambda(0) =c,\; \nu_0 = \frac{\nu(0)}{\beta(v_0)}, \;
\lambda_0 \nu_0 = c/ \beta(v_0)^2, \; \lambda_0 = 
\frac{c}{\beta(v_0)^2} \frac{\beta(v_0) }{\nu(0)} =
\frac{ \lambda(0)}{\beta(v_0)}   \; \;\;\;\;\;\; (47)  \]

\noindent where $ \lambda_0 $ is the wavelength of $ L $ at $ S $
and $ \lambda(0) $ is the wavelength of $ L $ corresponding
to G-velocity zero.  

Now consider two points $ P $ and $ Q $ such that $ L $
arrives at $ P $ and $ Q $ and the distance between $ A $
and $ P $ and $ A $ and $ Q $ are $ r_1 $ and $ r_2 $,
respectively, with $ r_0 < r_1 < r_2 $.  Let $ v_1 $ and $ v_2 $ be
G-velocities at $ P $ and $ Q $, respectively.  Then the light
velocity $ c(v_1) $ corresponding to G-velocity $ v_1 $ is
$ c/ \beta(v_1) ^2 $ and the corresponding $ c(v_2) $ is $
c/ \beta(v_2) ^2 $.  The time at $ P$ proceeds more slowly by
factor $ 1/ \beta(v_1) $ than the time corresponding to G-velocity
zero (the global time).

We can observe the following.  Let $ t_0 $ denote a unit time lapse
at G-velocity zero.  Then at $ P $ and $ Q $, the frequencies
$ \nu_1 $ and $\nu_2 $ measured during a time lapse $ t_0 $ are
still $ \nu_1 = \nu_2 = \nu(0) / \beta(v_0) $.
Now let $ \nu'_0 $ be the corresponding frequencies of
electromagnetic waves emitted from hydrogen atoms over the earth.
Let $ v_3 $ be the G-velocity on the surface of the earth.  Then
as above, it holds $ \nu'_0 = \nu(0) / \beta(v_3) $.
Thus

\[  \frac{ \nu'_0 - \nu_0}{\nu'_0} = \frac{\beta(v_0)
-\beta(v_3)}{\beta(v_0)} \approx \beta(v_0) - \beta(v_3)
\approx \frac{G M}{c^2 r_0} - \frac{G M}{c^2 r_3}
\approx \frac{G M}{c^2 r_0}  \; \;\;\;\;\;\; (48)   \]
\[ \frac{\lambda '_0 - \lambda_0}{\lambda'_0} = \frac{1/ \beta(v_3)
- 1/ \beta(v_0)}{1/ \beta(v_3)} \approx \beta(v_0) - \beta(v_3)
\approx \frac{GM}{c^2 r_0}   \;\;\;\;\;\;\;   (49)   \]

Here $ r_3 $ is the radius of the earth and $\lambda'_0$ is the wavelength
of electromagnetic waves emitted from hydrogen atoms over the earth.  
This is the red shift of light.  Note the acknowledgment the time on the
surface of $ A $ proceeds more slowly than the global time by factor
$ \beta(v_G) = 1+ GM_0/c^2 r_0 $ is also done by Einstein [4], and
the value $ GM/c^2 r_0 $ in (48) is obtained in a similar way
by Einstein [4,5].
On the other hand, since the velocities of $ L $ at $ P $ and $
Q $ are $ c/ \beta(v_1) ^2 $ and $ c/ \beta(v_2) ^2 $, respectively,
the wavlengths $ \lambda_1 $ and $ \lambda_2 $ at $ P $ and $ Q
$ of $ L $ satisfy

\[  \lambda_1 \nu_1 = c/ \beta(v_1) ^2, \; \lambda_1 =
\frac{c}{ \beta(v_1) ^2 \nu_1} = \frac{c \beta(v_0)}
{\beta(v_1) ^2 \nu(0)} = 
\frac{\beta(v_0)}{\beta(v_1) ^2} \lambda(0)    \]
\[ \lambda_2 = \frac{\beta(v_0)}{\beta(v_2) ^2}\lambda(0) 
\;\;\;\;\;\; \; (50)  \]

But when one measures the frequencies of $ L $ at $ P $ and $ Q $,
one uses atomic-clocks at $ P $ and $ Q $ which proceed more
slowly than the time at G-velocity zero by factor $ 1/ \beta(v_1) $
and $ 1/ \beta(v_2) $, respectively.  Thus the frequency
$ \nu'_1 $ of $ L $ at $ P $ measured during the time lapse
$ \beta(v_1)  t_0 $ is $ \beta(v_1) \nu(0) / \beta(v_0) $,
and the frequency $ \nu'_2 $ of $ L $ at $ Q $ measured
during the time lapse $ \beta(v_2) t_0 $ is $ \beta(v_2)
\nu(0) / \beta(v_0) $.
Thus we have the following equation.

\[  \frac{\nu'_1 - \nu'_2}{\nu'_1}=\frac{\beta(v_1)
- \beta(v_2)}{\beta(v_1)} \approx \beta(v_1) - \beta(v_2)
\approx \frac{G M}{c^2 r_1} - \frac{ G M}{c^2 r_2}   
\;\;\;\;\;\; \; (51)  \]

About wavelength, due to Subprinciple 2, we may consider that at
point $ P $, we choose a coordinate system $ K(v_1) $ such that
$ r'_1 = \beta(v_1) r $ and $ t'_1 = t / \beta(v_1) $.
Then the length of $ c/ \beta(v_1) $ at $ K $ (= a stationary
coordinate system at G-velocity zero with coodinates $ r $ and $ t $)
becomes $ c $ at $ K(v_1) $.  Then the corresponding wavelength $
\lambda'_1 $ of $ L $ at $ P $ becomes $ \lambda'_1 =
\beta(v_1) \lambda_1 = \beta(v_0) \lambda(0) / \beta(v_1) $.  In the same way,
if we consider a coordinate system $ K(v_2) $ with $ r'_2 =
\beta(v_2) r $ and $ t'_2 = t / \beta(v_2) $, then the corresponding
wavelength $ \lambda'_2 $ becomes $ \lambda'_2 = \beta(v_2) 
\lambda_2 = \beta(v_0) \lambda(0) / \beta(v_2) $.  Thus
we have

\[ \frac{\lambda'_1 - \lambda'_2}{\lambda'_1}=\frac
{ 1/ \beta(v_1) - 1/ \beta(v_2)}{1/ \beta(v_1)} = 1- 
\frac{\beta(v_1)}{\beta(v_2)} \approx
\beta(v_2) - \beta(v_1) \approx \frac{G M}{c^2 r_2}
-\frac{G M}{c^2 r_1}  \;\;\;\;\;\; (52) \]
\[ \lambda'_1 \nu'_1 = \lambda'_2 \nu'_2 = c   \; \;\;\;\;\;\; (53)  \]

The relations between $ \lambda'_1 $ and $ \nu'_1 $
and $ \lambda'_2 $ and $ \nu'_2$, respectively, may
be acknowledged to correspond to the constant light velocity principle
under the twofold metric principle.

In the above arguments, at any point, the frequency during a unit time
lapse of the global time remains $ \nu(0) / \beta(v_0) $ while
the corresponding wavelength depends on $ \beta(v_1) $.  If we consider
the emittance of light from hydrogen atom with K-velocity $ v_0 $
and G-vel0city zero, then the corresponding frequency may be $
\nu(0) / \beta(v_0) $, and the wavelength is $ \beta(v_0) \lambda(0) $
since the light velocity is $c$.  We
may admit the corresponding energy of these photons is $
h \nu(0) / \beta(v_0) $ (since the gravitational potential is zero)
which is proportional to the frequency.  Thus about the light emitted
from hydrogen with G-velocity $ v_0 $, the energy may be $h \nu _0 =
h \nu(0) / \beta(v_0) $ (not $hc/\lambda _0 $) independent of time.  Hence we
conclude the energy $ E $ of a photon with frequency $ \nu $
and wavelength $ \lambda $ such that $ \nu \lambda =
c / \beta(v) ^2 $ satisfies $ E = h \nu = hc/ \beta(v) ^2\lambda
$.  Since the mass of a photon is zero, the energy of a photon
may be independent from the gravitational field.

\vspace{2ex}

{\bf Remark 6} \hspace{5mm} We acknowledge the following : 

(*) For any star $ A $
with $ r/M $ at the surface $ S $ of $ A $, the time in $
A $ proceeds more slowly than the global time by factor $
\beta(v_G) \approx 1+ GM/c^2r \sim 1+ 2GM/c^2r $.

About the sun, $
GM_0/c^2r_0 \approx 6.37 \times 10^{-6} $.  Thus about any star with
$ r/M \approx 10^4 M_0/r_0 $, the corresponding $ \beta(v_T)
$ is about $ 1.064 $.  Thus the acknowledgment (*) does not seem
to produce any significant effects to the well known theory about
the life styles of many types of stars.

\vspace{2ex}

{\bf Remark 7} \hspace{5mm} If the Big Bang occurred really,
then at a very early
time $ t $, the corresponding $ \beta(v_T(t)) $ may have been
very large.  Then $ \nu(0) / \beta(v_T(t)) $ may be very small.
If this is true, then the observed microwave background emission may
have troubles with relation to the expansion of the universe.
To solve this problem, one possible hypothesis may be the following.

\vspace{2ex}

{\bf Hypothesis A} \hspace{5mm} Let $ E_T $ denote the total energy
of the universe which we assume constant (or the energy-dencity at the
beginning time of the universe, and in this case, the arguments should
be modified correspondingly in the following).  
When some energy $ E_0 $
is absorbed in the vacuum space, we admit $ E_0 $ is still contained
in $ E_T $.  In the rest of this remark, let $ t $ and $ t_0 $
denote the time variables which proceed with the same rate as the global
time over any stationary inertial system at present age.  We assume at
$ t=0 $, the Big Bang occurred.  So $ t $ may be used to 
represent an age of the universe, and may be used for  describing any
physical phenomenon.  We also assume the following (1)-(6).

\begin{enumerate}
\item The universe have continued expanding, and at time $ t $,
the energy density $ \rho(t) $ is denoted by $ E_T /
a(t) ^3 $, where $ a(t) $ may be admitted to represent the size
(or radious) of the universe at time $ t $.
\item  At any early time $ t $, $ \rho(t) $ is almost constant at
any point in the universe.
\item  The coefficient $ \beta(v_T(t)) $ at an early time $ t
$ is denoted by $ 1+ k \rho(t) ^{\alpha} $, where $ k $ and
$ \alpha $ may be constants which in this paper, we do not
know how to determine.
\item  The global time $ t_g(t) $ at an early time $ t $ proceeds more slowly
than $ t $ by factor $ 1/ \beta(v_T(t)) $.
\item  At an early time $ t $, any length $ l $ w.r.t. the present 
stationary inertial system plays a role as the length 
$ \beta(v_T(t)) l $.
\item Any physical phenomenon at an early time $ t $ can be described
of the same form (as the form of present age) over the coordinate 
system $ K(t) $ whose coordinates are of the form $(t_g(t),x(t),y(t),
z(t)) $, where $ t_g(t) = t_0/ \beta(v_T(t)), x(t) = \beta(v_T(t))x,
y(t) = \beta(v_T(t))y, z(t)= \beta(v_T(t))z $,
and $(x,y,z) $ is a coordinate variable over any stationary inertial
system of present age $ t_p $, where we admit $
\beta(v_T(t_P)) \approx 1 $.  

\end{enumerate}

In (4)-(6), at an early
time $ t $, a unit time lapse of $ t_g(t) $ plays a role
as a unit time lapse of the present age $ t_p $, and a unit length
over $ K(t) $ plays a role as a unitlength at present time $ t_p $.
Thus we may admit that at an early time $ t $ over $ K(t) $, the twofold
metric principle effect and the red shift effect may be almost negligible.
However we may also admit that in (6), " $\cdots$ can be described
of the same form $\cdots$ " are not used for excluding the possibility
of certain unifications of four forces (strong, electromagnetic,
gravitational and weak forces) which have been discussed often in the
litarature.

Then we can see that the frequency $ \nu(0) $ at an early time $ t $
could be still observed to be $ \nu(0) $ at present age if the space would
not have expanded.  
We also note that at an early time $ t $, $ a(t) $ plays a role as $
\beta(v_T(t)) a(t) $ at present age.  Thus in order to accord with Hubble's
law and the microwave background emission, it may hold

\[ \frac{\beta(v_T(t_p))a(t_p)}{\beta(v_T(t)) a(t)}= \frac
{a_0(t_p)}{a_0(t)}     \;\;\;\;\;\;\; (54) \]

\noindent where $ a_0(t) = \beta(v_T(t))a(t) $ ($a_0(t_p) $ also)
is the size (or radius) of the universe at time $ t $ so that
the frequency $ \nu(0) $ at time $ t $ is observed
as $ \nu(0) a_0(t)/a_0(t_p) $ at present age $ t_p $.
Thus under Hypothesis A, we may be able to develop reasonable arguments,
but this problem may be very difficult.
More detailed analysis will
be left for future studies.

\vspace{2ex}

Now we shall study how the acceleration due to the gravity of the earth
produce the total energy of a material body $ B $ with mass $ m $.
Let $ B $ be at point $ P $ with total energy $ \beta(v_1)
mc^2 \approx mc^2 + G m M / r_1 $ and falls to point
$ Q $ with total energy $ \beta(v_2) mc^2
\approx mc^2 + G m M / r_2 $ such that the mass of the earth is $ M $
and the distances between the center of the earth and $ P $ and $
Q $, respectivley, are $ r_1 $ and $ r_2 $, respectively, with
$ r_1 > r_2 $.  Then

\[ (\beta(v_2) -1)c^2 -(\beta(v_1) -1)c^2 \approx \frac{ G M}{r_2}
- \frac{G M}{r_1}   \; \;\;\;\;\;\; (55)  \]

Then Einstein[4] asserts that when the total energy of $ B $
is transformed into the corresponding photons' energy, we may observe
from (52) that the enrgy of a photon with frequency $ \nu'_1 $
at $ P $ (or $ \nu'_2 $ at $ Q $ , respectively)
in (53) has energy $ h \nu'_1 = hc / \lambda'_1 $ ( $ h \nu'_2 =
hc/ \lambda'_2 $, respectively). It seems to the author this assertion is
wrong since in the above, we conclude that photons' energy remains constant
over any gravitational field.  It seems that in the energy $ \beta(v)
mc^2$, only the energy $ mc^2 $ can be transformed into photons, and the
energy $(\beta(v) -1)mc^2 $ may be absorbed in the vacuum space
as the energy of an expanded spring is eventually absorbed in the vacuum space.

We may also remark the following.  When $ B $ is at $ P $, the
total energy of $ B $ and the earth is $ \beta(v_1) mc^2 +
\beta(v_0) M c^2 $, where $ \beta(v_0) M c^2 $ is the total energy
of the earth.  When $ B $ is at $ Q $, the total energy becomes

\[ \beta(v_2) mc^2 + \beta(v_0) M c^2 > \beta(v_1) mc^2
+ \beta(v_0) M c^2      \;\;\;\;\;\;\; (56) \]

Where does the difference energy $ \beta(v_2) mc^2 - \beta(v_1) mc^2$
come ?  It seems that this energy comes from the vacuum
space so that the vacuum space may be filled with a great amount of energy.
When $ B $ crashes over the surface of the earth, the kinematic
energy of $ B $ may be absorbed in the vacuum space as the result of
disappearing of thermal energy, moleculer decomposition energy, etc.
Some people assert that the total energy of the universe must be zero
so that the total energy consists of the equal (but sign opposite) energy
of particles and antiparticles.  These opinions seem to assert that our
universe was produced from the "naught".  However it is true that the
universe contains at least the space.  The above assertion may imply that
there must exist "antispace", and this assertion may be nonsense.
Thus it seems that the space is filled with
sufficiently large amount of
energy, and in order for the universe to be produced, at least
there existed the space and the principles and 
laws govorning the property of the
universe.

\vspace{2ex}

{\bf 5.4 An outer force and a gravitational field}  \hspace{5mm}
In this subsection,
we shall study the motion of a material body $ B $ with mass $ m $
over a gravitational field $ {\cal F} $ with potential 
function $ \Phi $ as in Subsection 5.2.  We also assume an outer
force $ \vec{F} = (F_x,F_y,F_z) $ works over $ B $.  Let $ E_G $
denote the potential energy of $ B $, $ E_{GK} $ denote the
kinematic energy due to $ {\cal F} $, and $ E_K $ be
the kinematic energy which is obtained due to
outher forces except the gravity.
Thus the total energy $ E_T = mc^2 + E_G + E_{GK} + E_K $. Now
let $ v_T $ denote the total-velocity of $ B $ so that $
\beta(v_T) mc^2 = E_T $.  We also let $ \vec{F_G} = (F_{Gx},F_{Gy},F_{Gz}) $
be the composition of $ \vec{F} $ and the gravitational force.  We shall
present an equation describing the motion of $ B $ which is similar
to the equation (5) in Subsection 5.1.

\vspace{2ex}

{\bf Subprinciple 4}  \hspace{5mm} Let $ K $ be a stationary
Cartesian coordinate system $ (t,x,y,z) $ over 
$ {\cal F} $, and $ B, m, \vec{F} 
$ and $ v_T $ be as above.  Let $ K' $ be a Cartesian coordinate
system  such that $ x' = \beta(v_T)  x, y' = \beta(v_T) y, z'
= \beta(v_T) z $ and $t' = t/ \beta(v_T) $.
Then it holds

\[ m\left(\frac{d^2 x'}{d t^{' 2}}, \frac{d^2 y' }{x t^{' 2}}, \frac
{d^2 z'}{d t'} \right) = (F_{Gx},F_{Gy},F_{Gz})   \;\;\;\;\;\;\; (57) \]

Thus it holds

\[ m \left(\frac{d^2x}{dt^2},\frac{d^2y}{dt^2},\frac{d^2z}{dt^2}
\right) = \vec{F_G}/\beta(v_T)^3    \;\;\;\;\;\;\; (58) \]

\vspace{3ex}

It seems difficult and will be left open to obtain the metric describing
the motion of $ B $.  We remark the metric $ A $ : $ c^2
d \tau ^2 = c^2dt^2 -(dx^2 + dy^2 + dz^2 ) $ in special relativity
is the metric over an inertial syatem where no outer force works.  Thus
when an outer force works over a material body moving in a space with
(almost) zero gravitational potential, the corresponding metric may
be different from the flat metric $ A $ above.  This problem will be also
left for future studies.  Here we observe  the metric may be
regarded to specify kinematic motions, rather than specify the carvature
of the space.

\vspace{2ex}

{\bf 5.5 Black holes.}  We shall discuss the possibility of the existence
of black holes.  We recall that the  Schwarzshild metric is of the
following form :

\[  c^2 d \tau ^2 = (1-2GM/c^2r)c^2 d t^2 -
(d r^2/(1- 2GM/c^2r) + r^2 d \theta ^2 + r^2 \sin ^2 \theta d
\phi ^2) \;\;\;\;\;\;\;\;   (22)  \]

The existence of black holes is usually predicted by considering in
the Schwarzshild metric the case $ 1-2 G M /c^2r \leq 0 $, i.e.,
$ r \leq 2 G M /c^2 $.  But from Subprinciple 2, we have derived the
towfold Schwartzshild metric and the twofold approximate Schwartzshild
metric both of which do not have the points $ r= 2GM/c^2 $ as their
singular points since $ \beta(v_T) $ may become very large, but
still remain always finite.
Since we acknowkedge that it 
always holds $ v<c $ and $ \beta(v) >0 $, it seems that
to study the case $ 1-2 G M /c^2r \leq 0 $ of the Schwarzshild
metric is meaningless.  We may need more reasonable arguments about
black holes.  We also acknowledge that the gravitational force
working over a material body $ B $ with total energy $
\beta(v) mc^2 $, $ m $ being mass, due to the gravity of a material
body with mass $ M $ is $ G m M /\beta(v) ^3 r^2 $,
where $ r $ is the diatance between $ A $ and $ B $.  This
acknowledgment is due to  Subprinciples 1,4.  From this observation,
it seems that we may need new arguments about
gravitational collapses.  Here we  recall an observation
that the nuclear force becomes a repulsive one when the diatance
becomes very small.

It seems that the sentence "the space is curved due to
gravity" is wrong.  In fact, light is deflected due to the gravity of
the sun.  However if we consider a spaceship $B$ having a very powerful rocket
engine $ E $ and consider the travelling of $ B $ near the sun, then
$ B $ travels along an almost "straight" line if $ E $ is sufficiently
powerful.  If the space is curved, then $ B $ may travel along a curved
line even when $ E $ is very powerful.  Moreover we may observe the
following.  If the space is curved due to gravity and a point $ P $ moves
from a place $ X$ to a place $Y$, then another point $ Q $ may occupy
the place $X$.  Thus it seems that at least a very large part
of our universe is 3-dimensional and Euclidian, the present theory about
the carvature of the universe is not sufficiently reasonable, and we do not
have yet a reasonable model of our universe.  Any reasonable
model may have to be at least
homeomorphic to our universe and be consistent with Hubble's law
since the law seems valid.

\vspace{2ex}

{\bf 5.6. The twofold metric principle.}  Based on the results in
Subsections 5.1-5.4, we propose the following principle.

\vspace{3ex}

{\bf The twofold metric principle}

\begin{enumerate}

\item The time at any stationary point with (gravitational) potential
(almost) zero proceeds with the same rate.  This time is called the global
time.
\item The time at a material body with T-velocity $ v_T $ 
(total velocity) proceeds more
slowly than the global time by factor $ 1/ \beta(v_T) $.
Any infinitesimal change of motion of any macro phenomenon $ E $
with T-velocity
$ v_T >0 $ can be described by the set of equations $ H(v_T) $ which
can be obtaind as follows.  First we choose a coordinate system $ K $
such that the direction of each spacial axis of $ K $ is a maximal
velocity-critical one or a zero velocity-critical one. Let $ H(0) $ be the
corresponding set of equations for describing the motion of $ E $
with T-velocity zero over $ K $.  Then $ H(v_T) $ is obtained
from $ H(0) $ by replacing , in each differential equation contained
in $ H(0)$,
(i) each infinitesimal time variable
$ dt $ with $ dt/ \beta(v_T) $, (ii) each infinitesimal maximal
velocity-critical direction length variable $ dr $ with
$ \beta(v_T) dr
$, and (iii) each infinitesimal zero velocity-critical direction length
variable $ dx $ with $ dx $.
\item  Subprinciples 1-4 hold.

\end{enumerate} 

\vspace{2ex}

{\bf Remark 8} \hspace{5mm}
In Subsection 5.3, we show that the light velocity over a gravitational
field is $ c/ \beta(v_G) ^2 $.  Since we acknowledge that the velocity of any
material body cannot exceed the light velocity, this may imply that in
New Subpriniciples, each occurrence of $ c $ should be replaced by $
c/ \beta(v_G) ^2 $.  About the Maxwell equations in Subsection 5.3,
this may be a little confuging.  However we acknowledge that in the Maxwell
equations, it must hold that $ \beta(v_G) $ in New Subprinciple 3 is
actually a value $ \beta'(v_G) \approx c/ \sqrt{c^2/ \beta(v_G) ^4 - v_G^2}
$ and from this $ \beta'(v_G) $, the correct light velocity
$ c/ \beta'(v_G) ^2 $ should be derived from (42).
These arguments may imply that the correct light velocity over a gravitational
field may be a little larger than $ c / \beta(v_G) ^2 $, but when $
v_G $ is sufficiently small w.r.t. $ c $, $ c/ \beta(v_G) ^2
$ can be used as an approximation.

\vspace{2ex}

{\bf 5.7 Einstein's (field) equations}  \hspace{5mm} Einstein's 
equations are of the following form.

\[  R^{\mu \nu} - \frac{1}{2}g^{\mu \nu} R = 
\kappa T^{ \mu \nu}    \;\;\;\;\;\;\; (59) \]

Here the cosmological term is omitted.  The Schwartzshild metric is
generally admitted to be derived by Einstein's equations.  However
we must remark this metric is derived by depending not only on Einstein's
equations, but also on the following two assumptions.

\vspace{2ex}

(1) Schwartzshild posutulated the metric should be of the following
form.

\[ c^2 d \tau ^2 = A(r)dt^2 - B(r)dr^2 - r^2d \theta ^2
- r^2 \sin ^2 \theta d \phi ^2    \;\;\;\;\;\;\; (60) \]

\noindent  so that $ x^0 = t, x^1 = r, x^2 = \theta, 
x^3 = \phi, g_{00}= A(r), g_{11}= -B(r), g_{22} = -r^2
, g_{33}= -r^2 \sin ^2 \theta $ etc. and $ A(r) \rightarrow 1 $ and $ B(r)
\rightarrow 1 $ as $ r \rightarrow \infty $.

(2) In a weak gravitational field (when $ r $ is sufficiently large),
$ A(r) $ should be $ A(r) = \eta_{00} + h_{00} = 1+ h_{00} $
and $ h_{00} = 2V/c^2 = -2GM/c^2r $.

\vspace{2ex}

These two assumptions may be regarded to be two boundary conditions.
In Einstein's equations, $ T^{\mu \nu} $ is the energy stress
tensor, and $ T^{\mu \nu}=0 $ at any point where the energy and
the mass are zero.  We also note the equations are of the symmetric
form w.r.t. all $ \mu, \nu $.  Thus when one solves
Einstein's equations, one may have to be very
careful, especially about boundary conditins. Oterwise one may obtain
meaningless solutions.

The problem whether Einstein's equtions are valid or not is beyond the scope
of this article.  However under the hypothesis that they are valid, it seems
that understanding what $ T^{\mu \nu} $ 
actually shoud be is important.
It seems photons have no mass, and their existence may not give any effect
to any gravitational field.  A gravitational field may be governed only
by the material bodies with nonzero mass existing over the field.
In this situation, mass and energy seem not equivalent.  Here we may 
observe the motion of a spring $ S $ caused by a force.
An outer force $ \vec{F} $ can make
$ S $ expanding, and the energy of the expanded $ S $ is eventually
absorbed in the vacuum space.  But the energy absorbed in the vacuum space
cannot produce the force $ \vec{F} $ by itself.  Thus the direction is
one-way, and in this 
situation, the motion of $ S $ and the corresponding kinematic 
energy may not be equivalent.

Now we shall present the following remarks about the notion of
scalar in tensor calculus.  The line element $ ds $ satisfies
the following.

\[ ds^2 = g_{\mu\nu} d x^{\mu} d x^{\nu}   \;\;\;\;\;\;\; (61) \]

Thus $ ds $ is admitted to be scalar since under a coordinate
transformation $ x^i \rightarrow  X ^i $, $ ds $ satisfies the
following.

\[ ds^2 = g'_{\mu\nu} d X^{\mu} d X^{\nu}   \;\;\;\;\;\;\; (62) \]

However here we must note the equation $ g'_{\mu\nu} d X^{\mu}
d X^{\nu} = g_{\mu\nu} d x^{\mu} d x^{\nu} $ holds because
we count the changes $ x^i \rightarrow X^i $.  For example, let
$ ds^2 = c^2 dt^2 - dx^2 - dy^2 - dz^2 $ and consider a
transformation $ T =2t, X =3x, Y = 4y $ and $ Z =5z $.
Then $ g'_{00} = 1/4, g'_{11} = 1/9, g'_{22} = 1/16 $ and $ g'_{33}
= 1/25 $.  Thus $ ds^2 = c^2d T ^2/4 - d X ^2/9
-d Y ^2/16 - d Z ^2/25 $.  Thus the metric 
is changed, and we must be always remembeing the relations $ T =
2t $ etc. for developing theory depending on the coordinate $
(T,X,Y,Z) $.  

In Subsection 5.1, we admit that over two parallel inertial systems
$ K $ and $ K' $ with relative constant velocity $ \vec{v} =(v_x,
v_y,v_z) $, the transformation $ {\cal T} $ satisfies
the following, where $ K $ is stationay with coordinate $ 
x^0 =ct, x^1=x, x^2=y,x^3=z $ and $ K' $ has coordinate $ X^0
=c t', X^1= x', X^2= y', X^3= z' $.

\[ X^0= c t' = ct/ \beta(v) = x^0 / \beta(v), \; X^1= x - v_x t = x^1
-v_x x^0 /c  \]
\[ x^1 = X^1 + v_x t = X^1 + \beta(v) v_x X^0/c, \; etc.   \;\;\;\;\;\;\;
(63) \]

If we admit the metric over $ K $ is the flat metric $ (1,-1,-1,-1)
$, then the metric over $ K' $ is of the following form.

\[ g'_{00} = \frac{dx^0}{d X ^0}\frac{dx^0}{d X ^0} \eta_{00} + 
\frac{dx^1}{d X ^0}\frac{dx^1}{d X ^0} \eta_{11} +\frac{dx^2}{dX^0}
\frac{dx^2}{dX^0} \eta_{22} \]
\[ + \frac{dx^3}{dX^0}\frac{dx^3}{dX^0} \eta_{33} = \beta(v) ^2 - 
\frac{\beta(v) ^2v^2}{c^2} = 1   \]

\[ g'_{ii} = \frac{dx^i}{dX^i}\frac{dx^i}{d X ^i}\eta_{ii}=-1 
\;\;\; (1 \leq i \leq 3)  \;\;\;\;\;\;\; (64) \]

Thus under the transformation $ {\cal T} $, the flat metric
is preseved.

Thus we conclude that in tensor calculus, we need a standard coordinate system
and the corresponding standard metric (if possible) when we begin
developing any theory.  In this context, we may conclude that in the
Schwartzshild metric, the space is 3-dimensional Euclidian, and the
procession of planatary orbits is predicted w.r.t. this coordinate system.
Thus the metric may be regarded to specify the motion of planets rather
than the carvature of the space.  It is not the space but the motion of
planets which is curved.

Now we turn to the problem of applying Einstein's equations to cosmology.
Generally in cosmology, Einstein's equations are applied to the universe by
assuming the universe consists of macroscopic fluid.  It seems
that this type of modelling has the following problems.

(3) In a macroscopic fluid (water or gas), two neighboring particles
are connected by a molecular force, or collide very often in a very
short time.  In the universe, two neighboring stars may collide very
scarcely, and the distance between them may remain  very large and the
same in a very long time.

(4) The present radious of the universe may be greater than $ 10^{10}
$ light years, and if the speed of propagation of a gravitational force
is that of light, then it may need more than $ 10^{10} $ years
for the propagation of gravitational forces existing very fay each other.
Consider a galaxy $ X $ and a star $ B $ in $X$ whose distance from the center
of $ X $ is $ l $, $ l $ being very large.  If the gravity
between $ X $ and $ B $ is due to  exchanging gravitons, then
how can each graviton $ \alpha $ know the future position of $ B $
which may be very far from the point at which $B$ exists when $ \alpha $
is emitted from $ X $ ?  

(5) From the obsevational facts, it seems that Hubble's law
plays more significant roles than the gravity in the expansion of the
universe.  Hubble's law seems to imply the space itself is expanding.
The expansion of the space itself seems to be suggested by observed
red shift of galaxies and the microwave background emission
which are generally admitted due to the expansion
of the universe.  If the receding velocity $ v $ of a galaxy from
the earth exceeds $ c $ and the space remains stationary, then $
\beta(v) $ is infinity and this is impossible from our acknowledgment
about the motions of material bodies.  Thus the space itself seems
to continue expanding.

From these remarks, it seems that the present universe is too vast to
be applied globally by Einstein's equations.  In developing theory,
probably we may have to count the effect of propagation speed of
gravity if the speed is finite, and also count the effect of Hubble's
law.  Or it may be more reasonable to admit that the gravitational force
is not due to exchanging gravitons, but the property of the space near
$ X $ is changed so that the propagation speed of gravity may be very
large ($>>c$).

In Subsection 5.2, we present Schwartzshild's method for deriving the
metric. In this method, the coordinate are $ (t,r,
\theta,\phi)$ the dimensions of whose compoments are not the same, and
correspondingly all $ \Gamma_{\mu \nu} $ (all 
$ g_{\mu \nu} $) do not have the same dimensions.  Components of
$\kappa T^{\mu \nu}$
have the dimension $ 1/ (length)^2 $ while
in the above deduction, for example, $ R_{22} $ is dimensionless. 

Another method deriving the Schwartzshild metric depends on assuming
$ ds^2 = A(dx^0)^2 -B \sum_{i=1}^{3}(dx^i)^2 $, where $ A $
and $ B $ should be determined, and are assumed to depend only
on $ r= [(x^1)^2 + (x^2)^2 + (x^3)^2)^{1/2} $ (see [16]).  In this
method, the solution is asserted to be the following.

\[ ds^2 = \frac{(1-a/r)^2}{(1+a/r)^2}(dx^0)^2 
-\left (\frac{a}{1+r}\right)^4
\sum_{i=1}^{3}(dx^i)^2    \]
\[ =  \frac{(1-a/r)^2}{(1+a/r)^2}(dx^0)^2 - \left(\frac{a}{1+r}
\right)^4((dr)^2 + r^2((d \theta)^2 + \sin ^2 \theta(\phi)^2))   
\;\;\;\;\;\;\; (65) \]

\noindent where $ a= G M /2c^2 $.  Then the method makes a
coordinate transformation $ x_0 ^1 = x_0, \theta ' = \theta,
\phi '= \phi $ and $ r' = d(1+a/r)^2 $, and asserts that the
Schwartzshild metric is derived.  However in (65), the metric does not
have singular points except $ r=0 $, and as we note often above,
coordinate transformations must preserve the properties of the original
coordinate system.  Thus this method seems also incomplete.

\section{Hubble's law}

Hubble's law seems to suggest that
the space itself continues expanding.  (If the space is "stationary" (not
expanding) and only the galaxies continue receding each other, then the
receding velocities remain the same or decrease due to gravity since no
outer forces do not seem to exist).  The observed red shifts due to the
receding galaxies also seem to support this scknowledgment as shown in
the following remark.  Let $ A $ and $ B $ be two galaxies such that
the distance $ l(t) $ between $ A $  and $B $
is very large for $ t \geq t_0 $, where $ t $ denotes
the global time.  We assume the velocitiy $ v_A $ of $ A$ (the velocity
$ v_B $ of $ B $) w.r.t. a stationary inertial system $ K $
( a stationary inertial system $ K' $) over which $ A $ exists
($ B $ exists) does not exceed $ c $ (since $ \beta(v_A),
\beta(v_B) < \infty $).  If the receding velocity $ v(t_0) $ of $ A $
at time $ t_0 $ from $ B $ exceeds $  2c $, then clealy
the light emitted from $ A $ at time $ t_0 $ never reaches $ B $.
Thus assume that $ v(t_0) < 2c $ and the light $L(t_0) $ emitted from
$ A $ at time $ t_0 $ can eventually reaches $ B $.  Now
assume $ L(t_0) $ arrives at $ B $ at time $ t_0 + \alpha $,
$ \alpha > 0 $.  Since the space is expanding, it holds $
\alpha > l(t_0) / c $.  Consider the next light signal $
L(t_0 + \lambda /c) $ is emitted at time $ t_0 + \lambda / c $,
where $ \lambda $ is the wavelength of the light.  Now let
$ P $ denote the "stationary" position at which $ B $ existed
when $ L(t_0) $ arrives at $ B $, so that the distance $ A $
and $ P $ is always $ c \alpha $. 
Then $ L(t_0 + \lambda /c) $ arrives at $ P $ at time $
t_0 + \lambda / c + \alpha $.  At this global time,
due to the expansion of the space, the diatence $ l_0 $
between $ B $ and $ P $ can be sufficiently large if
$ l(t_0) $ is sufficently large.  Then
$ L(t_0 + \lambda /c) $ arrives at $ B $ at time $ t_0
+ \lambda /c + \alpha + \beta $, where $ \beta >
l_0 /c $.  Here we assume the receding velocity of $ B $
from $ A $ is less than $ c $ so that $
L(t_0 + \lambda /c) $ can reach $ B $.  Since Hubble's constant
is very small, the above $ l_0 $ can be very large, and
we may observe a red shift $ \Delta \lambda / \lambda
> l_0 / \lambda - 1 $ which may be very large.

In the above arguments, we note that in the standard Doppler effect, one
considers the case where the space is "stationary", but in Hubble'law,
while the light is travelling, the space itself continues expanding
and the receding velocity is also increasing.  Thus in Hubble's law,
the receding velocity and the diatance should be accumulated
in the calculation.  The observed background microwave radiation
also seems to imply that for any global time $ t \geq t_1 >0 $, where
$t_1$ may be very small, the expansion
rates of the space were the same at all points in the universe.
We also remark that in the above expansion of the space, the space-time
transformation from $ A $ to $ B $ cannot be the Lorentz transformation
since if the light $ L $ emitted from $ A $ visits
$ B $ and returns to $ A $, then the wavelength of $ L $ becomes
very large as observed above.
In the rest of this subsection, we shall present
a remark about a relation possibly lying between the (present) age of
the universe and Hubble's constant under the hypothesis that Hubble's
law is valid.  (From observational facts and the above remarks, we may
perphapas need a modification of the law, but the modification may
be not very serious).  In fact, we shall present arguments for
supporting the following assertion.

Assertion A : the (present) age of the universe
is greater than the Hubble time $H_0^{-1}$.

In cosmology, the age of the universe is generally convinced
to be between $ 2 \times 10 ^{10}$ and $ 10^{10} $ years, 
or approximately $ 1.5 \times 10^{10} $ years.  Recently large
inhomogeneous structures called great walls were discovered whose
distances from the earth are estimated to be about $ 10^9 $ light
years (see [1]), and it has become a problem in dispute to study whether
it was possible for
such a large inhomogeneity to have been constructed in the estimated
age of the universe $ \approx 1.5 \times 10^{10} $ years.  It is
generally acknowledged that the age of the universe is smaller than the
Hubble time $ H^{-1}_{0} $.  Here $ H_0 $ is the Hubble constant to 
which the following cosmological observation is acknowledged.

Observation A.  For any galaxy $ N $ which is not too near nor 
too far, the receding velocity $ v$ of $ N $  w.r.t. the earth is 
related to the distance $ d $ between $ N $ and the earth by
the following equation :

\[  v = H_0 d  \;\;\;\;\;\;\;\; (1) \]

From many cosmological observations, the Hubble constant $ H_0 $
is generally acknowledged to be between $ 50 kms^{-1}Mpc^{-1} $ and
$ 100 kms^{-1}Mpc^{-1}$.
The (present) age $ t_p $ of the universe is assumed to be smaller
than $ H^{-1}_{0} $ by the following argument (see, e.g., [2]). 

First one
assumes that the Big Bang occurred, and the universe has continued
expanding after the Big Bang.   Let $ a(t) $ denote the so-called
scale factor at time $ t $ so that $ a(t) $ may denote the size or the radious
of the universe at time $ t $.
Then the history of the expansion
of the universe may be described by depicting the curve of $ a(t)
$.  Depict the figure of $ a(t) $ over the plane 
where the horizontal axis is the
time axis, t-axis, and the vertical axis is the a(t)-axis.
The origin $ O $ of axises corresponds to the beginning time of the
universe, i.e., the time when the Big Bang occurred.  Let $ P $
be the point on the t-axis such that $ OP$ denotes the present time
$ t_p $, i.e., the age of the universe.  Let $ Q $ denote the 
point on the curve $ a(t) $ corresponding to $ t_p $ so that 
$ PQ$ denotes $ a(t_p) $.  One assumes that the curve of $ a(t) $ is
concaved to the downward, i.e., $ \dot{a}(t) > 0 $ but $ \ddot{a}(t)
< 0 $, since $ \dot{a}(t) $ may decrease due to the gravitational
forces as $ t $ increases.  Let $ \alpha $ be the angle such that
the tangent at $ Q $ meets the t-axis at point $ T $ at the
angle $ \alpha $.  Then it holds

\[ \tan \alpha = PQ / PT = \dot{a}(t_p) \;\;\;\;\;\;\;\; (2) \]

so that

\[  OP < PT = PQ / \dot{a}(t_p) = a(t_p) / \dot{a}(t_p) =
             H^{-1}_{0} \;\;\;\;\;\;\; (3) \]

 Note that the relation
$ OP< H^{-1}_{0} $ is possible since one assumes the curve of $ a(t) $
is concaved
to the downward.  Thus  it may hold $ OP > H^{-1}_{0} $ in case where
(i) the curve of $ a(t) $ is concaved to the upward, i.e., it holds
$ \dot{a}(t) > 0 $ and $\ddot{a}(t) >0 $ , and (ii) $ a(t_p) / a(1) $
is sufficiently large, where $ 1 $ in $ a(1) $ is a unit global time
such that after $ t \geq 1 $, the expansion of the space has been
governed by Hubble's law (here $ H_0 $ may dpend on $ t $).
(We shall
present a more detailed condition for $ OP> H_0^{-1} $ later).  
Thus we may conclude the (present) age $ t_p $ of the universe depends not
only on $ H_0 $ but also on $ a(t_p) $ (or equivalently $ a(t_p)/
a(1)$ or $ a(t_p)/a(0) $).
Indeed, we shall develop arguments
for supporting the assertion that it holds $ \dot{a}(t) >0 $
and $ \ddot{a}(t) >0 $ by depending on Hubble's law.

First we consider an explosion of a material body $ M$ of huge mass
in the vacuum space at time $ t_2 $.  Each broken piece of $ M$
begins moving with initial velocity $ v $.  Then after a time lapse
$ t $ from $ t_2 $, the maximum distance $ d(t) $ between any
two broken pieces may be about $ 2 \times \max \{ vt \mid v $ is an
initial velocity of a broken piece of $ M$ \} when $ t $ is
small.  Then it may hold $ \ddot{d}(t) < 0 $ since $ \dot{d}(t) $ may
decrease as $ t $ increases due to the gravitational forces working
over all broken pieces of $ M$.  In the Big Bang theory, it seems 
that the scenario is quite different from the above material explosion
case since the expansion of the universe seems to occur homogeneously
and isotropically in the space, that is, the space itself 
continues
expanding.  Thus the distance between any two distinct spatial points
in the universe may increase as the time proceeds.  

Now we shall first present an elementsry argument for supporting
$ \ddot{a}(t) >0$.
Consider two distinct points $ P $ and $ Q $ in the
space.  We assume that at any point $ R $ in the space, the speed (rate)
of time lapse at $ R $ at global time $ t $ 
proceeds with  the
same speed since we assume the universe is homogeneous and isotropic.  
Then let $ t_1 \geq 1$ be any global time from the Big Bang, and let $l_1$ be
the distance between two distinct points $ P $ and $ Q
$ at time $ t_1 $.  Let $ t $ be a small time lapse
after time $ t_1 $, and let $ l_2 $ be the distance between $ P $
and $ Q $ at time $ t_1 +t $.  We call the value $ l_2/ l_1 $
the expansion rate for time lapse $t$ at time $ t_1 $.
Now let $ R $ and $ Z $ 
be two points whose distance is $ n l_1 $ ($ n \geq 2 $) at time
$ t_1 $.  Then since we assume that the space is homogeneous and
isotropic and the space itself continues expanding, the distance $
l_3 $ between $ R $ and $ Z $ at time $ t_1 + t $ may become
about $ n l_1 + n(l_2 - l_1)= n l_2 $.  Let $ m \geq 1 $ be an 
integer.  If $ t $
is sufficiently small and the expanding rate is almost constant
between time $ t_1 $ and time $ t_1 + mt $, then the distance
$ l_4 $ between $ R $ and $ Z $ at time $ t_1 + mt $ may 
become approximately $ l_4 \approx n l_1 (l_2 / l_1)^m $.

Then let $ e(t_1 +t)$ and $ e(t_1 +mt)$ be the expansion rates 
for time lapse $t$ and time lapse $mt$
at time $ t_1 $, respectively.  Then it
may hold

\[ e(t_1 +t)= l_2 / l_1, \;
 e(t_1 +mt) = nl_1(l_2/l_1)^m/nl_1 = (l_2/l_1)^m  \;\;\;\;\;\;\; (4) \]
   
Then it holds $ e(t_1 +mt)>e(t_1 +t)$ if $ m >1$.
If the equations (4) are valid, then the 
figure of the scale factor is
concaved to the upward, and in this case, it holds $ \ddot{a}(t) >0
$ for $ t \geq 1 $.

We shall develop a more formal argument
for supporting the assertion that for any $1< t \leq t_p$,
the scale factor  $ a(t) $ is approximately equal to an exponential function
of $ t $ so that $ \ddot{a}(t) >0 $ as follows.

Consider a galaxy $ N $ from which a light signal $S$ was emitted at
time $ t_p - t_1 -t_2$ for $ t_1 $ being about $ 
10^6 \sim 10^7 $ years and $ t_2 $ being small, and $S$ arrives
at the earth at $ t_p $ (present time).  We assume Hubble's law holds between
$ N $ and the earth.  We shall consider a sufficiently small time
lapse variable $ t>0 $ so that it holds $ 0<H_0 t << 1 $ and 
$ e^{H_0 t} \approx 1+ H_0 t $.  We also assume that for
$0 < t \leq t_2 $, it holds that $ 0<H_0 t< < 1 $ and 
 $ e^{H_0 t} \approx 1+ H_0 t $.  Then let $ L $ be the distance
between $ N $ and the earth at time $ t_p - t_1 - t_2 $.  We consider
the expansion rate $ b(t) $ for time lapse $t$ at time $t_p-t_1-t_2$
as a function of $ t $, where for
$ 0<t \leq t_2 $, $ b(t) $ denotes the expansion rate of
the universe when we set the initial time as  $ t_p - t_1 -
t_2 $.  Thus it holds $ a(t_p - t_1 - t_2 +t) \approx b(t)a(t_p - t_1 - t_2)
$.

Now let $0< t \leq t_2$ and 
$ \Delta t $ be a very small time lapse, and consider times
$ t_p - t_1 - t_2 +t $ and $ t_p -t_1 - t_2 + t + \Delta t $ 
at $ N $.  We assume that the distances
between $ N $ and the earth at time $ t_p - t_1 - t_2 +t $ and time 
$ t_p - t_1 - t_2 +t+ \Delta t  $ are
$  b(t) L$ and $  b(t+ \Delta t)L$, respectively.  
Let $ g(b(t) L)$ be the function of $ b(t) L $
such that the time lapse $ t_3 $ within which a light signal
$ S$ begins its travel from $N$ at time $ t_p -t_1 - t_2 +t$ and
ends its travel
at the earth at time $ t_p -t_1 - t_2 + t+ t_3$
 satisfies the following.

\[  ct_3=g(b(t) L) \;\;\;\;\;\;\; (5) \]

Then Hubble's law 
asserts the following.

\[  \displaystyle{\frac{g(b(t+ \Delta t)L)-g(b(t) L)}
   {\Delta t} \approx H_0 g(b(t) L), } \;\;\;\;\;\;\; (6) \]
   
\noindent where we wssume that $ H_0 g(b(t) L)$ 
and $ H_0 g(b(t+ \Delta t)L)$
are approximately equal.  From (6), we have

\[  g(b(t) L) \approx  k(L) e^{H_0 t} \;\;\;\;\;\;\; (7) \]

\noindent where $ k(L) $ is a function of $ L $ independent of $ t $.
Now we want to determine
$ b(t) $ and $ g(u) $ approximately.  To do this, from many well
known cosmological observations concerning Hubble' law, we may
assume that $ 1< g(u) /u << 2 $.  Thus we put approximately $ g(uL) =
(uL)^r $ for a positive number $ r $ with $ 1<r <<2 $.  Then
from (7), we have

\[  k(L)=L^r, \; b(t) L \approx L e^{H_0 t/r}  \;\;\;\;\;\;\; (8) \]

From (8),

\[  b(t)  \approx  e^{H_0 t/r} \;\;\;\;\;\;\; (9) \]

Since we assume that the space itself continues expanding, we
acknowledge that it holds approximately $ b(t+t') 
\approx b(t)b(t')$ for any
$ t,t' >0 $ with $ t,t'$ being small as above.
Note that if $b(t+t')
= b(t)b(t')$, then the following holds.

\[ \displaystyle{ \dot{b}(t) = \lim_{h\rightarrow 0}
\frac{b(t+h)
  -b(t)}{h} 
 = \lim_{h \rightarrow 0}\frac{
  b(t)(b(h)-1)}{h} =db(t), \; \; 
  b(t) = \alpha e^{dt}} \;\;\;\;\;\;\; (10) \]
  
\noindent where  $\lim _
{h \rightarrow 0}(b(h)-1)/h =d$, and $\alpha$ is a constant. 
  
If the expansion principle has not been changed so much from the early times
of the universe,  then $ a(t_p)
$ may be calculateds by accumulating $ b(t) $ in the following way.

\[  a(t_p) = a(1) e^{A(t_p)}, \; where \; A(t_p) = \int _1^{t_p}
(H_0/r) dt  \;\;\;\;\;\;\; (11) \]

The Hubble constant $ H $ may have been dependent on the global time,
but it seems that the principle has been the same.  Thus we consider the
case where the Hubble constant has been the same as $ H_0 $ and $ 1 \leq
r <<2 $ is small.  Thus we assume $ H_0/r \approx H_0 $.

From (11), it holds

\[ a(t_p) = a(1) e^{H_0 t_p}, \; t_p = H_0^{-1} \log (a(t_p)/a(1))  
\;\;\;\;\;\;\; (12) \]

Hence $ t_p $ depends not only on $ H_0^{-1} $ but also on $ a(t_p)
/a(1) $.  Since $ a(t_p)/a(1) $ seems very large, $ t_p $ may
be greater than $ H_0^{-1} $ by a large factor.
There may exist a possibility such that the Hubble constant is a function
$ H(t) $ and it holds $ \dot{H}(t)<0 $.  Then it may hold
$ a(t_p) = a(1) e^{B(t_p)}, \; B(t_p) = \int_1^{t_p}H(t)dt $, and
$ t_p $ may be smaller than $ H_0^{-1}\log (a(t_p)/a(1)) $.

The above arguments may have rough approximations, but it seems that the
age $t_p$ of the universe is greater than the Hubble time $H_0^{-1}$.

\section{The dark matter propblem}

In cosmology, the dark matter problem is of great concern (see [13]).
One of the main reasons the problem concerns is an observational fact 
that at a galaxy $ {\cal G} $, the rotationg speed of
a star $ S $ in $
{\cal G} $ is about $ 200 \sim 300 $ km/s seemingly independent
from the distance between $ S $ and the center of 
$ {\cal G} $.  It is generally
admitted this fact occurs because there exists a great amont of dark
(invisible) matter called invisible halo outside the visible part of
$ {\cal G} $.  The invisible halo is also admitted to exist in order
for $ {\cal G} $ to be stable.  We note the sun system is stable because
a great part of mass gathers at the sun.  Note also if the invisible
halo has great mass, then it may not be stationay and may round $
{\cal G} $ since otherwise it may fall into the center of $
{\cal G} $, during a very long history of $ {\cal G} $ which
seems impossible.  Thus if the above invisible halo exists, then $
{\cal G} $ may not be stable a fortiori. We shall present a new remark
(although simple but maybe important) about this problem in this section.

We consider a (x,y)-plane such that the sun is at the origin (0,0)
and a galaxy $ {\cal G} $ is at point $ P = (0,l) $, where $ l $
is the distance between the sun and $ {\cal G} $.  We assume $
{\cal G} $ is a disc $ C $ on the plane whose 
radius is $ r $.  To make the argument simple, we also assume the mass
$ M $ of $ {\cal G} $ gathers at point $ P $.  Let $ L $
denote the line segment parallel to the x-axis, passes through $P$
 and its endpoints are
at the periphery of $ C $.  We assume $ {\cal G} $
rotates at its own axis in the counterclockwise way.  Due to the Newtonian
mechanics, the rotating speed $ v(u) $ of a star with mass $ m $
whose distance from $ P $ is $ u $ satisfies $ m v(u) ^2
/u = G m M /u^2 $, i.e., $ v(u) 
= \sqrt{G M /u} $.

Let $ Q = (-r,l) $ be the left end point of $ L $.  Then the
rotating speed $ v(r) $ of a star at $ Q $ is $ \sqrt{G M /r}
$.  This value is observed about $ 200 \sim 300 km/s $.  Now let us
consider the rotating speed of a star whose distance from $ P $
is $ u $, $ 0<u<r $.  Let $ R $ be the point $ (-u,l) $ on
$ L $, and $ L_1 $ be the line segment parallel to the y-axis
whose end points are on the periphery of $ C $ and on which $R$ exists.
We note all the light
from all the stars on $ L_1 $ is observed to come from $ R $ on the
earth.  Now consider a star $ S $ at point $ T = (-u,l-b) $ on
$ L_1 $, where $ b=d/4 $ and $ d $ is the length of $ L_1 $.
Then the rotating speed of $ S $ is $ v(\sqrt{u^2 + b^2}) =
\sqrt{G M /\sqrt{u^2+b^2}} $.  Now let $ \alpha $ be the angle
between $ L_1 $ and $ L_2 $, where $ L_2 $ is the line 
segment connecting $ P $ and $ T $.  Then $\tan \alpha 
= b/u $.  We consider the ($ -y $)-direction of the rotating speed
$ v(\sqrt{u^2+b^2},-y) $ of $ v(\sqrt{u^2+b^2}) $
at point $T $.  It holds $ v(\sqrt{u^2+b^2},-y) =
v(\sqrt{u^2+b^2}) \cos \alpha = \sqrt{G M /\sqrt{u^2+b^2}}u/\sqrt{u^2
+b^2} < \sqrt{G M /\sqrt{u^2+b^2}}\sqrt{u^2+b^2}/r <
\sqrt{G M /\sqrt{u^2+b^2}}\sqrt{\sqrt{u^2+b^2}/r} =
\sqrt{G M /r} $.

Thus due to the Doppler effect, the ($ -y $)-direction of the rotating
speed of $ S $ may be observed smaller than $ \sqrt{G M /r} $
on the earth. The ($ -y $)-direction of the rotating speed of a star
at point $ (-u,l) $ may be $ \sqrt{G M /u} $ greater than $
\sqrt{G M /r} $.  But by counting the mean ($ -y $)-direction
rotating speed, we may observe the fact presented in the beginning part
of this section.  There may exist other unnoticed reasons about the
dark matter problem, but the above arguments seem worth studying.

\section{The principle of equivalence}

There exist in the litarature many interpretations of the principle
of equivalence.  One typical interpretation may be the assertion that
the gravitational mass is equivalent to the inertial mass.  Since it
turns out that the relativity principle is not valid, we must be
careful about the interpretation of the principle.  In this section,
we shall present comments about the following interpretation of the
principle of equivalence (see e.g. [1]).

A : ''In a small laboratory falling freely in a gravitational field,
the laws of physics are the same as those observed in a Newtonian
inertial system in the absence of gravitational field.''

We first remark the following experiment made by Hafele and Keating
(see [1]).
Consider three clocks $A$, $B$ and $C$ whose mechanisms are
equivalent, initially synchronized in setting and rate and at rest
on the earth's surface.  Clock $A$ remains at rest while $B$ is flown
over the earth by an eastward journey at height $h$ in an aircraft
whose speed relative to the ground is $v$, and $C$ is flown over
the earth by a westward journey at height $h$ in an aircraft
whose speed relative to the gound is also $v$.  After the
circumnavigations, $A$, $B$ and $C$ are compared, that is, the
proper time $\tau _A$ experienced by $A$ is compared with the
proper times $ \tau _B $ and $ \tau _C $ experienced by $B$ and
$C$, respectively.  

In 1971, Hafele and Keating made eastward and westward journies
round the earth on commercial jet flights, carrying caesium clocks
$B$ and $C$ ($B$ for the eastward journey and $C$ for the westward 
journey) which they later compared with clock $A$ which remained
at the US Naval Observatory in Washington (see [1]).  Then their experiment
showed that $ \tau _B - \tau _A = -59 \pm 10 $ and
$ \tau _C - \tau _A = 273 \pm 7$ (nanoseconds).  Now 
assume that the above principle of
equivalence is valid.  Consider a small local space $S$
very close to the ground.  Let $K$ be the Cartesian
inertial system falling freely in $S$ whose existence
is assured by the principle of equivalence.  Since $ \tau_B < \tau_C
$, due to the twofold metric principle, the eastward velocity
of $ K $ is greater than the westward velocity of $ K $ w.r.t.
the universe and $ K $ is not stationary w.r.t. the eastward and the
westward directions.  But then as in the proof of Theorem 2, 
the light velocity into
the eastward direction should be smaller than the light velocity into
the westward direction, which is a contradiction to the Michelson-Morley
experiment and our conclusion in Subsection 5.3.  We also recall that
over a gravitational field, the light velocity is $c/ \beta(v_G)^2$.
Thus we acknowledge
that any coordinate system over the gravitational field 
of the earth cannot be identified
with any inertial system moving with constant velocity in the vacuum space
where the gravitational potential is almost zero.  We also remark that
in the twofold metric principle, it is not the accelaration but the
velocity which plays a key role.  The above arguments show that a modified
version of the above principle $ A $ which is obtained from $ A $
by replacing the words "falling freely in a gravitational field" with
"falling freely with some velocity in a gravitational field" is also
not valid.
Hence we conclude that the above
version of the principle of equivalence is not valid.
From the twofold metric principle, we acknowledge that it holds
the kinematic energy of clock $B$ flying eastward is larger than that of
clock $C$ flying westward.

\section{Concluding remarks}

The special relativity principle asserts that all inertial systems are
equivalent (and the Lorentz transformation is valid and the light velocity
is $ c $ over any inertail system).  The general relativity principle
asserts all coordinate systems are equivalent.  In this paper, we show
that the special relativity principle is not valid by depending on
TI Lemma and TI Theorem (Section 4) and the general
relativity principle is not valid even
when we concern only tensor-based general relativity theory.
We propose a new principle called the twofold metric principle, and show
that in many applications, the new principle works well.

In Section 2, we present properties of the Lorentz transformation.
Section 3 presents an electromagnetic 
example which seems to be a strong evidence to the invalidity of
Special Relativity Principle by depending on the Maxwell equations. 
 In Section 4, we establish TI Lemma (Lemma 3) and TI Theorem (Theorem 1).
By depending on TI Theorem,
 we study properties of inertial systems.  We deduce that neither
Special
 Relativity Principle nor Strong Constant Light Velocity Principle
 is valid, and the Lorentz transformation is not the space-time
 transformation over two inertial systems 
 moving with nonzero relative constant
 velocity.   We  
deduce  that Einstein's transformation formulae of the Maxwell
equations are not valid.  In Section 5, we propose 
the energy-velocity equation and
the twofold metric principle.  Four Subprinciples are proposed, and
by depending on them, we present arguments for solving
(1) $ E = \beta(v)mc^2 $, (2) the travelling distance of a muon with
velocity $ 0.999c $, (3) a modified version of the Schwarzshild
metric called the twofold Schwartzshild metric,
(4) the Maxwell equations and the light velocity over the 
space with a gravitional field, (5) deflection and red shift of 
light due to gravity,
and (6) an explanation for the Michelson-Morley experiment.  Our
calculated value of procession of planatery orbits by depending on the
twofold Schwartzshild metric
is the same as the corresponding well known value
derived from the Schwarzshild metric.  But the twofold
Schwartzshild metric does
not have the points $ r=GM/c^2 $ as its singular points while the
Schwartzshild metric does.  In Section 6, we present arguments supporing
the assertion that Hubble's law implies the expansion of the space itself,
and the (present) age of the universe is greater than the Hubble time
$ H_0^{-1} $.
In Section 7, we present a new remark about
the dark matter problem.
In Section 8, we develop arguments
 from which we deduce that one well known interpretation of
 the principle of equivalence in general 
 relativity theory
 is not valid.
 
 
We remark the following.
 Dirac developed relativistic quantum mechanics on electrons by
 depending on the Lorentz transformations, and theoretically deduces
 the existence of positrons (see [2]).
 But in Dirac's arguments, the principle playing a key role
 may be said to be the symmetry
 principle rather than the relativity principle, where the symmetry principle
 may be a physical "philosohy" respecting symmetry in physics.  
 The theoretical effect of depending on the Lorentz transformation seems
 time dilation and length contraction, which may be observed also
 on the twofold metric principle.  The only
 space-time transformation $ {\cal T} $
 over two coordinate systems $ K $ and $ K' $ such that
$ {\cal T} $ is linear and symmetric w.r.t. $ K $ and $ K' $, and
time dilation and length contraction can be observed over $ {\cal T} $
may be the Lorentz transformation.  Since time dilation and length
contraction may be said to occur in the twofold metric principle, in a
sense, one may say that the Lorentz transformation can be derived by
depending on the twofold metric principle, the ev-equation
and the symmetry principle.
In Dirac's arguments, four-vectors also appear.  In any four-vector,
the temporal component may be related with the spacial components and the total
energy.  Thus four-vector arguments seem to be related with the twofold
metric principle.  We also recall that the well known arguments deriving
$\beta(v)mc^2$ by depending on four-vectors have in some part wrong
computation.
There may exist some correct results which can be
derived by depending on the Lorentz transformation.
Or many laws in physics may respect symmetry, and they may be "invaliant"
under the Lorentz transformation.  But
the Lorentz transformation does not preserve simultaneity, and produces
wrong results (such as time machines) sometimes.  Any result obtained
by depending on the Lorentz transformation may have to be checked of
its correctness by any other valid principles.  We note that sometimes
experimental tests may not be sufficient as in the case of the Schwartzshild
metric, where the metric predicts the orbit of Mercury correctly, but
also implies the existence of black holes (probably wrongly).

Inevitably this paper does not present a full analisis of special and
general relativity.  The author hopes it contains results and arguments
which will contribute to the development of physics in the new directions.
We may need both learning and pondering as an ancient Asian philosopher
Confucious said : "Learning without pondering may produce darkness,
and pondering without learning may produce crisis".

\section{Appendices}

{\bf Appendix 1} \hspace{5mm}
We shall show that the new synchronization relation introduced in Section 1
is an equivalence relation.  Clearly it is reflexive and symmetric.  We
shall prove its transitivity.  Let $P$,$Q$ and $R$ be three
points at rest on $K$, and $A$,$B$ and $C$ be three clocks which are 
at rest on $P$,$Q$ and $R$, respectively, and whose mechanisms are
equivalent. Assume that $A$ and $C$  and $B$ and $C$,
respectively, are synchronized.  As in Section 1, assume that a light sygnal
$L$ travels between $P$ and $Q$, and let $ t_1,t_2 $ and $t_3$ be the
times on $A$, $B$ and $A$, respectively, and let $l$ and
$r$ be as in Section 1.  
Now let 
$ t_2 = t_1 + l/r - b $, where $b$ is the bias of $A$ w.r.t.
$B$.  Since $A$ and $C$ are synchronized, it holds the bias of $A$
w.r.t. $C$ is zero.  The bias of $B$ w.r.t. $C$ is also zero.  Let
the distances between $P$ and $R$ and $R$ and $Q$ be $l_1$ and $l_2$,
respectively.  Let $r_1$ and $r_2$ be the velocities of light signals
for travelling from $P$ to $R$ and from $R$ to $Q$, respectively.
Now put $ e = l/r - (l_1 /r_1 + l_2 / r_2) $, and let a light signal
$L_2$ be emitted from $P$ at time $ t_1 + e $ 
on $A$, arrive at $R$ at time $ t_1 + e + l_1/r_1 $ on $C$, and finally arrive 
at $Q$ at time $ t_1 + e + l_1/r_1 + l_2/r_2 = t_1 + l/r $ on $B$.
It holds $b=l/r-(t_2-t_1),  
t_2 = t_1 + l/r - (e + l_1/r_1 + l_2 / r_2) + l/r =
t_1 + l/r $ and $ b = 0 $  
since (i) $A$, $B$ and $C$ are punctual, (ii) 
$ t_1 + l/r $ is the time at $ Q $ when $ L_2 $ arrives at $ Q
$, (iii) the time lapse for $ L_2 $ to travel from $ P $, via $
R $ to $ Q $ is $ l_1 / r_1 + l_2 / r_2 $, (iv) $ L_2 $ is
emitted from $ P $ at time $ t_1 + e $, and (v) the time lapse 
for $ L $ to travel from $ P $ to $ Q $ is $ l/r $. 
Thus the new synchronization 
relation is an equivalence relation.  Remark that for the synchronization
of $A$ and $B$, we need to know $ r $ and $ s  $.  If the 
clocks are not synchronized and the biases between them are not known
over an inertial system $K_0$, then one cannot even make any experiments
for checking whether or not the velocity of a light signal is constant
over $K_0$.

\vspace{2ex}

{\bf Appendix 2.} \hspace{5mm}
We shall study the twin paradox concerning the Lorentz transformation.
The standard argument about the
twin paradox proceeds as follows (see, e.g., [14]).
Let $K$ and $K'$ be two parallel inertial systems with relative constant
x'-axis velocity $ v> 0 $.
  Assume that there exist twins $A$
and $B$ , and let $P$ be a point at rest on the x-axis of $K$.  Twin $B$ starts
his travel from $P$ on the x-axis of $K$ with constant velocity $v > 0$
at time $t_1$ on $K$.  Then at time $ t_2 = t_1 + t$ $(t > 0) $ on $K$,
he arrives at a point $Q$ at rest on the x-axis of $K$.  
He changes his direction
toward $P$ and finally returns at $P$ at time $t_3$ on $K$.  Now
let $R$ be the point at rest on the x'-axis of $K'$ which 
coincides with $P$ at time
$t_1$ on $K$.  Then at time $t_2$ on $K$, $R$ coincides with $Q$.  Let
$t'_1$  and $t'_2$  be the times at $R$ on $K'$ corresponding to
$t_1$ and $t_2$, respectively.  By time dilation, it holds

\[ t_2 - t_1 = \beta(v) (t'_2 - t'_1) > t'_2 - t'_1 \]

Let $t'_3$ be the time on $B$ when $B$ returns at $P$.  Then as above,
one asserts that it holds 
$ t_3 - t_2 = \beta(v) (t'_3 - t'_2) $.  
Hence $ t_3 - t_1 = \beta(v) (t'_3 - t'_1) > t'_3 - t'_1 $.

Thus one asserts that if twin $A$ stays at $P$ during the travel of
twin $B$, then $B$ is younger than $A$ after his travel.  The above case
will be called the $(K,K')$ case.  By changing the roles of $K$ and $K'$
, we can consider the $(K',K)$ case as follows.  In this case, $A$ starts
his travel at point $R$ at rest on the x'-axis of $K'$ at time $t'_1$ on $K'$
with constant velocity $ -v $, arrives at point $S$ at rest on the x'-axis
of $K'$ at time $t'_2$ on $K'$, then changes his direction toward $R$
, and finally returns at $R$ at time $t'_3$ on $K'$.  Let $P$ and $Q$
be two points at rest on $K$ which coincide with $R$ and $S$, respectively,
when $A$ starts his travel at $R$ and returns at $R$, respectively.
We assert that while $A$ travels from $R$ to $S$, $A$ is at rest at $P$. Let
$ t_1,t_2 $ and $t_3$ be times at $P$,$P$ and $Q$ on $K$ corresponding
to $ t'_1,t'_2,t'_3 $, respectively.  As above, one asserts it holds

\[ t'_3 - t'_1 = \beta(v) (t_3 - t_1) > t_3 - t_1 \]

Then the twin paradox asserts that due to the special relativity principle, the
$(K,K')$ case and the $(K',K)$ case should be equivalent, and the relation
between $ (t_3 - t_1) $ and $ (t'_3 - t'_1) $ should be the same
in both cases, a contradiction.

We note the following two remarks.

\begin{enumerate}
\item In the $(K,K')$ case, the initial clock condition is such that all the
clocks on $K$ indicate the same time, and no two distinct clocks on
$K'$ indicate the same time, while in the $(K',K)$ case, the initial
condition is the converse of the $(K,K')$ case.
\item In the $(K,K')$ case, when $B$  changes his direction at $Q$ toward
$P$ and travels from $Q$ to $P$, the coordinate system in which the time
at $B$ is measured is not in fact $K'$, 
contrary to our assumption above,
but the coordinate system
which is parallel to $K$ and its origin moves on the x-axis of 
$K$ with velocity $ -v $.  Thus the times on $B$ are not measured
by the same clock while the time on $P$ is
measured by the same clock.  About the $(K',K)$ case, 
 the same argument holds.
\end{enumerate}

Due to (1) and (2), the $(K,K')$ case and the $(K',K)$ case are two distinct
events occurring on $K$ and $K'$ as far as we acknowledge that the
Lorentz transformation is valid.  The Lorentz transformation
is a bijection from $CO(K)$  onto $CO(K')$.  For any coordinate $ 
(x,y,z,t) \in CO(K) $, there exists
one and only one coordinate $ (x',y',z',t') \in CO(K')
$ which is the image of $ (x,y,z,t) $ under the Lorentz transformation
.  Thus for any event $E$ occurring on $K$, there exists the unique
event $ E' $ occurring on $K'$ which is the image of $E$ under the
Lorentz transformation, and vice versa.  This means that we cannot
find any contradictions in the Lorentz transformation as far as
we concern only  the Lorentz transformation by discarding  all
physical understandings about the universe.  Some people assert that
the twin paradox occurs since we do not count the time lapse needed
for acceleration.  But this explanation is meaningless since $B$
in the $(K,K')$ case (or $A$ in the $(K',K)$ case) can travel with constant
velocity as long as he wishes so that the effects by the acceleration
become negligible.  This can be seen as follows.
Consider the $(K,K')$-case.
Twin $ B $ begins his travel from $ P $ at time $ t= t' =0
$ and accelerates, and  at time $ t'_1 $ on his clock,
his velocity w.r.t. $ A $ becomes $ v>0 $.  Then $ B $ continues
his travel with velocity $ v $ until time $ t'_2 $ on his clock.
By time dilation, on  $A$'clock, at least the time lapse $ \beta(v)
(t'_2 - t'_1) $ passes.  Then $ B $ decelerates and his velocity
w.r.t. $ A $ becomes $ 0 $ at time $ t'_3 $ on his clock.
$ B $ changes his direction toward $ A $ and accelerates,
and his velocity w.r.t. $ A $ becomes $ -v $ at time $ t'_4
$ on his clock.  $ B $ continues his travel with velocity
$ -v $ until time $ t'_5 $ on his clock.  Then by time
dilation, at least the time lapse $ \beta(v)(t'_5 - t'_4) $ passes
on $A$'s clock.
Then $ B $ decelerates and returns at $ P $ at time $
t'_6 $ on his clock.  We note that $ t'_1,t'_4 - t'_2 $ and $
t'_6 - t'_5 $ may depend only on $ v $, and $ B $ can travel
as long as he wishes so that $ t'_2 - t'_1 + t'_5 - t'_4 -
(t'_1 + t'_4 - t'_2 + t'_6 - t'_5) $ is arbitrarily large.
Thus the time lapse at $ A $ during the $B$'travel is at least
$ \beta(v)(t'_2 - t'_1 + t'_5 - t'_4) $ which can become arbitrarily
larger than the time lapse $ t'_6 $ at $ B $.
From results in Section 4, we conclude that
the twin paradox is not a paradox, but theoretically it is wrongly
produced due to the invalidity of the Lorentz transformation.

\vspace{2ex}

{\bf Appendix 3}  \hspace{5mm}
We shall present the following remarks.

(1) Scientists often talk about the (present) age of the universe.
It is not
an iertial system that ages, but our universe itself continues aging.
Now consider any inertial system $ K' $, and the travelling of
an imaginary signal $ S(K',u_2) $ with $ u_2 >0 $ being very large
from point $ S $ to point $ R $ both being at rest over $ K' $ as
in the proof of TI Theorem.  If $ S(K',u_2) $ goes back into a past hystory
over $ K' $, then this means that $ S(K',u_2) $ goes back into a past
time $ t_0 $ at $ R $ in our universe.  But this must be impossible
since the time $ t_0 $ at $ R $ in our universe already disappeared.
In the above travelling of
$ S(K',u_2) $, even when $ u_2 $ is arbitrarily large, $ u_2 $ is still
finite and $ S(K',u_2) $ reaches a very far point $ a_0 $ over $ K' $
within a second, but the time at $ a_0 $ in our universe
when $ S(K',u_2) $ arrives at $ a_0 $ must be greater than the
time at $ R $ in our universe when signal $ S(K,u_1)
$ is emitted from $ R $ over $ K $.
As a Gedankenexperiment,
we can imagine the travelling of $ S(K',u_2) $ and conclude that our
universe continues aging and consists of the entire only one connected
universe,
and the travelling of $ S(K',u_2) $ needs a positive time lapse in our
universe.  We also note that we have not observed any time period
in our universe which was repeated.

(2) Consider Andromeda galaxy whose distance from the earth is acknowledged
to be about $ 2.5 \times 10^6 $ light years.  We can observe Andromeda
galaxy, but actually we receive light signals emitted from Andromeda
galaxy $ 2.5 \times 10^6 $ years ago. (We denote this time by $ t_0 $
as the corresponding universe time $=$ the global time in Section 5).
This means that Andromeda galaxy
at time $ t_0 $ already disapeared in our universe, and at the present
time of the universe, Andromeda galaxy of the present time $ t_p $
exists there.  Time $ t_p $ may be about $ t_0 + 2.5 \times 10^6 $
years.  Let $ K $ be an inertial system wich may be admitted to exist
over the solar system.
Then if we send imaginary signal $ S(K,u_1) $ with $ u_1 $
being arbitrarily large, signal $ S(K,u_1) $ will arrive at the point $
P $ in the universe, at which Andromeda existed at time $ t_0 $,
at time $ t_p + \alpha $, $ \alpha >0 $, and it is impossible for
$ S(K,u_1) $ to arrive at $ P $ before any time $ t_1 \leq t_p $.
Signal $ S(K,u_1) $ cannot go into any past hystory in the universe even if $
u_1 $ is arbitrarily large.  The earth of yesterday does not exist
over any synchronized coordinate system of today.

(3) Assume that in Proposition 3, it holds $ \alpha_2 \alpha_4 \neq 0
$.  Let $ E_1,E_2, E_3 $ and $ E_4 $ be the travelling of
imaginary signals $ S(K,u_1) $ and $ S(K',u_2) $ between $ R $
and $ S $ as in Definition 1, where $ u_1 $
and $ u_2 $ are sufficiently large.  Now consider events $ 
k(E_2 + E_3) + k(E_4 + E_1) $ and $k(E_1 + E_4) + k(E_3 + E_2)
$ for $ k \geq 1 $ (recall the third proof of (3) of TI Lemma).  
Then the more $ k,u_1 $ and $ u_2 $ are large,
the more vastly imaginary signals can go into a future (can go
into a past, respectively) while the event $ k(E_2 + E_3)
+k(E_4 + E_1) $ occurs (while the event $ k(E_1 + E_4) + 
k(E_3 + E_2) $ occurs, respectively).  
In the event $ k(E_1 + E_4) + k(E_3 + E_2)$ for large $ k $,
imaginary signals can go into the time when the Big Bang occured
under the assuption that the Big Bang really occured.
This means that the universe in
which event $ k(E_2 + E_3) + k(E_4 + E_1) $ occurs (or event $k(E_1
+ E_4) + k(E_3 + E_2) $ occurs) must be distinct from the 
universe where we live.  Thus if TI Lemma does not hold, then there
should exist our infinitely many universes, and for any person $ B $,
"all the replicas" of $ B $ from his birth to his death would always
exist somewhere in the universe which we conclude is impossible.  

(4) We observe the sun, stars, galaxies and great walls whose distances
from the earth range in a very large scope.  We acknowledge
this implies that we
have only one connected universe, this unique universe has continued aging
from  a very old past to the present and the time proceeds from the past
to the future, and not in the backward way.  We acknowledge that the
travellings of imaginary signals $ S(K,u_1) $ and $ S(K',u_2) $ above with
$ u_1 $ and $ u_2 $ being arbitrarily large occur in our unique
universe and cannot go into other universes since they occur over
inertial systems $ K $ and $ K' $ which exist in our universe.

\vspace{2ex}

{\bf Appendix 4.} \hspace{5mm}
We acknowledge that 
the proof of TI Lemma is valid.
But moreover we can observe the following odd properties of the Lorentz
transformations and those transformations presented in Proposition 3
with $\alpha_2 \neq 0$.
We first consider the Lorentz transformation case.  Recall the
proof of TI Theorem for the assumption $\alpha_2 \neq 0 $.
When $ v=0 $, one cannot observe that
any imaginary signal $ S(K',u_2) $ goes into a past history over
$ K' $.  Thus one may expect that the larger $ v $ is, the more
deeply imaginary signal $ S(K',u_2) $ over $K'$ with $ u_2 $ being
sufficiently large can go into a past history over $ K' $.  This
is exactly true as we can see in the following.
Let $ K $
and $ K' $ be two parallel coordinate systems 
with relative constant x'-axis velocity $ v \geq 0$ such that the
space-time transformation from $ K $ onto $ K' $ is the Lorentz
transformation.  Let $H=<K,C>$ and $J=<K',D>$ be the cc systems of
$K$ and $K'$, respectively.
Let $ R $ and $ S $ be two points at rest
on the x'-axis of $ K' $ whose x'-coordinates are
$ x'_1 $ and $ x'_2 $ with $ l= x'_2 - x'_1 >0 $.
As in the proof of TI Theorem, we consider travellings of
imaginary signals from $ R $ to $ S $ and from $ S $ to $ R $.
Consider time $ t>0 $ over $ K $, and assume that the origin $
O $ of $ K $ coincides with $ R $ at time $ t $ on
$ C(O) $, and imaginary signal $ S(K,u_1) $ over $K$ is
emitted from $ O $ to $ S $ at time $ t $ on $ C(O) $,
and arrives at $ S $ at time $ t+a $ on $ C(Q) $
with $ a $ being very small, where $ Q $ is the point
at rest on the x-axis of $ K $ which coincides with $ S $
when $ S(K,u_1) $ arrives at $ S $.  Let $ t'_1 $
and $ t'_2 $ be times on $ D(R) $ and $ D(S) $, respectively, such that
$ O $ and $ R $ coincides at time $ t'_1 $ on $ D(R) $,
and $ Q $ and $ S $ coincide at time $ t'_2 $ on $ D(S) $.
One can see easily that the x-coordinate $ x_1 $ of $ Q 
$ equals $ \beta(v)(l+ v(t'_2 - t'_1))$.
Now assume that imaginary signal
$ S(K',u_2) $ over $K'$ is emitted from $ S $ at time $ t'_2 $ on 
$ D(S) $ and arrives at $ R $ at time $ t'_3 $ on $
D(R) $, where $ u_2 >0 $ is sufficiently large w.r.t. $ l $.
By the Lorentz transformation, the following hold :

$ t'_1 = \beta(v)t, \; t'_2 \simeq \beta(v)(t- v \beta(v)(l+ v(t'_2 - t'_1))
/c^2), \;
 t'_3 = t'_2 + l/ u_2 \simeq t'_2 $,

$t'_2 - t'_1 \simeq \beta(v)(t- v \beta(v)(l + v(t'_2 - t'_1))/c^2)
-\beta(v)t$,

$(t'_2 - t'_1)(1+ \beta(v)^2 v^2/c^2) \simeq - \beta(v)^2 vl/c^2 $

$t'_3 - t'_1 \simeq t'_2 - t'_1 \simeq -\beta(v)^2 vl /(c^2(1+
\beta(v)^2 v^2/c^2)) = -vl/c^2 $

Then one can observe that it holds the larger $ v $ is, the larger $ t'_1
- t'_3 $ is.  This observation implies that the larger
$ v $ is, the more deeply imaginary signal $
S(K',u_2) $ can go into a past history over $ K' $.
Note also that $ \lim _{v \rightarrow c} t'_1(v) - t'_3(v) \simeq 
l/c $.  This is contradictory to our acknowledgement that
the property of imaginary signal $ S(K',u_2) $
should depend only on $ K' $ and $ u_2 $, and should be
independent from $ v $.

Now we consider the situation presented in Proposition 3.  Assume that
$ \alpha_2 \neq 0 $.  We consider the case $ \alpha_2 <0 $ as in the 
proof of TI Theorem.  The case $ \alpha_2 >0 $ can be considered
similarly.  Let $ S,R,t,O,P,a,t'_1,t'_2,t'_3,u_1,u_2,l $ be as
above.  As in the proof of TI Theorem, the following
hold :

$ t'_1 = \gamma_2 t, \; t'_2 \simeq \alpha_2(\alpha_3 l + \gamma_3(t'_2
-t'_1)+ \gamma_2 t, \;
t'_3 = t'_2 + l/ u_2 \simeq t'_2 $,

$t'_3 - t'_1 \simeq \alpha_2 \alpha_3 l/(1- \alpha_2 \gamma_3) <0 $

We need continuity of $ \alpha_2 $ so that for $ v=0 $, $\alpha_2(v)$
must be zero, and we conclude that $ \alpha_2 $ is not a constant and
a function of $ v $.  
Then $ t'_3 - t'_1 $ varies as $ v $ varies, and $t'_3 - t'_1$ depends
on $l$. 
We observe again similar contradictory phonomena
as in the above Lorentz transformation case.
We also note that in the above cases, $ l $ can be arbitrarily small.
For example, for $ l= 1 mm $, the above phenomena can be still
observed.

\vspace{2ex}

{\bf Appendix 5} \hspace{5mm}

{\bf Theorem 6} \hspace{5mm}
Let $ K $ and $ K' $ be two parallel inertial systems with relative
constant x'-axis velocity $ v>0 $.  Assume that the light velocity
over $ K $ is $ c $ independent of its direction.  Then the
velocity of a light signal over $K'$ is uniquely determined if
its direction is fixed.

{\bf Proof.}   Let $P$ and $Q$ be two points at rest on 
$K$.  Let $(x_1,y_1,z_1),(x_2,y_2,z_2)$ be the (x,y,z)-coordinates of $P$
and $Q$, respectively, and put $x_2 - x_1 = x, y_2 - y_1 = y$, and $
z_2 - z_1 = z $.
  We shall prove the assertion for the
case $x > 0$.  The other cases can be handled in the same way.
Consider time $t_1$ on $K$, and let $R$ and $S$ be two points at
rest on  $K'$ which coincide with $P$ and $Q$,
respectively, at time $t_1$ on $K$.  Assume that a light signal $L$ is 
emitted from $P$ at time $t_1$ on $K$, and arrives at $Q_1$ at time $
t_2 = t_1 + t$ $(t > 0)$ on $K$, where $Q_1$ is the point at rest
on  $K$ and $Q_1$ coincides with $S$ at time $t_2$
on $K$.  Then the direction from $P$ to $Q$ can be denoted
by $ (y/x,z/x) =(a,b) $ and the direction from $R$ to $S$ can be
denoted by $ (\alpha _2 y/\alpha _1 x,\alpha _3z/\alpha _1x)
 = (a \alpha _2 / \alpha _1, b \alpha _3 /\alpha _1) $, where $\alpha_1
 $-$ \alpha_4$ are as in Theorem 5.  Now
$t$ satisfies the following :

\[ c^2t^2 = (x + vt)^2 + y^2 + z^2 = (x + vt)^2 + a^2x^2 + b^2x^2 \]

Thus

\[ t = (xv + x\sqrt{v^2 + (1 + a^2 + b^2)(c^2 - v^2)})/(c^2 - v^2) \]

Now let $t'$ be the corresponding time lapse on $K'$ within
which $L$ travels from $R$ to $S$.  Then

\[ t' = \alpha _4 t = \alpha _4 x(v + \sqrt{v^2 + (1 + a^2 + b^2)(c^2 - v^2)})
   (c^2 - v^2) \]
   
Now the distance $l'$ from $R$ to $S$ satisfies

\[ l' = \sqrt{\alpha _1 ^2x^2 + \alpha _2 ^2y^2 + \alpha _3^2z^2} = 
   x\sqrt{\alpha _1 ^2 + a^2 \alpha _2 ^2 + b^2 \alpha _3^2} \]
   
Thus the velocity $c'$ of $L$ for travelling from $R$ to $S$
on $K'$ is

\[ c' = l'/t' = \sqrt{\alpha _1 ^2 + a^2 \alpha _2 ^2 + b^2 
   \alpha _3^2}(c^2 - v^2)/
   (\alpha _4 (v + \sqrt{v^2 + (1 + a^2 + b^2)(c^2 - v^2)})) \]
   
Thus $c'$ is uniquely determined by $\alpha _1,\alpha _2,\alpha _3
,\alpha _4,v,a$ and $b$.  $\Box$

\vspace{2ex}

Theorem 6 implies that an observer at rest on $K'$ can determine his
velocity w.r.t. the universe (or at least w.r.t. a local space)
by observing the velocities of light
signals into many directions over $K'$ if he knows equations for determining
constants $\alpha
_i, 1 \leq i \leq 4 $.  In Section 5, we present  Subprinciples
1-3 from which one may determine the values of $\alpha_i, 1 \leq i \leq 4 $.

\vspace{2ex}

\end{document}